\documentclass[12pt,prd,floatfix,nofootinbib,a4paper,superscriptaddress]{revtex4-2}
\pdfoutput=1

\usepackage{amsmath, amssymb, amsfonts, amsthm, latexsym, epsfig, mathrsfs, xcolor, bbm, slashed, braket, cancel}

\usepackage[inline]{enumitem}

\usepackage{setspace}
\usepackage[marginal, multiple]{footmisc}

\usepackage[T1]{fontenc}
\usepackage[utf8]{inputenc}
\usepackage{lmodern}

\usepackage[colorlinks, allcolors=blue!70!black, linktocpage]{hyperref}

\numberwithin{equation}{section}

\usepackage{cleveref}

\usepackage{microtype}

\usepackage{floatrow}
\let\OLDtableofcontents\tableofcontents
\renewcommand\tableofcontents[1]{%
    {\baselineskip 0.5ex %
	\OLDtableofcontents{#1}}%
}

\let\OLDthebibliography\thebibliography
\renewcommand\thebibliography[1]{%
	\setstretch{1.079} 
	\OLDthebibliography{#1}%
	\small %
	\setlength{\itemsep}{0.2\baselineskip} 
}

\let\OLDfootnote\footnote
\renewcommand\footnote[1]{%
	\setlength{\footnotesep}{0.75\baselineskip}%
	{\footnotesize \OLDfootnote{#1}}%
}

\usepackage{enumitem}

\setlist[enumerate]{itemsep=2pt, label=(\arabic*), ref=(\arabic*)}

\newlist{condlist}{enumerate}{2}
\setlist[condlist,1]{noitemsep, topsep=0pt, label=(\arabic*), ref=(\arabic*)}
\setlist[condlist,2]{noitemsep, label=(\alph*), ref=(\arabic{condlisti}.\alph*)}
\crefname{condlisti}{condition}{conditions}
\crefname{condlistii}{condition}{conditions}

\newlist{propertylist}{enumerate}{1}
\setlist[propertylist,1]{noitemsep, topsep=0pt, label=(\arabic*), ref=(\arabic*)}
\crefname{propertylisti}{Property}{Properties}

\renewcommand\thesection{\arabic{section}}
\renewcommand\thesubsection{\arabic{subsection}}

\makeatletter
\def\p@subsection{\thesection.}
\def\p@subsubsection{\thesection.\thesubsection.}
\makeatother 

\theoremstyle{plain}
\newtheorem{thm}{Theorem}

\theoremstyle{definition}

\theoremstyle{remark}

\crefname{section}{sec.}{sec.}
\crefname{appendix}{Appendix}{Appendices}
\crefname{table}{Table}{Tables}

\crefname{definition}{Def.}{Defs.}
\crefname{prop}{Prop.}{Props.}
\crefname{lemma}{Lemma}{Lemmas}
\crefname{corollary}{Cor.}{Cors.}
\crefname{thm}{Theorem}{Theorems}
\crefname{remark}{Remark}{Remarks}

\crefname{ass}{Assumptions}{Assumptions}
\crefname{property}{Properties}{Properties}

\newcommand{\be}{\begin{equation}\begin{aligned}}
\newcommand{\ee}{\end{aligned}\end{equation}}

\newcommand{\lb}{\left}
\newcommand{\rb}{\right}

\newcommand{\lra}{\leftrightarrow}

\newcommand{\mc}{\mathcal}

\newcommand{\ms}{\mathscr}
\newcommand{\mf}{\mathfrak}
\newcommand{\bb}{\mathbb}

\renewcommand{\emptyset}{\varnothing}

\newcommand{\eqsp}{\, ,\quad} 

\definecolor{indigo(dye)}{rgb}{0.0, 0.25, 0.42}

\newcommand{\Lie}{\pounds} 
\newcommand{\defn}{\mathrel{\mathop:}=} 

\newcommand{\union}{\cup} 

\newcommand{\norm}[1]{\lb\Vert\, #1 \,\rb\Vert}		

\newcommand{\inn}{\textrm{in}}

\newcommand{\out}{\textrm{out}}

\newcommand{\pb}[2]{ \big\{#1, #2 \big\}  }

\newcommand{\op}[1]{\boldsymbol{#1}}
\newcommand{\1}{\op{1}}

\newcommand{\Alg}{\mathscr{A}}

\newcommand{\Algex}{\mathscr{A}_{\rm in, Q}}

\newcommand{\Algexqp}{\mathscr{A}_{\rm in, QP}}

\newcommand{\cop}{\op{\mc{Q}}}

\newcommand{\currop}{\op{\mc{J}}}

\newcommand{\s}{\omega}

\newcommand{\Nalg}{\mathscr{A}^{\GR}}

\newcommand{\Hilb}{\mathscr{H}}

\newcommand{\antiHilb}{%
\hspace{4pt} 
  \vbox{%
    \hrule height 0.5pt
    \kern0.25ex
    \hbox{%
      \kern-0.3em
      \ifmmode\Hilb\else\ensuremath{\Hilb}\fi
      \kern0em
    }
  }
}

\newcommand{\Fock}{\mathscr{F}}

\newcommand{\EM}{\textrm{EM}}

\newcommand{\KG}{\textrm{KG}}

\newcommand{\YM}{\textrm{YM}}

\newcommand{\GR}{\textrm{GR}}

\newcommand{\Aut}{\mathfrak{a}}

\newcommand{\dS}{d\Omega}

\newcommand{\LGT}{\lambda}

\let\oldsetminus\setminus
\renewcommand{\setminus}{\!\oldsetminus\!} 

\newcommand{\td}[2][]{\frac{d #1}{d #2}} 

\let\oldint\int
\renewcommand{\int}{\oldint\limits}

\let\oldlim\lim
\renewcommand{\lim}{\oldlim\limits}

\renewcommand{\bar}{\overline}

\newcommand{\hyp}{\mathcal{H}}

\newcommand{\scri}{\ms I}

\newcommand{\nfrac}[2]{{{}^#1\!\!/\!_#2}}

\newcommand{\half}{\nfrac{1}{2}}

\begin{document}
\setstretch{1.2}


\title{Infrared Finite Scattering Theory in Quantum Field Theory and Quantum Gravity}


\author{Kartik Prabhu}
\email{kartikprabhu@ucsb.edu}
\affiliation{Department of Physics, University of California, Santa Barbara, CA 93106, USA.}

\author{Gautam Satishchandran}
\email{gautamsatish@uchicago.edu}
\affiliation{Enrico Fermi Institute and Department of Physics\\ The University of Chicago\\ 5640 South Ellis Avenue, Chicago, IL 60637, USA.}

\author{Robert M. Wald}
\email{rmwa@uchicago.edu}
\affiliation{Enrico Fermi Institute and Department of Physics\\ The University of Chicago\\ 5640 South Ellis Avenue, Chicago, IL 60637, USA.}

\begin{abstract}
It has been known since the earliest days of quantum field theory (QFT) that infrared divergences arise in scattering theory with massless fields. These infrared divergences are manifestations of the memory effect: At order $1/r$ a massless field generically will not return to the same value at late retarded times ($u \to + \infty$) as it had at early retarded times ($u \to - \infty$). There is nothing singular about states with memory, but they do not lie in the standard Fock space. Infrared divergences are merely artifacts of trying to represent states with memory in the standard Fock space. If one is interested only in quantities directly relevant to collider physics, infrared divergences can be successfully dealt with by imposing an infrared cutoff, calculating inclusive quantities, and then removing the cutoff. However, this approach does not allow one to treat memory as a quantum observable and is highly unsatisfactory if one wishes to view the $S$-matrix as a fundamental quantity in QFT and quantum gravity, since the $S$-matrix itself is undefined. In order to have a well-defined $S$-matrix, it is necessary to define ``in'' and ``out'' Hilbert spaces that incorporate memory in a satisfactory way. Such a construction was given by Faddeev and Kulish for quantum electrodynamics (QED) with a massive charged field. Their construction can be understood as pairing momentum eigenstates of the charged particles with corresponding memory representations of the electromagnetic field to produce states of vanishing large gauge charges at spatial infinity. (This procedure is usually referred to as ``dressing'' the charged particles.) We investigate this procedure for QED with massless charged particles and show that, as a consequence of collinear divergences, the required ``dressing'' in this case has an infinite total energy flux, so that the states obtained in the Faddeev-Kulish construction are unphysical. An additional difficulty arises in Yang-Mills theory, due to the fact that the ``soft Yang-Mills particles'' used for the ``dressing'' contribute to the Yang-Mills charge-current flux, thereby invalidating the procedure used to construct eigenstates of large gauge charges at spatial infinity. We show that there are insufficiently many charge eigenstates to accommodate scattering theory. In quantum gravity, the analog of the Faddeev-Kulish construction would attempt to produce a Hilbert space of eigenstates of supertranslation charges at spatial infinity. Again, the Faddeev-Kulish ``dressing'' procedure does not produce the desired eigenstates because the dressing contributes to the null memory flux. We prove that there are no eigenstates of supertranslation charges at spatial infinity apart from the vacuum. Thus, analogs of the Faddeev-Kulish construction fail catastrophically in quantum gravity. We investigate some alternatives to Faddeev-Kulish constructions but find that these also do not work. We believe that if one wishes to treat scattering at a fundamental level in quantum gravity --- as well as in massless QED and Yang-Mills theory --- it is necessary to approach it from an algebraic viewpoint on the ``in'' and ``out'' states, wherein one does not attempt to ``shoehorn'' these states into some pre-chosen ``in'' and ``out'' Hilbert spaces. We outline the framework of such a scattering theory, which would be manifestly infrared finite.
\end{abstract}

\maketitle
\tableofcontents

\section{Introduction}
\label{sec:intro}

The seminal work of Lehmann, Symanzik and Zimmerman (LSZ) \cite{LSZ}, Haag and Ruelle \cite{PhysRev.112.669,Ruelle1962}, and others established that conventional scattering theory should be well-defined in the case of massive quantum fields. In particular, for massive fields, it should be possible to obtain a unitary $S$-matrix relating the standard ``in'' and ``out'' Fock spaces of asymptotic states. However, in four spacetime dimensions, when one has massless quantum fields, one encounters severe difficulties in carrying out this program \cite{mott_1931,Sommerfeld_1931,fierz,Bloch_1937}. Classical massless fields that interact with massive fields or undergo suitable self-interactions will generically undergo a {\em memory effect} wherein, at order $1/r$ in null directions, the field at late retarded times will not return to the value it had at early retarded times.\footnote{In spacetime dimension $d$, the memory effect occurs at Coulombic order, $1/r^{d-3}$, whereas radiation decays as $1/r^{d/2 -1}$ \cite{Satishchandran_2019}. For $d=4$, both occur at order $1/r$, so memory directly affects the quantization of the ``in'' and ``out'' radiation. For $d > 4$, the memory effect does not lead to infrared divergences in the quantized radiation. The discussion of this paper is restricted to $d=4$.} Thus, at order $1/r$, the Fourier transform of a solution with memory will diverge as $1/\omega$ at low frequencies. In the quantum theory, the one-particle norm of the positive frequency part of such a solution is infinite. Consequently, if one tries to express a quantum state corresponding to a classical solution with memory as a vector in the standard Fock representation, it will have an infinite number of ``soft'' (i.e. arbitrarily low frequency) massless quanta and its norm will be “infrared divergent.'' In other words, although states with memory are entirely legitimate quantum field states that necessarily arise in scattering processes, they cannot be accommodated in the standard Fock space. Consequently, the $S$-matrix cannot be defined as a map taking ``in'' states in the standard Fock representation to ``out'' states in the standard Fock representation, and infrared (IR) divergences will arise if one attempts to do so. 

The most common way of dealing with such infrared divergences is to initially impose an infrared cutoff (so that the ``out'' state can be expressed as an ordinary Fock space vector), calculate inclusive processes that sum over all possible states of the low frequency massless quanta in the cutoff state, and then remove the cutoff \cite{Bloch_1937,PhysRev.140.B516,Yennie:1961ad}. As a practical matter, this procedure works quite successfully if one is interested in obtaining typical quantities of direct relevance for accelerator experiments, such as (inclusive) cross-sections for the scattering of ``hard'' particles. However, the infrared cutoff removes the memory effect, so one cannot even ask questions about memory as a quantum observable, as has been of particular recent interest (see e.g. \cite{Hawking:2016msc,Strominger:2017zoo,Ashtekar:2018lor} and references therein). Furthermore, even if one is interested only in ``hard'' particles, this approach cannot properly deal with issues such as the entanglement of ``hard'' and ``soft'' particles, which should result in decoherence of the ``hard'' particles \cite{Carney_2017,Semenoff_2019}. More significantly, this approach is highly unsatisfactory if one wishes to view the $S$-matrix as a fundamental quantity in the formulation of quantum field theory and quantum gravity, since the $S$-matrix itself is undefined.\footnote{We note that there are at least two notions of an ``infrared finite'' $S$-matrix in the literature. The notion that we are concerned with in this paper is to construct appropriate Hilbert spaces of ``in'' and ``out'' states and to obtain the $S$-matrix as a well-defined map between these Hilbert spaces. An alternative notion is to develop a procedure for rendering the standard (infrared divergent) $S$-matrix amplitude finite (see e.g. \cite{Hannesdottir_2019,Hannesdottir:2019rqq}). While such a procedure then can be used to calculate ``inclusive quantities'' or determine formal properties of the $S$-matrix amplitudes \cite{Kapec:2017tkm,Choi:2017ylo}, there is no actual ``out'' state (with memory) constructed by this procedure.}

In order have a well-defined $S$-matrix, it clearly is necessary to construct Hilbert spaces of ``in'' and ``out'' states such that the ``in'' states evolve to the ``out'' states. As we have just indicated, this is not the case if one takes the ``in'' and ``out'' Hilbert spaces to be the standard Fock spaces, since a generic state ``in'' Fock space state will evolve to an ``out'' state with a nonvanishing probability for nonzero memory, which cannot be accommodated in the ``out'' Fock space. Thus, if we wish to have a well-defined $S$-matrix, we must make alternative choices of the ``in'' and ``out'' Hilbert spaces that contain states with nonvanishing memory. In order to have a satisfactory scattering theory, these ``in'' and ``out'' Hilbert space constructions should satisfy the following properties:
\begin{enumerate}
\item
The ``in'' and ``out'' Hilbert spaces are obtained by the ``same construction.'' More precisely, if we identify the algebra of ``out'' field observables with the algebra of ``in'' field observables via a change of the time orientation of the bulk spacetime, we require that the ``in'' and ``out'' Hilbert space representations of these algebras be unitarily equivalent.  \label{enum:S1}

\item
Dynamical evolution maps all ``in'' states to ``out'' states and vice-versa, so that one has a unitary $S$-matrix. \label{enum:S2}

\item
The ``in'' and ``out'' Hilbert spaces should admit a natural, continuous action of the Poincar\'e group.\footnote{In the gravitational case we require that the ``in'' and ``out'' Hilbert spaces should admit a natural, continuous action of the BMS group.} \label{enum:S3}

\item
The ``in'' Hilbert space should be large enough to contain incoming states representing all ``hard'' scattering processes. \label{enum:S4}

\item
The ``in'' and ``out'' Hilbert spaces should be separable, so that they are not ``too large''.\footnote{A nonseparable Hilbert space was previously studied in \cite{Kibble1,Kibble2,Kibble3,Kibble4} which considered the direct sum over all memory representations. We discuss the deficiencies of this direct sum in \cref{sec:NKFreps}.} \label{enum:S5}

\end{enumerate}

For the case of quantum electrodynamics (QED) with a massive charged field, a satisfactory construction of ``in'' and ``out'' Hilbert spaces was given many years ago by Faddeev and Kulish \cite{Kulish:1970ut} based on the earlier work of \cite{Dollard,Chung_1965,Greco:1967zza}. However, the main purpose of our paper is to show that a similar construction does {\em not} work in a satisfactory way for QED with massless charged particles and for Yang-Mills theory. Furthermore, we will show that such a construction does not work at all in quantum gravity. We argue that in these cases, at a fundamental level, scattering theory should be formulated at the level of algebraic states, without attempting to ``shoehorn'' all the states into a single, separable Hilbert space.

Since many of our arguments and constructions require a considerable amount of technical machinery, we now provide a brief sketch of all of the key results of the paper, so that a reader can obtain the gist of our arguments without having to delve into the details that we will provide in due course in the body of the paper. We begin by describing the Faddeev-Kulish construction for QED with a massive charged field. In order to understand the relevant ingredients of their construction, it is necessary to reformulate it in the language of the memory effect and the related symmetries and charges. In this section, for ease of explanation, we will work in the bulk spacetime --- introducing ``null coordinates'' $(u,r,x^A)$, where $u=t-r$ and $x^A$ denotes angular coordinates on the sphere --- and work to appropriate orders in $1/r$. However, in the remainder of this paper it will be more convenient and conceptually clearer to express both the classical and quantum theory in terms of the conformal completion of Minkowski spacetime.  

The classical memory effect at future null infinity for an electromagnetic field corresponds to having the angular components, $A^{(1)}_A(u,x^{A})$, of the vector potential at order $1/r$  asymptote to different values at early and late retarded times, $u \to \pm \infty$. Since the electric field at order $1/r$ is given by 
\be 
E^{(1)}_{A}=-\partial_{u}A_{A}^{(1)}
\ee 
it follows that there will be a nontrivial memory effect if and only if at order $1/r$ the electric field satisfies $\int_{-\infty}^{\infty}du~ E^{(1)}_A  \neq 0$. Since $\int_{-\infty}^{\infty}du~ E^{(1)}_A $ is proportional to the integrated force on a test particle placed at a large distance from the source of radiation, this fact allows one to give a physical interpretation of the memory effect in terms of a charged test particle receiving a net momentum kick at order $1/r$ due to the passage of the radiation \cite{Staruszkiewicz_1981,Bieri_2013EM}. Since we assume that $E^{(1)}_A \to 0$ at early and late retarded times, $A^{(1)}_A$ is ``pure gauge'' at early and late retarded times, but the electromagnetic memory
\be
\label{eq:memE}
\Delta^{\out}_A \defn - \int_{-\infty}^{\infty} du~E^{(1)}_{A}  =  A^{(1)}_A|_{u=+\infty} - A^{(1)}_A|_{u=-\infty}
\ee 
is gauge invariant, as is obvious from the fact that it is given by an integral of the electric field. In \cref{eq:memE}, we have appended the superscript ``out'' to $\Delta^{\out}_A$ to distinguish the electromagnetic memory of the outgoing radiation from the electromagnetic memory, $\Delta^{\inn}_A$, of incoming radiation. 

The gauge transformations relevant for changing the angular components of the vector potential at order $1/r$ are the so-called ``large gauge transformations''
\be
A_\mu \to A_\mu + \nabla_\mu \LGT
\label{largegauge}
\ee
 where $\LGT = \LGT(x^A)$ is a function of purely the angular coordinates $x^A$. Under such a gauge transformation, we have
\be
\label{eq:Agauge}
A^{(1)}_A \to A^{(1)}_A + {\mathscr D}_A \LGT
\ee
where ${\mathscr D}_A$ is the derivative operator on the unit sphere. In fact, the ``gauge transformations'' \cref{largegauge} are actually ``symmetries'' in the sense that they have nonvanishing symplectic product with other solutions, i.e., they are not degeneracies of the symplectic form. There are charges and fluxes associated with these symmetries. The charge ${\mathcal Q}_{u}(\LGT)$ associated with the symmetry $\LGT$ at retarded time $u$ is given by
\be
\label{eq:QEMl}
{\mathcal Q}_{u}(\LGT) = \frac{1}{4\pi} \int_{S(u)} d\Omega ~ \LGT(x^A) F^{(2)}_{ur} (u,x^A) 
\ee
where $S(u)\cong \bb{S}^{2}$ is an asymptotic sphere at fixed retarded time $u$, $F_{\mu \nu}$ is the electromagnetic field tensor, and the superscript ``$(2)$'' denotes the order $1/r^2$ part of the field as $r \to \infty$ at the given value of $u$. The difference of the charge ${\mathcal Q}_{u}(\LGT)$ at two retarded times $u_{1}$ and $u_{2}$ is determined by a corresponding flux between these retarded times  associated with the symmetry $\lambda$ 
\be 
\label{eq:QEMllux}
\mc{Q}_{u_{2}}(\lambda) - \mc{Q}_{u_{1}}(\lambda) = \int_{u_{1}}^{u_{2}}du \int_{\bb{S}^{2}}d\Omega~ \lambda(x^{A})\bigg(J^{(2)}_{u}(u,x^{A}) - \frac{1}{4\pi}\ms{D}^{A}E^{(1)}_{A}(u,x^{A})\bigg)
\ee 
where $J^{(2)}_{\mu}$ is the charge-current at order $1/r^2$, which can be nonvanishing only if there are massless charged fields. In the limit as $u_{1}\to -\infty$ and $u_{2}\to +\infty$, this first term corresponds to the total flux of charge-current, which we denote as 
\be 
\label{eq:JEM}
\mc{J}^{\out}(\lambda) \defn \int_{-\infty}^{\infty}du \int_{\bb{S}^{2}}d\Omega~\lambda(x^{A}) J^{(2)}_{u}(u,x^{A})
\ee
where, again, the ``out'' corresponds to the ``outgoing'' flux of massless charge-current. The second term on the right-hand-side of \cref{eq:QEMllux} in this limit is proportional to the divergence of memory $\ms{D}^{A}\Delta^{\out}_{A}(x^{A})$ smeared with $\lambda(x^{A})$ on the sphere. Finally, in this limit, the charges $\mc{Q}_{u_{2}}(\lambda)$ and $\mc{Q}_{u_{1}}(\lambda)$ approach future time-like infinity $i^{+}$ and spatial infinity $i^{0}$, respectively. Therefore, in the case where $u_{1}=-\infty$ and $u_{2}=+\infty$, \cref{eq:QEMllux} yields that the charges given by \cref{eq:QEMl} and the flux of null charge-current given by \cref{eq:JEM} are related to the electromagnetic memory by\footnote{\label{magpar} Memory can be decomposed into electric and magnetic parity parts via $\Delta_{A}=\ms{D}_{A}\alpha + \epsilon_{A}{}^{B}\ms{D}_{B}\beta$. \Cref{memch} only involves the electric part, $\alpha$. Magnetic parity charges similar to \cref{eq:QEMl} can be defined (although they are not associated with large gauge transformations) and an analog of \cref{memch} (without the current term) then holds. The constructions of our paper could be straightforwardly extended to include magnetic parity memory. However, magnetic parity memory does not arise in usual scattering processes starting with states of vanishing memory --- although it can occur in certain processes (see \cite{Satishchandran_2019} for an example in the gravitational case). We will focus entirely on electric parity memory in this paper.}
\be\label{memch}
\frac{1}{4\pi}\int_{\bb{S}^{2}}d\Omega~\Delta^{\out}_{A}(x^{A})\ms{D}^{A}\lambda = \mc{Q}_{i^{+}}(\lambda) - \mc{Q}_{i^{0}}(\lambda)+ \mc{J}^{\out}(\lambda)
\ee 
where the charges are defined as limits as $u\to \pm \infty$. The difference of charges on the right hand side of \cref{memch} is known as the ``ordinary memory effect'' and the contribution due to the total charge-current flux of massless charged fields is known as the ``null memory effect'' \cite{Bieri_2013EM}. 

Similar charges and fluxes associated with the symmetry $\LGT$ can be defined at past null infinity, wherein we replace retarded time $u$ in the above formulas by advanced time $v = t + r$. In a scattering situation, there is, in general, no direct relation between the memory $\Delta^{{\rm out}}_A$ of the electromagnetic field at future null infinity and the memory $\Delta^{{\inn}}_A$ at past null infinity. Indeed, as we have already indicated, if the incoming electromagnetic field has vanishing memory, the outgoing electromagnetic field will generically have nonvanishing memory. However, there is a matching of the incoming and outgoing charges as one approaches spatial infinity \cite{Strominger:2013jfa,CE,Henneaux:2018gfi,KP-EM-match,Mohamed_2021}. Specifically, we have
\be
 {\mathcal Q}^{\rm out}_{i^{0}}(\LGT) = {\mathcal Q}^{\inn}_{i^{0}}(\LGT \circ \Upsilon)
\label{chcons}
\ee
where we have used ``in/out'' to denote that the limit is taken from past/future null infinity to spatial infinity and  $\Upsilon$ is the antipodal map on a sphere, so that $(\LGT \circ \Upsilon) (\theta, \varphi) = \LGT(\pi - \theta, \varphi + \pi)$. 

The ``conservation law'' \cref{chcons} is the key to enabling one to define ``in'' and ``out'' Hilbert spaces satisfying conditions \ref{enum:S1}--\ref{enum:S5} in the case of QED with massive charged fields. If we restrict all of the incoming states to have fixed, definite large gauge charges at spatial infinity for all $\LGT$, then the outgoing states will have the corresponding charges given by \cref{chcons}. Hence, if we can construct ``in'' and ``out'' Hilbert spaces of definite values of all charges at spatial infinity, it should be possible to satisfy properties \ref{enum:S1} and \ref{enum:S2} above. However, since these charges are not invariant under Lorentz transformations, it will not be possible to have the Poincar\'e group have a continuous action on a space of incoming or outgoing states of definite charges except in the case where all charges (including the ordinary total electric charge) vanish\footnote{One could start with a Hilbert space with nonvanishing charges and obtain a new space that admits an action of the Poincar\'e group by taking the direct sum of the continuous family of Hilbert spaces with charges equal to the action of the Lorentz group on the original charges. However, this direct sum Hilbert space would be non-separable. Furthermore, there would be no infinitesimal action of the Lorentz group on the direct sum Hilbert space, so, in particular, the angular momentum operator would not be defined \cite{Buchholz86,Gervais_1980}.} at spatial infinity \cite{Gervais_1980,Buchholz86,Frohlich:1978bf,Frohlich:1979uu}. Thus, in order to satisfy property \ref{enum:S3}, we restrict the incoming states --- and, therefore, the outgoing states by \cref{chcons} --- to have vanishing charges at spatial infinity. It may appear that the requirement of vanishing total electric charge will violate condition \ref{enum:S4} of our above requirements on the ``in'' Hilbert space, since it will allow us to consider only scattering processes with an equal number of charged particles and antiparticles. However, if we wish to consider the scattering of, say, two electrons, we can simply add two positrons ``behind the moon'', i.e., in incoming states which do not interact significantly with the electrons or with each other \cite{Frohlich:1979uu}. Thus, arguably, the restriction to states of vanishing charges does not preclude having representatives of all ``hard'' scattering processes. 

As discussed in more technical detail in \cref{sec:MaxMKG}, a separable Hilbert space of ``in'' and ``out'' states of vanishing charges for QED with massive charged particles can be constructed as follows. For the construction of the ``in'' Hilbert space, we note that the charges at past timelike infinity, $ {\mathcal Q}_{i^{-}}(\LGT)$, are determined by the incoming state of the massive charged particles. The (improper) incoming Fock space state $|p_1 \dots p_n; q_1 \dots q_n \rangle$ consisting of $n$ incoming charged particles and $n$ incoming antiparticles with definite momenta $p_1, \dots, p_n$ and $ q_1, \dots, q_n$, respectively, has vanishing total ordinary electric charge and can be seen to be an eigenstate of the charge operator $ \op{{\mathcal Q}}_{i^{-}}(\LGT)$ for all $\LGT$. We denote its eigenvalue as $\mc{Q}_{i^{-}}(\LGT;p_{1}\dots q_{n})$. By the corresponding version of \cref{memch} for past null infinity with ${\mathcal J}^{\inn}=0$ (since we are considering only massive charged particles) we can obtain an (improper) state for which all large gauge charges vanish at spatial infinity ($v \to + \infty$) by pairing $|p_1 \dots p_n; q_1 \dots q_n \rangle$ with any incoming electromagnetic field state that lies in the representation with memory $\Delta^{\inn}_A$ determined by\footnote{This relation uniquely determines the electric parity memory (see \cref{magpar}).} 
\be\label{memch0}
\frac{1}{4\pi}\int_{\bb{S}^{2}} d \Omega~ \Delta^{\inn}_A \ms{D}^A \LGT  = - {\mathcal Q}_{i^{-}}(\LGT;p_{1},\ldots, q_n) \,
\ee
for all $\LGT(x^{A})$. In other words, we can take the tensor product of the one-dimensional Hilbert space spanned by the improper state $|p_1 \dots p_n; q_1 \dots q_n \rangle$ with the Fock space representation of the electromagnetic field with memory $\Delta^{\inn}_A $ given by \cref{memch0}. The pairing of the charged particle state $|p_1 \dots p_n; q_1 \dots q_n \rangle$ with electromagnetic states in the representation $\Delta^{\inn}_A $ is usually referred to as ``dressing'' the charged particles with a corresponding ``cloud of soft photons.''\footnote{Note that one does not have to ``dress'' the charged particles with a specific state , i.e., \emph{any} electromagnetic state with the required memory is allowed. The ``cloud of soft photons'' refers to any state in the representation with the required memory. A charged particle ``dressed'' with an infrared ``cloud of soft photons'' is sometimes referred to as an ``infraparticle'' \cite{Schroer_1963,Buchholz86}.}  We can then obtain a Hilbert space with arbitrary proper Fock space states of $n$ charged particles, $n$ antiparticles and arbitrary ``hard'' photon states by taking a direct integral over $p_1, \dots, p_n, q_1, \dots, q_n$. We then take the direct sum over $n$. This yields a separable Hilbert space that has representatives of all incoming states of the massive charged particles with vanishing total electric charge and all incoming ``hard'' photon states. This construction is equivalent to the one given by Faddeev and Kulish. All states in this ``in'' Hilbert space are eigenstates with eigenvalue $0$ of all of the large gauge charges at spatial infinity.\footnote{The relationship between the Faddeev and Kulish ``dressed states'' and eigenstates of $\cop_{i^{0}}(\lambda)$ has been previously discussed in \cite{Kapec:2017tkm,narain81}.} By conservation of charges at spatial infinity, these states should evolve to states in the similarly constructed ``out'' Hilbert space. Indeed, finiteness of the Faddeev-Kulish ``S-matrix amplitudes'' has been verified to all orders in perturbation theory \cite{Gabai:2016kuf}. These results are also supported by recent, rigorous analyses of perturbative QED \cite{Duch_2021} as well as non-perturbative studies of nonrelativistic QED \cite{Chen_2009_II,Chen_2009_I}.\footnote{Similar analyses have also been done in the case of infrared divergences arising from the scattering of a massless scalar field coupled to a massive scalar field (sometimes referred to as the ``Nelson model'')  \cite{Frolich_1973,Frolich_1974,Pizzo_2003}. } Consequently, all of the above properties \ref{enum:S1}--\ref{enum:S5} should be satisfied.

Thus, the Faddeev-Kulish construction provides definitions of ``in'' and ``out'' Hilbert spaces that enable one to have a well-defined $S$-matrix.\footnote{As has been pointed out in \cite{Dybalski_2017} the original Faddeev-Kulish construction did not include the ``virtual photons'' associated to the Coulomb fields of the outgoing electrons. In this paper, the "Coulomb fields" of any incoming/outgoing particles are automatically included through the constraints arising from Maxwell's equations at past/future timelike infinity (see \cref{subsec:chmem}).}
However, it should be noted that this construction has a number of unpleasant features. First, it allows only states of vanishing total ordinary electric charge. As already mentioned above, this can be dealt with by putting any excess charges ``behind the moon.'' A more unpleasant feature is that it requires the incoming massive charged particles to be ``dressed'' with incoming electromagnetic states with the corresponding memory. This dressing is quite unnatural, since --- although the incoming massive charged field and incoming electromagnetic radiative field are completely independent degrees of freedom --- it requires the incoming electromagnetic radiative state to ``know'' the exact state of the incoming charged field. Furthermore, since each state in the Faddeev-Kulish Hilbert space has an extremely high degree of entanglement between the state of the massive charged field and the state of the electromagnetic field, one cannot have a coherent superposition of incoming charged particle states of different momenta \cite{Carney_2017}. Thus, the Faddeev-Kulish Hilbert space appears to artificially exclude many states that one might wish to consider. Nevertheless, by restricting consideration to the states in the Faddeev-Kulish Hilbert space, one should obtain a genuine $S$-matrix, with no infrared divergences.

We turn now to differences that occur if we consider QED with a massless charged field, as will be discussed in detail in \cref{sec:MaxZMKG}. Since there are no incoming massive particles, the charges ${\mathcal Q}_{i^{-}}(\LGT)$ at timelike infinity vanish in the massless case. However, since there are incoming massless particles, the charge-current flux ${\mathcal J}^{\inn}(\LGT)$ at null infinity will not vanish. One can perform a construction of ``in'' and ``out'' Hilbert spaces that is completely analogous to the Faddeev-Kulish construction as follows: In the massless case, the (improper) incoming Fock space states $|p_1 \dots p_n; q_1 \dots q_n \rangle$ of the charged field are eigenstates of the charge-current operator $\op{{\mathcal J}}^{\inn}(\LGT)$ for all $\lambda$. Therefore, one can again pair $|p_1 \dots p_n; q_1 \dots q_n \rangle$ with the incoming electromagnetic field states that lie in the representation with memory $\Delta^{\inn}_A$ chosen so as to give vanishing charges at spatial infinity. In this case, by \cref{memch}, the required $\Delta^{\inn}_A$ is determined by 
\be
\frac{1}{4\pi}\int_{\bb{S}^{2}}d \Omega~ \Delta^{\inn}_A \ms{D}^A \LGT  =  -{\mathcal J}^{\inn}(\LGT;p_{1},\dots,q_{n}) \, .
\label{memch1}
\ee
where $ {\mathcal J}^{\inn}(\LGT;p_{1},\dots,q_{n})$ denotes the eigenvalue of $\op{{\mathcal J}}^{\inn}(\LGT)$ in the state $|p_1 \dots p_n; q_1 \dots q_n \rangle$. By taking a direct integral over $p_1, \dots, p_n, q_1, \dots, q_n$ and a direct sum over $n$, we will again get a separable Hilbert space of states with vanishing charges at spatial infinity, so the ``in'' Hilbert space should unitarily map to the similarly constructed ``out'' Hilbert space under dynamical evolution. This yields a direct analog for massless charged fields of the Faddeev-Kulish construction for massive charged fields.

However, although the Faddeev-Kulish construction can be carried out in close analogy with the massive case, a truly significant difference arises in the nature of the resulting states. For massive charged fields, the charges $ {\mathcal Q}_{i^{-}}(\lambda;p_{1}\dots q_{n})$ for the state $|p_1 \dots p_n; q_1 \dots q_n \rangle$ correspond to a smooth function on the sphere. Consequently, the corresponding memory $\Delta^{\inn}_A$ determined by \cref{memch0} is smooth, and the corresponding memory representation of the electromagnetic field has a dense set of nonsingular states. By contrast, in the massless case, the flux ${\mathcal J}^{\inn}(\LGT;p_{1},\dots,q_{n})$ for the state $|p_1 \dots p_n; q_1 \dots q_n \rangle$ has $\delta$-function angular singularities in each of the directions $x^A_i$ of the momenta $p_1, \dots, p_n, q_1, \dots, q_n$. It follows that the memory $\Delta^{\inn}_A (x^B)$ determined by \cref{memch1} will have angular singularities of the form $1/|x^B - x^B_i|$ in the vicinity of $x_i^B$. These additional angular singularities occurring in the massless case correspond to what are referred to as \emph{collinear divergences}. If one is interested in calculating inclusive cross-sections, they merely give rise to an additional nuisance in that one must introduce a further angular cutoff in addition to the usual infrared cutoff when performing calculations \cite{K_1962,LN_1964}. But they give rise to a fatal difficulty for the usefulness of the ``in'' and ``out'' Hilbert spaces constructed above. The angular singularities in the memory are such that the memory is not square integrable over a sphere. This implies that the expected electric field $\langle E^{\inn}_A \rangle$ in any state in the memory representation paired with $|p_1 \dots p_n; q_1 \dots q_n \rangle$ cannot be square integrable over null infinity. By further arguments, it can be seen that the total energy flux of the electromagnetic field in any state in this memory representation is infinite. In other words, in the massless case, the required ``soft photon dressing'' of the charged particles always carries infinite energy. Thus, {\em all} of the allowed states of the electromagnetic field in this construction are physically unacceptable. Although we should be able to obtain a well-defined scattering theory between states in the ``in'' and ``out'' Hilbert spaces, none of the scattering states are of any physical relevance.

We now turn to Yang-Mills theory with a compact, semi-simple Lie group, which will be discussed in more detail in \cref{subsec:YM}. The Yang-Mills fields occurring in nature are strongly coupled to other fields and do not behave as free fields at asymptotically early and late times. However, we can consider the scattering theory of ``pure'' Yang-Mills theory (with no coupling to other fields) as a toy model that has features similar to both electromagnetism and gravity. Collinear divergences similar to massless QED occur in Yang-Mills theory. Consequently, as in massless QED, the ``dressing'' required by the Faddeev-Kulish construction will be singular. However, an additional --- and, in some respects, even more serious --- difficulty arises in the Yang-Mills case, due to the fact that the Yang-Mills field acts as its own source. The analog of \cref{memch} in the Yang-Mills case is
\be
\frac{1}{4\pi}\int_{\bb{S}^{2}} d \Omega~\Delta^{\! \YM, \out}_{A,j} \ms{D}^A \LGT^j  = {\mathcal Q}^{\YM}_{i^{+}}(\LGT) - {\mathcal Q}^{\YM}_{i^{0}}(\LGT) + {\mathcal J}^{\YM, \out}(\LGT)\label{memchym}
\ee
where $j$ denotes a Lie algebra index, such indices are lowered and raised with the Cartan-Killing metric, and the charges are defined by a natural generalization of \cref{eq:QEMl} where the Lie algebra valued field strength is now integrated with $\lambda^{i}(x^{A})$. Since there are no massive sources, the charges at timelike infinity vanish, ${\mathcal Q}^{\YM}_{i^{+}}(\LGT) = 0$. The charge-current flux in the Yang-Mills case is
\be
 {\mathcal J}^{\YM, \out}(\LGT) = \frac{1}{2\pi}\int_{-\infty}^{\infty} du\int_{\bb{S}^{2}}\dS~ c^{i}{}_{jk}q^{AB}\LGT_{i}A_{A}^{(1)j}E_{B}^{(1)k}
\ee
where $c^{i}{}_{jk}$ denote the structure constants of the Lie algebra. Similar charges and fluxes associated to the symmetry $\lambda$ can be defined at past null infinity. The analog of \cref{memch1} for obtaining eigenstates of vanishing charge\footnote{The charges at spatial infinity satisfy the commutation relations $\left[{\mathcal Q}_{i^{0}}^{\YM} (\LGT_{1}), {\mathcal Q}_{i^{0}}^{\YM} (\LGT_{2}) \right] = {\mathcal Q}_{i^{0}}^{\YM} ([\LGT_{1},\LGT_{2}])$ where $[\LGT_{1},\LGT_{2}]^{i}=c^{i}{}_{jk}\lambda^{j}_{1}\lambda^{k}_{2}$. For a semisimple Lie group it is impossible to have an eigenstate of all charges unless all of the charge eigenvalues vanish.} at spatial infinity for the Yang-Mills field is
\be
\frac{1}{4\pi}\int_{\bb{S}^{2}} d \Omega~\Delta^{\! \YM {\rm ,in}}_{A,j} \ms{D}^A \LGT^j  = - {\mathcal J}^{\YM, \inn}(\LGT) \, .
\label{memchym1}
\ee
The key difference with massless QED is that the ``hard'' and ``soft'' quanta now correspond to the same field. Thus, we must use ``soft'' Yang-Mills quanta to ``dress'' (via the memory, $\Delta^{\! \YM {\rm ,\inn}}_{A,j}$) the ``hard'' Yang-Mills quanta. But these ``soft'' quanta will then make additional contributions to the current flux, so we will not get an eigenstate of charges at spatial infinity by choosing the memory to satisfy \cref{memchym1}, with $\mathcal J^{\YM, \inn}(\LGT)$ the flux of the ``hard'' Yang-Mills quanta. Thus---in addition to the fact that, as in the case of massless QED, this soft ``dressing'' is singular and therefore the corresponding Yang-Mills current flux is infinite---one cannot get states of vanishing charges at spatial infinity by attempting to pair flux eigenstates with corresponding memory representations. 

Thus, in order to obtain an analog of the Faddeev-Kulish Hilbert space in the Yang-Mills case, one must find some other means to obtain a suitable Hilbert space of eigenstates of vanishing charges at spatial infinity. However, there are insufficiently many such states, as can be seen from the fact that the charge ${\mathcal Q}^{\YM}_{i^{0}} (\LGT)$ at spatial infinity acts on the ``in'' and ``out'' states as an infinitesimal generator of the large gauge transformation associated with $\LGT^j$. Thus, an ``in'' state with vanishing charges at spatial must be gauge invariant with respect to all large gauge transformations, which is a strong constraint on the $n$-point functions of the Yang-Mills electric field. In particular this implies that the $1$-point function must vanish, the $2$-point function must be proportional to the Cartan-Killing metric and, more generally, all $n$-point functions must be proportional to Casimirs of the Lie algebra. Although there exist states that satisfy these conditions, these conditions are far too restrictive to allow one to satisfy condition \ref{enum:S4} of our requirements on the ``in'' Hilbert space.

We now turn to general relativity, which will be considered in detail in \cref{sec:Grav}. We introduce Bondi coordinates $(u,r,x^A)$, and let $C_{\mu \nu}(u,x^{A})$ denote the deviation of the spacetime metric from the asymptotic Minkowski metric $\eta_{\mu \nu}$ at leading order in $1/r$ in these coordinates.\footnote{Again, we state our main results in this section in Bondi coordinates in the bulk spacetime, but in \cref{sec:Grav} we will work at null infinity in the conformally completed spacetime.} The classical memory effect at future null infinity in the gravitational case corresponds to having the angular components, $C_{AB}$, at order $1/r$ asymptote to different values at early and late retarded times, $u \to \pm \infty$. This will occur if and only if at order $1/r$ the Bondi News $N_{AB}=-\partial_{u}C_{AB}$ satisfies $\int_{-\infty}^{\infty}du~N_{AB} \neq 0$. In the gravitational case, the presence of memory physically corresponds to an array of test particles initially at rest receiving a permanent relative displacement at order $1/r$ due to the passage of gravitational radiation \cite{Zeld_Poln}. 

If the Bondi News goes to zero at early and late retarded times, $C_{AB}$ will be ``pure gauge'' at early and late retarded times, but the gravitational memory
\be
\Delta^{\! \GR,\rm out}_{AB} \defn\frac{1}{2}\int_{-\infty}^{\infty} du~N_{AB} = -\frac{1}{2}\big(C_{AB}|_{u=+\infty} -C_{AB}|_{u=-\infty}\big)
\ee 
is gauge invariant. The relevant gauge transformations in the gravitational case are the supertranslations whose infinitesimal action is given by 
\be
C_{AB} \to C_{AB} - fN_{AB} - 2\bigg( \ms{D}_A \ms{D}_B f - \frac{1}{2} q_{AB} \ms{D}^C \ms{D}_C f\bigg)
\ee
where $f = f(x^A)$ is an arbitrary function on the sphere and $q_{AB}$ is the metric on the unit sphere. The supertranslations are, in fact, ``symmetries,'' i.e., they are not degeneracies of the symplectic form. Again, there are charges and fluxes associated with these symmetries. The charge $\mc{Q}_{u}^{\GR} (f)$ associated with the supertranslation $f$ at retarded time $u$ is given by
\be
\label{eq:superch}
\mc{Q}_{u}^{\GR}(f) =-\frac{1}{8\pi} \int_{S(u)} d \Omega~ f(x^A) \left[C^{(3)}_{urur} (u, x^A) -\frac{1}{4}N^{ AB} C_{AB} \right]   
\ee
where $S(u)\cong \bb{S}^{2}$ is an asymptotic sphere at fixed retarded time $u$, $C_{\mu \nu \rho \sigma}$ is the Weyl tensor and the superscript ``$(3)$'' denotes the order $1/r^3$ part as $r \to \infty$ at fixed $u$. The difference of the charge ${\mathcal Q}^{\GR}_{u}(f)$ at two retarded times $u_{1}$ and $u_{2}$ is determined by a corresponding flux between these retarded times  associated with the symmetry $f$
\be 
\label{eq:QGRu1u2}
\mc{Q}_{u_{2}}^{\GR}(f) - \mc{Q}^{\GR}_{u_{1}}(f) = -\frac{1}{32\pi} \int_{u_1}^{u_2}du \int_{\bb{S}^{2}}d\Omega~f\big(N^{AB}N_{AB} + 2 \ms{D}^{A}\ms{D}^{B}N_{AB}\big).
\ee 
If other massless fields are present and if their stress energy $T_{\mu \nu}$ satisfies the dominant energy condition then \cref{eq:QGRu1u2} is modified by the substitution $N^{AB}N_{AB}\to N^{AB}N_{AB}+32\pi T_{uu}^{(2)}$ where the ``(2)'' denotes the order $1/r^{2}$ part as $r\to \infty$ at fixed $u$. In the gravitational case, \cref{eq:QGRu1u2} with $u_1 \to - \infty$ and $u_2 \to + \infty$ directly yields an analogous formula to \cref{memch} relating charges, fluxes and memory\footnote{This equation determines the electric parity part of the memory. The ``magnetic parity'' part of the memory is determined by $\epsilon^{CA}\ms{D}_{C}\ms{D}^{B}\Delta^{\! \GR}_{AB}$, which can be expressed in terms of the difference of magnetic parity charges \cite{Satishchandran_2019} with no null memory contribution. All of the analysis of this paper could be straightforwardly generalized to include magnetic parity memory. However, as in the electromagnetic case (see \cref{magpar}), we shall focus entirely on electric parity memory in this paper.}
\begin{equation}
\label{memchgrav}
-\frac{1}{8\pi}\int_{\bb{S}^{2}}d\Omega~\Delta^{\!\GR,\out}_{AB}\ms{D}^{A}\ms{D}^{B}f = \mc{Q}^{\GR}_{i^{+}}(f) - \mc{Q}^{\GR}_{i^{0}}(f) + \mc{J}^{\GR,\out}(f).
\end{equation} 
The terms on the right hand side of \cref{memchgrav} involving the difference of charges is referred to as ``ordinary memory'' \cite{Zeld_Poln} and the term involving the flux is usually referred to as ``null memory'' or ``nonlinear memory'' \cite{Bieri_2013GR,Christ_nonlin_mem} given by 
\be 
 \mc{J}^{\GR,\out}(f) = \frac{1}{32\pi}\int_{-\infty}^{\infty}du\int_{\bb{S}^{2}}d\Omega~fN^{AB}N_{AB}.
\ee 
Similar formulas hold for the ``in'' memory in terms of difference of charges at past timelike infinity and spatial infinity as well as incoming null memory. As in the electromagnetic case, there is a matching of the incoming and outgoing charges as one approaches spatial infinity as originally conjectured by Strominger \cite{Strominger:2013jfa}
\be 
\label{eq:supermatch}
\mc{Q}_{i^{0}}^{\GR,\out}(f)=\mc{Q}_{i^{0}}^{\GR,\inn}(f \circ \Upsilon).
\ee 

As is well known, significant difficulties arise in the formulation of a quantum theory of gravity in the bulk spacetime. However, as Ashtekar has emphasized, no such difficulties arise in the asymptotic quantization of the radiative degrees of freedom of the gravitational field at null infinity \cite{PhysRevLett.46.573,asymp-quant,Ashtekar:2018lor}. Thus, the notion of asymptotic states of the quantum gravitational field in asymptotically flat spacetimes is well-defined, irrespective of the details of the bulk theory of gravity. In view of the classical memory effect, it is not possible that ``in'' states with vanishing memory (i.e., states in the standard ``in'' Fock representation) will generically evolve to ``out'' states with vanishing memory. Thus, infrared divergences similar to those occurring in QED must arise if one attempts to define an $S$-matrix with the conventional choices of ``in'' and ``out'' Hilbert spaces. One may ask whether there exist alternative choices satisfying conditions \ref{enum:S1}--\ref{enum:S5} above.

In linearized gravity with matter sources, massive fields will contribute to ordinary memory and massless fields will contribute to null memory, in close analogy with QED. In the case of QED, the vanishing of the charges $\mc{Q}_{i^{0}}(\lambda)$ at spatial infinity -- including the ordinary electric charge -- was required to have a Lorentz group action. As discussed above, in QED, it can be argued that the requirement of vanishing total electric charge is not a problem for obtaining representatives of all ``hard'' scattering processes because one can always put additional charges ``behind the moon''. However, in linearized gravity with massive/massless sources, the  analogous requirement of vanishing total $4$-momentum is a serious problem, since the ordinary vacuum state is the only state that satisfies this requirement --- there is no way to ``cancel'' the $4$-momentum if a state by adding particles. Therefore the Faddeev-Kulish construction fails for this elementary reason at this initial stage. 

Nevertheless, one could give up on having a well-defined action of the Lorentz transformations and attempt to construct states of definite, non-vanishing charges $\mc{Q}^{\GR}_{i^{0}}(f)$ at spatial infinity \cite{Choi:2017bna}. For linearized gravity with a massive field source, one can straightforwardly carry out an analog of the construction of \cref{subsec:MaxMKGKFreps} for massive QED. For linearized gravity with a massless field source, one can carry out an analog of the construction of \cref{subsec:MaxZMKGKFrep} for massless QED. Indeed, the situation for linearized gravity with a massless field source is somewhat better than massless QED in that the singularities of the memory are less severe \cite{Akhoury_2011}. In linearized gravity, for incoming momentum eigenstates of massless particles, the corresponding flux in \cref{memchgrav} will again have $\delta$-function angular singularities. However, on account of the presence of two derivatives on the left side of \cref{memchgrav} (as compared to the one derivative in \cref{memch1}), the corresponding collinear divergence singularities of $\Delta^{\!\GR}_{AB}$ will be of the form $\log |x^A - x^A_i|$. Although still singular, this is square integrable and does not imply an infinite energy flux of soft gravitons. Thus, arguably, in linearized gravity, the ``dressed states'' in the analogously constructed Faddeev-Kulish Hilbert space are physically acceptable, although since there is no a well-defined action of the Lorentz group, the states obtained in this construction do not have a well-defined angular momentum. 

However, we show in \cref{subsec:GravKFreps} that the Faddeev-Kulish type of construction fails catastrophically in nonlinear gravity. The fundamental problem is that, as in the Yang-Mills case, the ``soft gravitons'' that must be used to dress the ``hard gravitons'' will contribute their own flux, thereby invalidating any attempt to pair flux eigenstates of hard gravitons with corresponding memory representations. Thus, as in the Yang-Mills case, in order to obtain an analog of the Faddeev-Kulish Hilbert space, one must find some other means to obtain eigenstates of the supertranslation charges. In the Yang-Mills case, charge eigenstates must have vanishing charges and thus the $n$-point functions of any eigenstate of all large gauge charges must be invariant under all large gauge transformations. Although this is a highly restrictive condition, there do exist some invariant states besides the vacuum state. However, as we show in \cref{thm:eigensupcharge} of \cref{subsec:GravKFreps}, the corresponding condition in quantum gravity is that the $n$-point functions of the news must be invariant under supertranslations. However, this requirement is incompatible with the fall-off requirements on states. Thus, apart from the vacuum state, there are no eigenstates whatsoever of the supertranslation charges. Thus there is no analog of the Faddeev-Kulish Hilbert space in nonlinear gravity. 

Thus, if one is to obtain ``in'' and ``out'' Hilbert spaces in quantum gravity that satisfy properties \ref{enum:S1}--\ref{enum:S5}, one will have to do so by a very different means than by the Faddeev-Kulish construction. We explore some possibilities in \cref{sec:NKFreps} involving direct integrals with respect to Gaussian measures of Fock representations with memory. We find that these also do not work. Of course, our analysis does not exhaust all possibilities, but we do not see any further avenues of approach that appear promising. Thus, we believe that for gravity (as well as for Yang-Mills theory and massless QED), no satisfactory Hilbert space construction of ``in'' and ``out'' states can be given.  

What does this mean for scattering theory? There is no problem defining ``in'' and ``out'' states that should accommodate all scattering processes, allowing arbitrary incoming and outgoing ``hard'' particle states and arbitrary memory. The difficulties arise entirely from the attempt to ``shoehorn'' all states relevant to scattering theory into a single, separable Hilbert space. It is our view that there is no need to try to do this. An ``in'' state can be defined in the algebraic viewpoint as a positive linear function on the algebra of ``in'' observables. In this viewpoint one would specify an ``in'' state by giving the complete list of the correlation functions of the ``in'' fields --- where this list must satisfy positivity requirements. Any state in any Hilbert space construction gives rise to a state in this sense, since one can compute all the correlation functions and they will automatically satisfy the positivity requirement. Conversely, the Gel'fand-Naimark-Segal (GNS) construction shows that any state in the algebraic sense can be realized as a vector in some Hilbert space, so one does not get entirely new objects by considering states in the algebraic sense. But by considering states in the algebraic sense, one is freed from the necessity of choosing in advance a particular Hilbert space in which it lies. Thus, one may consider any ``in'' state that one wishes, without placing any ``dress requirements'' on the state. If one evolves the chosen ``in'' state through the bulk, one will get some ``out'' state, defined, again, as a list of correlation functions of the ``out'' fields. There is no reason to impose an a priori restriction as to which Hilbert space this ``out'' state will lie in --- and one will get infrared divergences if one selects the wrong one. As discussed in \cref{sec:AlgScatt}, we see no difficulty of principle in describing scattering theory in this framework. Of course, if one is interested in calculating quantities relevant to collider physics, we are not suggesting that there would be any advantage to taking such an approach over the usual approach of working in the standard Fock space and imposing infrared cutoffs. However, if one wishes to treat scattering at a fundamental level, we believe it is necessary to approach it from such an algebraic viewpoint on ``in'' and ``out'' states.

The structure of the remainder of the paper is as follows. In \cref{sec:classphase} we briefly review the classical phase space of a free scalar field in an asymptotically flat spacetime and give a precise notion of ``observable'' on this phase space. We also define local field observables at null infinity and determine their Poisson brackets. In \cref{sec:revhad}, we review the algebraic viewpoint on quantization and the formulation of free field theory in this framework. We briefly review the notion of Hadamard states in the bulk and consider their limit to null infinity in the case of massless fields. Although we are, of course, interested in the scattering theory of interacting fields, these interacting fields are assumed to behave like free fields in the asymptotic past and future, so the results of this section provide the tools needed to define the asymptotic quantization of interacting fields. In \cref{sec:MaxMKG} we consider QED with a massive, charged Klein-Gordon field. In \cref{subsec:MaxMKGclass} we construct the asymptotic algebra and Hadamard states of the massive scalar field at timelike infinity and the electromagnetic field at null infinity. In \cref{subsec:chmem} we consider the extension of the field algebras to include charges and Poincaré generators. In \cref{subsec:MaxMKGreps} we obtain Fock representations of the field algebras. The standard Fock representation of the massive Klein-Gordon field provides all of the necessary asymptotic states of that field, but we need all of the memory representations of the electromagnetic field to have an adequate supply of asymptotic electromagnetic states for scattering theory. In \cref{subsec:MaxMKGKFreps} we use these representations to construct the Faddeev-Kulish representation in massive QED by pairing momentum eigenstates of the Klein-Gordon field with corresponding memory representations of the electromagnetic field, thereby ``dressing'' the charged particles. In \cref{sec:MaxZMKG} we consider massless scalar QED. In \cref{subsec:MaxZMKGclassquant} we construct the asymptotic algebra of field observables at null infinity for a massless, charged Klein-Gordon field, and we extend this algebra to include charges and Poincaré generators in \cref{subsec:MaxMlessKGext}. In \cref{subsubsec:HadstatesFockExtAlg} we obtain the Fock representations of the massless scalar field. In \cref{subsec:MaxZMKGKFrep} we construct the analog of the Faddeev-Kulish representations for massless QED and point out the serious problems arising from the singular nature of the required memory representations. In \cref{subsec:YM} we consider source-free Yang-Mills theory and discuss the new serious difficulty that arises from the fact that the Faddeev-Kulish ``dressing'' also contributes to the charge-current flux. In \cref{sec:Grav} we consider general relativity. In \cref{sec:GR-quant} we provide the asymptotic algebra of observables in vacuum gravity. This algebra is extended in \cref{sec:GR-quant2} to include the BMS charges. In \cref{subsec:GravKFreps} we prove the non-existence of Faddeev-Kulish representations in quantum gravity. Some alternatives to Faddeev-Kulish representations are explored in \cref{sec:NKFreps} but none are found to be satisfactory. Finally, in \cref{sec:AlgScatt} we advocate for the development of an ``algebraic scattering theory,'' wherein one does not attempt to ``shoehorn'' all of the asymptotic states of scattering theory into pre-chosen ``in'' and ``out'' Hilbert spaces. Such a formulation of scattering theory would be manifestly infrared finite.

\subsection*{Notation and conventions}
\label{subsec:notation}
We work in natural units ($G=c=\hbar=1$) and will use the notation and sign conventions of \cite{Wald-book}. In particular, our metric signature is mostly positive and our sign convention for the curvature is such that the scalar curvature of the round sphere is positive. Greek indices $(\mu,\nu,\ldots)$ correspond to tensors in the ``bulk'' spacetime\footnote{We also will use Greek indices in several places in \cref{sec:classphase} to denote tensors on phase space.} $\mc{M}$ and we will use $y$ to denote arbitrary coordinates on $\mc{M}$. 

We will generally use the symbol $\ms{A}$ to denote a $\ast$-algebra of observables, $\s$ to denote a state on the algebra, $\ms{H}$ to denote a Hilbert space and $\ms{F}$ to denote a Fock space. Algebras of local field observables in the asymptotic past and future will be denoted as $\ms{A}_{\inn}$ and $\ms{A}_{\rm out}$, respectively. We will append a superscript ``in'' or ``out'' on various other quantities to distinguish between quantities defined in the asymptotic past or future, but we will omit this superscript when the context is clear. We will use superscripts ``EM'', ``KG'', ``KG0'', ``YM'', and``GR'' on quantities to distinguish between the particular cases of the electromagnetic, massive Klein-Gordon, massless Klein-Gordon, Yang-Mills, and gravitational fields, respectively. Thus, for example, $\Alg^{\rm EM}_{\inn}$ denotes the algebra of local electromagnetic field observables in the asymptotic past. We will append subscripts ``$Q$'' and ``$P$'' to denote the extensions of algebras of local field observables to include large gauge charges and Poincare generators respectively. Thus, for example $\Alg^{\rm KG}_{{\inn}, Q}$ denotes the extension of the algebra of local field observables of a massive scalar field in the asymptotic past to include large gauge charges.

Quantum observables will be denoted by the boldfaced version of the symbol for the corresponding classical observable; for example, the quantum observable corresponding to a classical scalar field \(\phi\) is denoted by \(\op\phi\).

We will work with the Penrose conformal completion (see, e.g., \cite{Wald-book,Geroch-asymp}) of flat spacetime (for QED and Yang-Mills theory) and asymptotically flat spacetimes (for gravity). The conformal boundaries \(\scri^\pm\) denote future/past null infinity, \(i^0\) denotes spatial infinity and \(i^\pm\) denotes future/past timelike infinity. The conformal factor will be denoted by \(\Omega\) and without loss of generality we impose the Bondi condition \(\nabla_\mu\nabla_\nu \Omega = 0\) at null infinity \(\scri^\pm\). The null normal to \(\scri^\pm\) will be denote \(n^\mu = \nabla^\mu \Omega\). 

We will frequently encounter down index tensors on $\scri^\pm$ that are orthogonal to $n^\mu$ in each index. We will denote such tensors with capital Latin letters $(A,B,\dots)$. For example, the pullback of the electric field $E_\mu = F_{\mu \nu} n^\nu$ to $\scri^\pm$ is such a tensor and it will be denoted as $E_A$. Similarly, the (degenerate) metric on $\scri^\pm$ (obtained from the pullback of the conformal spacetime metric) will be denoted as $q_{AB}$. We also will use capital Latin letters to denote equivalence classes of ``up'' index tensors on $\scri^\pm$, where two such tensors are equivalent if they differ by a multiple of $n^\mu$ in any index. (Such ``up'' index tensors are dual to the corresponding down index tensors.) The metric $q_{AB}$ acts non-degenerately on such equivalence classes of vectors, so it has an inverse, which we will denote as $q^{AB}$. We will use $q_{AB}$ and $q^{AB}$ to lower and raise capital Latin indices. Most of our analysis will be done with incoming fields on past null infinity \(\scri^-\) and we will use coordinates \(x = (v,x^A)\) on \(\scri^-\), where \(v\) is the advanced Bondi time coordinate and \(x^A\) are arbitrary coordinates on a \(2\)-sphere. Note that the index on the coordinates \(x^A\) should not be confused with a tensor index as described above.


\section{Classical phase space: Observables and asymptotic description}
\label{sec:classphase}

Our interest in this paper is in interacting quantum field theories, specifically, QED, Yang-Mills theory, and quantum gravity. However, we will be concerned only with the description of states at asymptotically early and late times, where it will be assumed that the states correspond to states of ``in'' and ``out'' free field theories. Thus, in essence, for the considerations of this paper, we need only be concerned with the structure of free field theory. The quantum theory of a free field is based on the phase space structure of the classical theory. In this section, we will review the phase space structure relevant for our considerations and explain the notion of ``observable'' that we shall use. For the case of a massless field, we will relate the ``bulk'' description of the field to its asymptotic description at null infinity. 

Since the phase space of a field theory is infinite dimensional, it would take some effort to define a mathematically precise Fr\'{e}chet space or other structure of phase space (see \cite{AS-symp}) that would properly incorporate the smoothness and fall-off conditions of the fields and provide a suitable topology on these fields. We believe that this could be done but we shall not attempt to do so here. Thus, we will freely use terms like ``smooth vector field on phase space'' in our discussion below without attempting to give a mathematically precise meaning to such terms. 

The basic structure of the classical phase space of a linear field theory is well illustrated by the case of a real scalar field $\phi$. Since we want to apply our constructions to the asymptotic behavior of the gravitational field in general relativity, it would not be reasonable to assume more structure than would be present on a globally hyperbolic, asymptotically flat, curved spacetime. Thus, we will take as our model system a real scalar field $\phi$ on a globally hyperbolic spacetime $(\mc{M},g)$, with Lagrangian
\be
    \mc L = - \frac{1}{2} \left[\nabla^\mu \phi \nabla_\mu \phi + m^2 \phi^2 + \xi R \phi^2 \right] 
\label{KGlag}
\ee
where $m$ denotes the mass, $\xi$ is an arbitrary constant, and $R$ is the Ricci scalar. Then $\phi$ satisfies 
\be
\label{ccKG}
    \bigg(\Box - m^2 - \xi R\bigg)\phi = 0.
\ee

As discussed in detail in \cite{Lee_1990,Iyer_1994} and many other references, the Lagrangian \cref{KGlag} endows the theory with a symplectic form, which thereby provides the space of initial data for solutions with a phase space structure. For the scalar field \cref{KGlag}, the points of phase space $\mathcal P$ can be taken to be the quantities $(\phi, n^\mu \nabla_\mu \phi)$ on a spacelike Cauchy surface $\Sigma$, where $n^\mu$ denotes the unit normal to $\Sigma$. In general, the symplectic form, $\Omega$, is a $2$-form on $\mathcal P$, i.e., at each point of $\mathcal P$ it maps a pair of tangent vectors into a number. However, in the case of a linear theory as considered here, $\mathcal P$ has a vector space (or, more generally, an affine space\footnote{As we shall see, in electromagnetism and gravity, the presence of large gauge transformations implies that there are many points of phase space that have vanishing gauge invariant field strengths. Any of these points could serve as an ``origin''.}) structure, and we can identify tangent vectors with points of $\mathcal P$. Consequently, we can view $\Omega$ as a bilinear map on $\mathcal P$. The symplectic product of two solutions $\phi_1, \phi_2$ is given by
\be
\Omega_{\Sigma}^{\KG} (\phi_1, \phi_2) = \int_\Sigma  \sqrt{h} d^3 x~\left[\phi_1 n^\mu \nabla_\mu \phi_2 - \phi_2 n^\mu \nabla_\mu \phi_1 \right]
\label{symprod}
\ee
where $\sqrt{h} d^3 x$ is the proper volume element on $\Sigma$. This symplectic product is conserved, i.e., it is independent of the choice of Cauchy surface $\Sigma$.

If $\mathcal P$ were finite dimensional, the nondegeneracy of the symplectic form $\Omega_{\alpha \beta}$ would imply that it has an inverse $\Omega^{\alpha \beta}$, where the Greek indices here represent tensor indices on phase space. A classical observable $F$ on $\mathcal P$ could then be taken to be an arbitrary smooth map $F: {\mathcal P} \to \mathbb{R}$. The inverse symplectic form would then allow us to define the Poisson bracket of any two such observables $F_1, F_2$ to be the observable on phase space given by
\be
\pb{F_1}{F_2} = \Omega^{\alpha \beta} \nabla_\alpha F_1 \nabla_\beta F_2.
\label{pbfd}
\ee
However, on an infinite dimensional phase space, the symplectic form is only weakly nondegenerate and its inverse will not be defined on all one-forms on the phase space. Thus, we cannot use \cref{pbfd} to define the Poisson bracket of arbitrary smooth functions on phase space. 

Nevertheless, on a general phase space, we can define the Poisson bracket on a particular class of smooth functions $F$. Namely, suppose $F$ is such that there is a smooth vector field $X^\alpha$ on $\mathcal P$ with the property that for all smooth curves $z(\alpha)$ on phase space, we have
\be
\delta F = \Omega(\delta z, X)
\label{Fgen}
\ee
where
\be
    \delta z \defn \lb. \td{\alpha} z(\alpha) \rb\vert_{\alpha = 0} \eqsp 
    \delta F \defn \lb. \td{\alpha} F( z(\alpha) ) \rb\vert_{\alpha = 0} \, .
\ee
Formally, \cref{Fgen} corresponds to $X^\alpha = \Omega^{\alpha \beta} \nabla_\beta F$, but \cref{Fgen} is expressed in a way that avoids the introduction of the inverse symplectic form. If \cref{Fgen} holds, we say that the function $F$ {\em generates} the vector field $X^\alpha$. Given two functions \(F_1\) and \(F_2\) on phase space that generate vector fields \(X^\alpha_1\) and \(X^\alpha_2\), respectively, we can define the Poisson bracket of $F_1$ and $F_2$ by
\be
    \pb{F_1}{F_2} \defn -\Omega(X_1, X_2).
\label{pbdef}
\ee
Formally, this corresponds to \cref{pbfd} because, formally, $\Omega(X_1, X_2) = \Omega_{\alpha \beta} X^\alpha_1 X^\beta_2 =  \Omega_{\alpha \beta}  \Omega^{\alpha \gamma} \nabla_\gamma F_1 \Omega^{\beta \eta} \nabla_\eta F_2 =  \Omega^{\eta \gamma} \nabla_\gamma F_1 \nabla_\eta F_2$. However, \cref{pbdef} avoids introducing the inverse symplectic form and is well-defined. For the case of an infinite dimensional phase space, we define an {\em observable} to be a smooth function $F$ on phase space that satisfies \cref{Fgen}. By construction, the Poisson bracket of any two observables is well-defined. It is only for classical observables in this sense that we can hope/expect to have quantum representatives. 

The situation with regard to obtaining observables simplifies considerably in the case of a phase space $\mathcal P$ with vector space structure, as considered here. Consider a vector field $X$ corresponding to an infinitesimal displacement at each $\phi$ of the form of an affine transformation
\be
\phi \mapsto \phi + \epsilon(L \phi + \chi_0)
\label{Xaffine}
\ee
where $\chi_0$ is a constant ($\phi$-independent) displacement and $L$ is a linear map on phase space. Suppose that $L$ satisfies 
\be
\Omega(\psi, L \phi) = - \Omega(L \psi, \phi)
\label{Bsym}
\ee
for all $\phi, \psi \in {\mathcal P}$.
Then it is straightforward to verify that the function $F: {\mathcal P} \to \mathbb{R}$ defined by 
\be
F(\phi) = \Omega(\phi, \chi_0) + \frac{1}{2} \Omega(\phi, L \phi)
\label{Flinobs}
\ee
satisfies \cref{Fgen}. Thus, any function $F$ on phase space of the form \cref{Flinobs} with $L$ satisfying \cref{Bsym} is an observable on phase space.

An important class of observables on $\mathcal P$ are the local field observables. Let $f : \mc{M} \to \mathbb{R}$ be a test function on spacetime, i.e., a smooth function of compact support. Let $F_f$ be the linear function on phase space given by 
\be
F_f(\phi) = \phi(f) \defn  \int \sqrt{-g} d^4 y~\phi(y) f(y) .
\label{4smear}
\ee
where $y$ denotes arbitrary coordinates on $\mc{M}$. Then $F_f$ can be shown to be an observable as follows. Let 
\be
E f = Af - Rf
\ee
where $Af$ denotes the advanced solution to \cref{ccKG} with source $f$ and $Rf$ denotes the retarded solution with source $f$. Then $Ef$ a smooth, source-free solution to \cref{ccKG} with initial data of compact support, so it corresponds to a point in $\mathcal P$. By lemma 3.2.1 of \cite{Wald_1995}, for any solution $\phi$ we have\footnote{Note that our convention for the symplectic form in \cref{symprod} has the opposite sign compared to the one used in \cite{Wald_1995}.}
\be
\phi (f) = \Omega_{\Sigma}^{\KG}(\phi, Ef) .
\label{321id}
\ee
Thus, $F_f$ is of the form \cref{Flinobs} with $\chi_0 = Ef$ and $L = 0$.
Thus, the ``smeared fields'' $\phi(f)$ are observables.\footnote{Note that the field evaluated at a point, $\phi(x)$, is too singular to be considered to be an observable. The associated vector field $X^\alpha$ would correspond to an infinitesimal displacement in the direction of the singular solution given by the advanced minus retarded solution with delta function source at $x$, which does not lie in the phase space.} It is not difficult to see that the Poisson bracket of smeared fields is given by
\be
\pb{\phi(f_1)}{\phi(f_2)} = E(f_1, f_2) 1
\label{pblocfield}
\ee
where ``$1$'' denotes the constant function on $\mathcal P$ that maps all points to $1$, and 
\be
E(f_1, f_2) = \int\sqrt{-g} d^4 y~ f_1(y) Ef_2 (y)  \, .
\ee

For any Cauchy surface $\Sigma$ and any test function $s$ on $\Sigma$, we may define the linear function $F_{\Sigma, s}$ on phase space by
\be
F_{\Sigma, s} (\phi) = \phi_\Sigma (s) \defn  \int_\Sigma \sqrt{h} d^3 x~ \phi(x) s(x) .
\ee
We may similarly define $(n^\mu \nabla_\mu \phi)_\Sigma (s)$. These ``3-smeared'' fields are also observables, which can be seen from \cref{321id} to be equivalent to the ``4-smeared'' observables \cref{4smear}. Namely, we have
\be
\phi_\Sigma (s) = \phi(f)
\ee
where $f$ is a test function on spacetime such that the initial data for $Ef$ on $\Sigma$ is $[Ef]_\Sigma = 0$ and $[n^\mu \nabla_\mu (Ef)]_\Sigma = s$. A similar formula holds for $(n^\mu \nabla_\mu \phi)_\Sigma (s)$.

Our main interest in this paper is to characterize the states of quantum fields in asymptotically flat spacetimes at asymptotically early and late times. It therefore will be important to have a description of the phase space and observables that characterizes the behavior of the field at asymptotically early and late times. We shall now explain how this can be done for massless fields. The corresponding asymptotic quantization will be described in the next section. The classical and quantum asymptotic description of massive fields will be given in \cref{sec:MaxMKG}.

For massless fields, we assume that past null infinity, $\scri^{-}$, and future null infinity, $\scri^{+}$, can be treated as Cauchy surfaces, so that initial data at $\scri^{-}$ or $\scri^{+}$ uniquely determines a solution.\footnote{This is true in Minkowski spacetime but is an assumption in a general asymptotically flat spacetime. It would not hold in spacetimes with a black hole or white hole, but one could presumably then supplement the asymptotic description of states at null infinity by including states on the horizon of the black hole or white hole.} For a massless field with the Cauchy surface taken to be $\scri^{-}$, initial data for solutions consists of the specification of the conformally weighted scalar field, $\Phi$, on $\scri^{-}$ 
\be
\Phi \defn  \lim_{\scri^{-}} \Omega^{-1} \phi \, 
\label{phitil}
\ee
where $\Omega$ is a conformal factor, which, in Bondi coordinates, can be chosen to be $\Omega = 1/r$. We assume that the solutions in $\mathcal P$ are such that $\partial \Phi/\partial v = O(1/|v|^{1+\epsilon})$ for some $\epsilon > 0$ as $v \to \pm \infty$. This will ensure that all integrals below will converge. Note, however, that we do {\em not} assume that $\Phi \to 0$ as $v \to \pm \infty$, as this would exclude the memory effect. Although we could, of course, restrict consideration to initial data at $\scri^-$ satisfying $\Phi \to 0$ as $v \to \pm \infty$, if interactions occur in the bulk, such initial data will generically evolve to fields at $\scri^+$ that do not satisfy $\Phi \to 0$ as $u \to \infty$. Since we wish to treat $\scri^-$ and $\scri^+$ on an equal footing in scattering theory, we do not require $\Phi \to 0$ as $v \to \pm \infty$ at $\scri^-$.

In terms of the initial data \cref{phitil}, the symplectic product \cref{symprod} is given by
\be
\Omega_{\Sigma}^{\KG 0} (\phi_1, \phi_2) = \int_{\scri^{-}} dv d \Omega \left[ \Phi_1 \frac{\partial \Phi_2}{\partial v} -  \Phi_2 \frac{\partial \Phi_1}{\partial v} \right] 
\label{symprodscri}
\ee
where we have inserted an extra ``0'' in the superscript ``KG0'' on $\Omega$ to indicate that this formula holds only for the case of a massless scalar field. It is convenient to define 
\be
\Pi\defn \partial_{v}\Phi
\ee
on $\scri^-$, since this quantity will arise in many formulas below.
It follows from \cref{symprodscri} that for any test function $s$ on $\scri^-$, we have
\be
\Pi (s) \defn  \int_{\scri^{-}} dvd\Omega~ \frac{\partial \Phi}{\partial v} (v,x^{A}) s(v,x^{A}) = \frac{1}{2} \Omega_{\Sigma}^{\KG 0} (Ef, \Phi)= - \frac{1}{2} \phi(f)
\label{scriobs}
\ee
where $f$ is a function on spacetime such that on $\scri^-$ we have $\lim_{\scri^{-}}\Omega^{-1}Ef=s$.
Thus, the smeared field quantities $\Pi (s)$ on $\scri^-$ are observables on phase space that are essentially equivalent\footnote{\label{essequiv}We say ``essentially equivalent'' because if $f$ is of compact support on spacetime, then in a curved spacetime --- where Huygens' principle does not hold for the wave equation (\ref{ccKG}) --- $Ef$ will not be of compact support on $\scri^{-}$ and vice-versa, so the test function spaces do not align precisely. We will ignore this issue here. Except for the case of nonlinear gravity, our applications are to Minkowski spacetime, where Huygens' principle does hold and the correspondence is exact.} to the bulk field observables $\phi(f)$. The Poisson brackets of these observables at $\scri^-$ are given by
\be
\pb{\Pi (s_1)}{ \Pi(s_2)} &= \frac{1}{4} \pb{\phi(f_1)}{ \phi (f_2)} \\
&= \frac{1}{4} E(f_{1},f_{2}) 1 \\
&= \frac{1}{4} \Omega_{\Sigma}^{\KG 0}(Ef_1, Ef_2) 1 \\
&= \frac{1}{4} \int_{\scri^{-}} dv d \Omega \left( s_2 \frac{\partial s_1}{\partial v} - s_1 \frac{\partial s_2}{\partial v} \right) 1.
\label{scricomm0}
\ee
Here, the third line was obtained by writing
\be
E(f_1, f_2) = \int\sqrt{-g} d^4 y~ f_1(y) Ef_2(y)  = - \Omega_{\Sigma}^{\KG 0}(Ef_2, Ef_1)
\ee
where \cref{321id} with $f=f_1$ and $\phi = Ef_2$ was used.

Finally, note that if $w$ is a test function of the form $w = \partial s/\partial v$ for some test function $s$, then
\begin{eqnarray}
\Phi (w) &\defn & \int_{\scri^{-}} dv d \Omega~\Phi (v, x^A) w(v, x^A) =  \int_{\scri^{-}} dv d \Omega~\Phi  \frac{\partial s}{\partial v}  \nonumber \\
&=& -  \int_{\scri^{-}} dv d \Omega~\frac{\partial \Phi}{\partial v}  s  = -\Pi (s).
\end{eqnarray}
Thus, $\Phi (w)$ for $w = \partial s/\partial v$ is equal to $-\Pi (s)$ and hence is well-defined and corresponds to a local observable in the bulk. However, if $w$ is not of this form --- i.e., if $\int dv~w(v, x^A) \neq 0$ for some $x^A$ --- then $\Phi (w)$ does not correspond to a local observable in the bulk. 

\section{Algebraic viewpoint: Quantization of free fields and asymptotic quantization of massless interacting fields}
\label{sec:revhad}

Since we are concerned in this paper with the possible choices of a Hilbert space of ``in'' and ``out'' states in scattering theory, it is essential to have a notion of the structure of the theory prior to a choice of Hilbert space. The algebraic approach provides such a notion. The purpose of this section is to review the key ideas ideas in the algebraic approach and describe the asymptotic quantization of massless fields corresponding to the asymptotic characterization of phase space given at the end of the previous section. For further discussion of the algebraic viewpoint we refer the reader to \cite{Wald_1995,Hollands:2014eia,Witten_2021}. 

In the algebraic approach, one assumes that the quantum field observables have the structure of a $*$-algebra $\Alg$. States are then defined as positive linear functions on the algebra, i.e., a state, $\omega$, is simply a linear map $\omega: \Alg \to \bb C$ such that $\omega(a^* a) \geq 0$ for all $a \in {\Alg}$. If we take $\Alg$ to be generated by local (smeared) field observables, then an arbitrary element $a \in \Alg$ would be a sum of products of local field observables, so a specification of $\omega$ would be equivalent to providing the complete list of the correlation functions of the field observables.

Our interest in this paper is in interacting quantum field theories, specifically, QED, Yang-Mills theory, and quantum gravity. There are many nontrivial and still unanswered questions about the formulation of interacting quantum field theories. However, in this paper, we will be concerned only with the behavior of these fields at asymptotically early and late times. As is normally done in scattering theory, we will simply \emph{assume} that states of the theory behave at asymptotically early and late times like states of the corresponding ``in'' and ``out'' free field theories, i.e., that the interactions can be neglected at asymptotically early and late times. Of course, the determination of the relationship between the ``in'' and ``out'' states requires knowledge of the interacting quantum field theory, but our analysis in this paper will be exclusively concerned with the nature of ``in'' and ``out'' states and whether suitable Hilbert spaces of such states can be defined. Thus, as previously stated at the beginning of \cref{sec:classphase}, for the considerations of this paper, we need only be concerned with the structure of free field theory. 

The structure of the quantum theory of a free field is well illustrated by the case of a real scalar field $\phi$, \cref{KGlag}. The classical phase space structure of the real scalar field was described in \cref{sec:classphase}. The quantum theory of $\phi$ is defined by specifying an algebra, $\Alg$, of quantum observables. We obtain $\Alg$ by starting with the free algebra of the smeared fields $\op{\phi}(f)$, their formal adjoints $\op{\phi}(f)^{\ast}$ and an identity $\1$ where $f$ is a real-valued, smooth function on $\mc{M}$ with compact support. The algebra $\Alg$ is then obtained by factoring this free algebra by the following relations: 
\begin{enumerate}[label=(A.{\Roman*})]
\label{KGalg}
\item $\op{\phi}(c_{1}f_{1}+c_{2}f_{2})=c_{1}\op{\phi}(f_{1})+c_{2}\op{\phi}(f_{2})$ for any $f_{1},f_{2}$ and any $c_{1},c_{2}\in \mathbb{R}$, i.e., the smeared field is linear in the test function \label{A21}
\item $\op{\phi}((\Box - m^2 -\xi R)f)=0$ for all $f$, i.e., $\op{\phi}$ satisfies the field equation in the distributional sense  \label{A22}
\item $\op{\phi}(f)^{\ast}=\op{\phi}(f)$ for all  $f$, i.e., the field is Hermitian \label{A23}
\item $[\op{\phi}(f_{1}), \op{\phi}(f_{2})]=iE(f_{1},f_{2})\op{1}$, i.e., the field satisfies canonical commutation relations (see \cref{pblocfield})\label{A24}
\end{enumerate}

As already mentioned above, a state is a linear map $\s:\Alg \to \bb{C}$ that satisfies $\s(a^{\ast}a)\geq 0$ for all algebra elements $a\in \Alg(\mc{M},g)$. We further require the normalization condition $\s(\1)=1$. A state is thus determined by specifying its smeared ``$n$-point correlation functions'' $\s(\op{\phi}(f_{1})\dots \op{\phi}(f_{n}))$. If we have a Hilbert space $\Hilb$ on which the smeared fields are represented as operators satisfying \ref{A21}--\ref{A24}, then any normalized vector $\ket{\Psi} \in \Hilb$ gives rise to a state via $\omega(a) = \braket{\Psi | \pi(a) |\Psi}$ for all $a \in \Alg$, where $\pi(a)$ is the operator representative of $a$. More generally, any normalized density matrix $\rho$ on $\Hilb$ gives rise to a state via $\omega(a) = {\rm tr} (\rho \pi(a))$. Conversely, by a remarkably simple construction due to Gel'fand, Naimark and Segal (GNS), given an algebraic state $\s:\Alg \to \bb{C}$, one can obtain a representation, $\pi$, of $\Alg$ on a Hilbert space $\Hilb$ and a vector $\ket{\Psi} \in \Hilb$ such that $\omega(a) = \braket{\Psi | \pi(a) |\Psi }$ for all $a \in \Alg$. The GNS construction consists of starting with the vector space $\Alg$ and using $\s$ to define an inner product on $\Alg$. One then completes $\Alg$ in this inner product and factors out any degenerate elements to get a Hilbert space $\Hilb$. By construction, $\Hilb$ contains a dense set of vectors $\ket{a}$ corresponding to elements $a \in \Alg$. We obtain a representation, $\pi$, of $\Alg$ on $\Hilb$ by the formula $\pi(a) \ket{b} = \ket{ab}$ for all $a,b \in \Alg$. The vector $\ket{\Psi} \in \Hilb$ corresponding to $\s$ is simply $\ket{\1}$. Note that $\ket{\1}$ is cyclic, i.e., the action of $\pi(a)$ on $\ket{\1}$ for all $a \in \Alg$ generates a dense subspace of states. Note further that this construction uses only the $*$-algebra structure of $\Alg$.

A state is called pure if it cannot be written as a sum of two other states with positive coefficients; otherwise the state is referred to as mixed. The GNS construction will represent a mixed state as a vector (rather than density matrix) in $\Hilb$, but for a mixed state the GNS representation will be reducible. In particular, for a state that corresponds to a density matrix on a Hilbert space $\tilde{\Hilb}$ that carries an irreducible representation of $\Alg$, the GNS construction will suitably enlarge $\tilde{\Hilb}$ to a Hilbert space $\Hilb$ on which the state is represented as a vector.\footnote{For example, the GNS construction represents a thermal state as a vector in its ``thermofield double''.} 

An important class of states are known as ``Gaussian states'' (also referred to as ``quasi-free states'' or ``vacuum states''). By definition, for Gaussian states, the $n$-point functions for $n > 2$ are given by formulas in terms of the $1$- and $2$-point functions that are analogous to the formulas for the $n$-th moments of a Gaussian probability distribution. This can be described by saying that the ``connected $n$-point functions'' (also known as ``truncated $n$-point functions'') vanish for all $n>2$ (see e.g. \cite{Hollands:2014eia}). For example, for a Gaussian state of the Klein-Gordon field the $3$-point function is given by 
\be
\label{eq:gauss3pt}
\s(\op{\phi}(y_{1})\op{\phi}(y_{2})\op{\phi}(y_{3}))=&\s(\op{\phi}(y_{1}))\cdot  \s(\op{\phi}(y_{2})\op{\phi}(y_{3})) + \s(\op{\phi}(y_{2}))\cdot  \s(\op{\phi}(y_{3})\op{\phi}(y_{1}))\\
&+\s(\op{\phi}(y_{3}))\cdot \s(\op{\phi}(y_{1})\op{\phi}(y_{3}))-2\s(\op{\phi}(y_{1}))\cdot \s(\op{\phi}(y_{2}))\cdot \s(\op{\phi}(y_{3}))
\ee 
where all ``unsmeared'' formulas here and below should be interpreted as holding distributionally. The GNS Hilbert space of a Gaussian state $\s$ has a natural Fock space structure
\be 
\label{Fock}
\Fock(\Hilb_{1}) = \mathbb{C}\oplus \lb[ \bigoplus_{n\geq 1}\underbrace{\big(\Hilb_{1} \otimes_{S}\dots \otimes_{S}\Hilb_{1} \big)}_{n \text{ times}} \rb].
\ee 
where $\otimes_{S}$ is the symmetrized tensor product, and the inner product on the ``one-particle Hilbert space'' $\Hilb_1$ is determined\footnote{More precisely on the space of smooth functions $f$ of compact support we define the inner product $\braket{f_{1}|f_{2}}=\s(\op{\phi}(f_{1})^* \op{\phi}(f_{2}))$ (see \cite{Kay:1988mu} for details).} by the $2$-point function $\s(\op{\phi}(y_{1})\op{\phi}(y_{2}))$. In Minkowski spacetime, the Poincar\'e invariant vacuum state $|0 \rangle$ is a Gaussian state and the Fock space \cref{Fock} is the standard choice of Hilbert space for free field theory.

The general definition of a state given above admits many states with singular ultraviolet behavior --- too singular for nonlinear field observables to be defined. It is therefore necessary to impose an additional restriction on the short distance behavior of states. In most treatments of quantum field theory in Minkowski spacetime, this issue is not highlighted because the vacuum state $|0 \rangle$ has the required ultraviolet behavior, as do all states in the corresponding Fock space \cref{Fock} with smooth $n$-particle ``mode functions''. Thus, the states that are normally considered in usual treatments satisfy the required condition on ultraviolet behavior. However, in this paper, we seek alternative choices of Hilbert spaces --- since, as explained in \cref{sec:intro}, the standard Fock space of ``in'' and ``out'' states cannot accommodate the states that arise in scattering processes --- so it is essential that we explicitly impose the condition that states have the required ultraviolet behavior. This additional restriction on states is given by the Hadamard condition, which requires that the short distance behavior of the $2$-point function of any allowed state be of the form
\begin{equation}
\label{2pthad}
    \s(\op{\phi}(y_{1})\op{\phi}(y_{2}))=\frac{1}{4\pi^{2}}\frac{U(y_{1},y_{2})}{\sigma+i0^{+}T}+V(y_{1},y_{2})\log(\sigma+i0^{+}T)+W(y_{1},y_{2}).
\end{equation}
Here $\sigma$ is the squared geodesic distance between $y_1$ and $y_2$, $T= t(y_1) - t(y_2)$ with $t$ a global time function on spacetime, $U$ and $V$ are smooth, symmetric functions that are locally constructed via the Hadamard recursion relations \cite{DeWitt:1960fc}, and $W$ is also smooth and symmetric. The Hadamard condition can be very usefully reformulated in terms of microlocal spectral conditions on the distribution $\s(\op{\phi}(y_{1})\op{\phi}(y_{2}))$ \cite{Radzikowski:1996pa}, but we shall not need this reformulation here. The Hadamard condition \cref{2pthad} together with the positivity condition on states implies that the connected $n$-point functions for $n\neq 2$ of a Hadamard state are smooth and symmetric \cite{Sanders:2009sw}.

We conclude this section by giving the asymptotic quantization of $\op{\phi}$ in the massless case. Again, we assume that past null infinity, $\scri^{-}$, and future null infinity, $\scri^{+}$, can be treated as Cauchy surfaces. We cannot proceed by starting with the bulk theory and taking limits of correlation functions to $\scri^{-}$ or $\scri^{+}$, since the quantum fields are distributional on spacetime and cannot straightforwardly be restricted to a lower dimensional surfaces such as $\scri^{-}$ or $\scri^{+}$. However, we can proceed by working with the asymptotic description of the classical phase space given at the end of the previous section. 

For the asymptotic quantization on $\scri^-$, we take the observables on phase space to be $\op{\Pi}(s)$ (\cref{scriobs}), where $s$ is an arbitrary test function on $\scri^-$ with conformal weight $-1$. We define the algebra $\Alg_{\inn}$ by starting with the free algebra generated by $\op{\Pi} (s)$, $\op{\Pi}(s)^*$ and $\op{1}$ and factoring it by relations corresponding to \ref{A21}--\ref{A24}. Conditions \ref{A21} and \ref{A23} translate straightforwardly to $\Alg_{\inn}$. There is no condition corresponding to condition \ref{A22} since we are now smearing $\op{\Pi}$ with free data for solutions. The commutation relation \ref{A24} translates to
\be
\left[ \op{\Pi} (x_1), \op{\Pi} (x_2) \right] =  \frac{i}{2} \delta'(v_1, v_2) \delta_{\mathbb{S}^{2}}(x_1^A, x_2^A) \op{1}
\label{scricomm}
\ee
(see \cref{scricomm0}) where $x=(v,x^{A})$ are coordinates on $\scri^{-}$ and this equation is to be understood as a distributional relation on $\scri^{-}$. This completes our specification of the algebra $\Alg_{\inn}$. The algebra $\Alg_{\rm out}$ is defined similarly. 

The algebra $\Alg_{\inn}$ constructed in this manner is essentially equivalent\footnote{We say ``essentially equivalent'' for the reason stated in \cref{essequiv}.} to the bulk free field algebra $\Alg$. For an interacting theory, the bulk algebra, of course, is no longer a free field algebra, but the central assumption of scattering theory is that states on the bulk algebra asymptote to states on the free field algebras $\Alg_{\inn}$ and $\Alg_{\rm out}$ at early and late times, respectively.

We now impose regularity conditions on states on $\Alg_{\inn}$. For the bulk theory, we imposed the Hadamard condition \cref{2pthad} on states on $\Alg$, and we wish to express this condition as a corresponding condition on states on $\Alg_{\inn}$. For the conformally invariant case ($\xi = 1/6$) in a spacetime with a regular timelike infinity, it has been shown \cite{Moretti:2005ty} that Hadamard states on $\Alg$ correspond to states on $\Alg_{\inn}$ whose $2$-point function is of the form\footnote{A similar result holds for any field (including massive fields) on a Killing horizon \cite{Kay:1988mu}.}
\begin{equation}
\label{sympsmear}
   \s \left( \op{\Pi} (x_1) \op{\Pi} (x_2) \right) =-\frac{1}{\pi} \frac{\delta_{\mathbb{S}^{2}}(x_1^A, x_2^A)}{(v_1 - v_2 -i0^{+})^{2}} +  S(x_1 ,x_2) 
\end{equation}
where $S(x_1,x_2)$ is a smooth function on $\scri^{-} \times \scri^{-}$.
In particular, this result holds in Minkowski spacetime (for an arbitrary $\xi$, since $\xi$ does not enter the equations of motion in that case). We assume that this form holds generally for massless fields, i.e., with no restriction to $\xi = 1/6$ or to spacetimes with a regular timelike infinity. Thus, we impose \cref{sympsmear} as the ultraviolet regularity condition on states on $\Alg_\inn$. Note that in Minkowski spacetime, the $2$-point function of the Poincar\'e invariant vacuum state $\omega_0$ takes the form \cref{sympsmear} with $S=0$, i.e.,
\begin{equation}
\label{vac2pt}
   \s_0 \left( \op{\Pi} (x_1) \op{\Pi} (x_2) \right) =-\frac{1}{\pi} \frac{\delta_{\mathbb{S}^{2}}(x_1^A, x_2^A)}{(v_1 - v_2 -i0^{+})^{2}} .
\end{equation} 

In addition to the ultraviolet regularity condition on states, we impose the following decay conditions on states, analogous to the classical decay conditions mentioned below \cref{phitil}: We require that $S$ and all connected $n$-point functions for $n\neq 2$ decay for any set of $|v_{i}|\to \infty$ as $O((\sum_{i}v_{i}^{2})^{-1/2-\epsilon})$ for some $\epsilon>0$.

In the subsequent sections, we will assume that the quantization of the ``in'' and ``out'' electromagnetic and gravitational fields are given by a direct analog of our construction of $\Alg_{\inn}$ above, and we will impose ultraviolet regularity (Hadamard) conditions on states given by the direct analog of \cref{sympsmear}, as well as the analogous decay conditions.

Finally, we note that we have included only observables that are linear in the field $\phi$ in our algebras, $\Alg$ and $\Alg_{\inn}$, of local field observables. For the case of the bulk theory, $\Alg$ can be extended to include smeared polynomial quantities (``Wick polynomials'') in the field by a ``Hadamard normal ordering'' procedure (see \cite{Hollands:2014eia}). However, an analogous procedure does {\em not} work for $\Alg_{\inn}$, as Hadamard normal ordering produces quantities that are too singular in the angular directions. Thus, we cannot extend $\Alg_{\inn}$ to include polynomial local field observables. Nevertheless, quantities that are quadratic in the fields can be defined as quadratic forms by Hadamard subtraction, using \cref{vac2pt} for the subtraction. In particular, for any Hadamard state $\omega$, we may define the expected value of $\op{\Pi}^2$ by
\be\label{quadform}
   \s \left( \op{\Pi}^2 (x) \right) &= \lim_{x' \to x} \left[  \s \left( \op{\Pi} (x)\op{\Pi} (x') \right) -  \s_0 \left( \op{\Pi} (x) \op{\Pi} (x') \right) \right] \\
   &= S(x,x).
\ee
We can use this notion to define expected values of observables that are quadratic in the fields. However, higher powers of $\op{\Pi}(x)$ cannot even be defined as quadratic forms. In particular, since the stress-energy flux through $\scri^{-}$ is $T_{vv}=\Pi^{2}$ this implies that the local energy flux cannot be defined as an operator and is only well-defined as a quadratic form (i.e. only its expected value is well-defined). This result is in accord with arguments given in \cite{Bousso_2017}.

\section{QED with a massive, charged Klein-Gordon field}
\label{sec:MaxMKG}

In this section we consider massive scalar QED, i.e., the theory of a Maxwell field $A_\mu$, coupled to a charged massive complex Klein-Gordon scalar field $\varphi$ in Minkowski spacetime. The Lagrangian for this theory is
\be
    \mc L = - \frac{1}{4} F^{\mu \nu} F_{\mu \nu} - \frac{1}{2} D^\mu \bar\varphi D_\mu \varphi - \frac{1}{2} m^2 \bar \varphi \varphi
\label{lqed}
\ee
where $F_{\mu \nu} = \partial_\mu A_\nu - \partial_\nu A_\mu$ and $D_\mu$ is the gauge covariant derivative operator
\be
D_\mu \varphi \defn \partial_\mu \varphi - i q A_\mu \varphi \eqsp D_\mu \bar\varphi \defn \bar{D_\mu \varphi} \, .
\ee
The theory is invariant under the action of gauge transformations
\be
A_\mu \mapsto A_\mu + \partial_\mu \LGT \eqsp \varphi \mapsto e^{iq \LGT} \varphi \, .
\label{gaugetrans}
\ee

In \cref{subsec:MaxMKGclass}, we give the asymptotic quantization of the massive Klein-Gordon and electromagnetic fields. In \cref{subsec:chmem}, we extend the algebra of asymptotic observables to include large gauge charges and Poincar\'e generators. In \cref{subsec:MaxMKGreps}, we construct Fock representations of the extended algebra of asymptotic observables with arbitrary choices of memory. Finally, in \cref{subsec:MaxMKGKFreps} we construct the Faddeev-Kulish Hilbert space.

\subsection{Asymptotic quantization of QED with a massive Klein-Gordon field}
\label{subsec:MaxMKGclass}

We wish to provide a characterization of the states in QED in terms of free field states in the asymptotic past and asymptotic future. For definiteness, we will focus upon the asymptotic past; exactly the same procedure is used for the asymptotic future. We assume that in the asymptotic past, classical solutions approach solutions to the free Klein-Gordon and Maxwell equations.\footnote{This assumption is supported by rigorous studies of the classical behavior of the QED fields \cite{Fang_2019,Klainerman_2018,Psarelli1_1999,Psarelli2_1999}.} Correspondingly, in the asymptotic past, states in QED should approach ``free field `in' states,'' i.e., states on the tensor product
\be
\Alg_{\inn} = \Alg^{\rm EM}_{\inn} \otimes \Alg^{\rm KG}_{\inn} \, .
\ee
of the asymptotic algebra,  $\Alg^{\rm EM}_{\inn}$, of the free electromagnetic field with the asymptotic algebra, $\Alg^{\rm KG}_{\inn}$, of the free massive Klein-Gordon field. Thus, our task in this subsection is to obtain the free field algebras $\Alg^{\rm EM}_{\inn}$ and  $\Alg^{\rm KG}_{\inn}$. 

The strategy for obtaining $\Alg^{\rm EM}_{\inn}$ was presented in the previous section. The electromagnetic field is conformally invariant, so, classically, for solutions with appropriate fall-off at spatial infinity, one can choose a gauge\footnote{Note that the vector potential \(A_\mu\) is \emph{not} smooth at null infinity in the Lorenz gauge when there is a non-vanishing total charge (see Remark~4 and eq.~(52) of \cite{Satishchandran_2019}). Nevertheless one make other choices that yield a smooth $A_\mu$. \label{fn:lorenz-A}} so that the vector potential $A_\mu$ extends smoothly to $\scri^-$. We may further choose a gauge for which $n^\mu A_\mu \vert_{\scri-} = 0$ where $n^\mu \defn \partial_v$ is the null normal. In this gauge, the pullback of $A_\mu$ to $\scri^-$ is a down index tensor on $\scri^-$ that is orthogonal to $n^\mu$, so in accord with the notational conventions stated at the end of \cref{subsec:notation}, we denote it as $A_A$.

The points of the classical phase space are given by the specification of $A_A$ on $\scri^-$. This is the analog of the specification of $\Phi$ on $\scri^-$ in the scalar field case. The analog of the observable $\Pi=\partial_{v}\Phi$ on $\scri^-$ is the \emph{electric field} 
\be
    E_A = - \Lie_n A_A = - \partial_v A_A
\ee
which is the pullback to $\scri^-$ of $F_{\mu \nu}n^\nu$. Note that $E_A$ is gauge invariant. The symplectic form is given by
\be 
\Omega_{\scri}^{\EM}(A_{1},A_{2})=-\frac{1}{4\pi}\int_{\scri^{-}}d^{3}x ~\big[E_{1A}A^{A}_{2}-E_{2A}A^{A}_{1}\big].
\ee 
The local field observables for the Maxwell field on $\scri^-$ are 
\be
E(s) = \int_{\scri^-} d^{3}x~E_A(x)s^A(x)
\label{esmear}
\ee
where $s^A$ is test vector field on $\scri^-$, with no conformal weight, and the capital Latin index is in accord with the notational conventions stated at the end of \cref{subsec:notation} because \cref{esmear} depends only on the equivalence class of the vector field. Note that the observable $E(s)$ generates the infinitesimal affine transformation \(A_A \to A_A - 2\pi \epsilon s_A\). The Poisson brackets are 
\be 
\label{eq:E-Poisson}
\{E(s_{1}),E(s_{2})\}=-4\pi^{2}\Omega^{\EM}_{\scri}(s_{1},s_{2})1
\ee 
where, for test functions $s^{A}_{1},s^{A}_{2}$ we have that 
\be 
\Omega_{\scri}^{\EM}(s_{1},s_{2})=-\frac{1}{4\pi} \int_{\scri}d\Omega dv~\big[s_{1}^{A}\partial_{v}s_{2A} - s_{2}^{A}\partial_{v}s_{1A}\big]=-\frac{1}{2\pi}\int_{\scri^{-}}d\Omega dv~s_{1}^{A}\partial_{v}s_{2A}.
\ee 
In exact parallel with the asymptotic quantization of the massless scalar field given in \cref{sec:revhad}, the algebra $\Alg^{\rm EM}_{\inn}$ is defined to be the free algebra generated by the smeared fields $\op{E}(s)$, their formal adjoints $\op{E}(s)^{\ast}$, and an identity $\op{1}$ --- where $s^A (x)$ are real test vector fields on $\scri^{-}$ --- factored by the following relations: 
\begin{enumerate}[label=(B.{\Roman*})]
\label{NewsMaxAlg}
    \item \label{A31} $\op{E}(c_{1}s_{1}+c_{2}s_{2})=c_{1}\op{E}(s_{1})+c_{2}\op{E}(s_{2})$ for any $s_{1}^A,$ $s_{2}^A$ and any $c_{1},c_{2}\in \bb{R}$
    
    \item \label{A32} $\op{E}(s)^{\ast}=\op{E}(s)$ for all $s^{A}$ 
    
    \item \label{A33} $[\op{E}(s_{1}),\op{E}(s_{2})] = -i 4\pi^2  \Omega^{\textrm{EM}}_{\scri}(s_{1},s_{2})\op{1}$ for any $s_{1}^A,$ $s_{2}^A$ 
\end{enumerate}
 
We shall denote states on the algebra $\Alg_{\inn}^{\EM}$ as $\s^{\EM}$. The Hadamard regularity condition on asymptotic states of the electromagnetic field analogous to \cref{sympsmear} is that the $2$-point function has the form 
    \be 
    \label{hadformSR3}
   \s^{\EM}(\op{E}_{A}(x_{1})\op{E}_{B}(x_{2}))= -\frac{q_{AB}\delta_{\mathbb{S}^{2}}(x^{A}_{1},x^{A}_{2})}{(v_{1}-v_{2}-i0^{+})^{2}} + S_{AB}(x_{1},x_{2})
    \ee
where $S_{AB}$ is a (state-dependent) smooth bi-tensor on $\scri^{-}$ that is symmetric under the simultaneous interchange of $x_{1},x_{2}$ and the indices $A,B$.  Furthermore, we require that the connected $n$-point functions for $n\neq 2$ of a Hadamard state on $\scri^{-}$ are smooth.\footnote{It is possible that, in analogy with the bulk theory \cite{Sanders:2009sw}, this requirement actually a consequence of \cref{hadformSR3}. However, we have not investigated whether this is the case.} The $2$-point function of the Poincar\'e invariant vacuum state $\s^{\EM}_{0}$ is given by \cref{hadformSR3} with $S_{AB}=0$. 

Finally, we impose a decay condition on states to ensure that all fluxes are well-defined. We require states to be such that $S_{AB}$ and all connected $n$-point functions for $n\neq 2$ decay for any set of $|v_{i}|\to \infty$ as $O((\sum_{i}v_{i}^{2})^{-\half-\epsilon})$ for some $\epsilon>0$.  This completes the specification of $\Alg^{\rm EM}_{\inn}$ and the allowed states on $\Alg^{\rm EM}_{\inn}$.

We turn now to the asymptotic quantization of a massive complex scalar field \(\varphi\) in Minkowski spacetime. We follow the same basic strategy of finding an appropriate asymptotic surface that can be treated as a Cauchy surface. We then obtain the asymptotic description of the classical phase space by finding appropriate initial data on the asymptotic surface and we express the symplectic form in terms of this initial data. We then use \cref{321id} to obtain observables involving this initial data that correspond to local observables in the bulk, and we obtain the Poisson brackets of these observables. This enables us to define $\Alg^{\rm KG}_{\inn}$.

\begin{figure}[h!]
	\centering
	\includegraphics[width=0.4\textwidth]{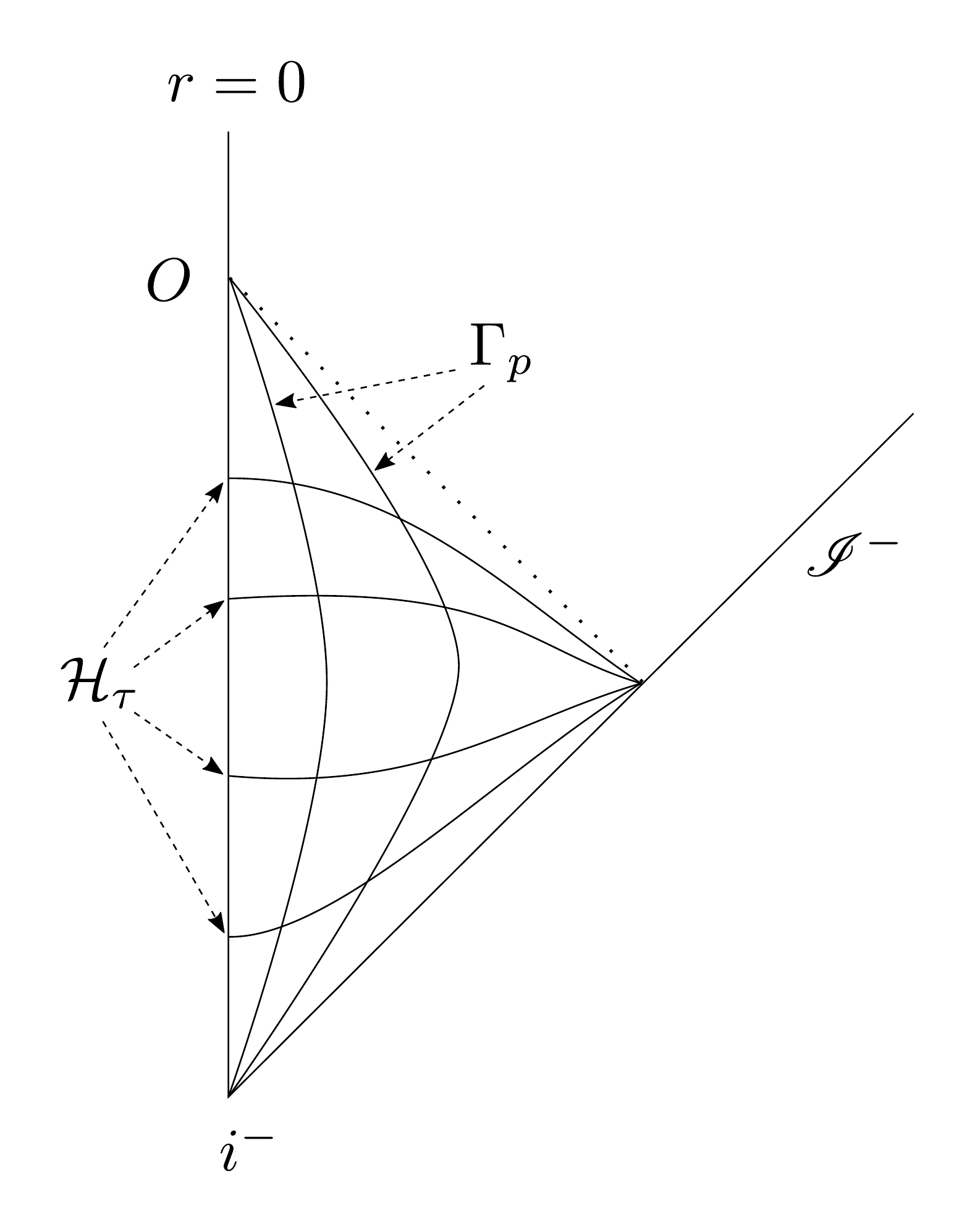}
	\caption{A schematic picture (with angular dimensions suppressed) of the family of hyperboloids \(\hyp_\tau\) used to take the limits to timelike infinity \(i^-\). The vertical line labeled \(r=0\) is the axis of rotational symmetry, \(O\) is an arbitrary choice of origin in Minkowski spacetime with past light cone depicted by a dotted line, and \(\scri^-\) denotes past null infinity. The \(\hyp_\tau\) are \(3\)-dimensional hyperboloids of \(\tau = \text{constant}\) with \(\tau \to -\infty\) corresponding to the limiting hyperboloid \(\hyp^-\) at \(i^-\). The \(\Gamma_p\) denote curves of constant \(p = (\rho, x^A_p)\) along which the limit to past timelike infinity is taken.}
    \label{fig:hyperboloids}
\end{figure}

For massive fields, the appropriate asymptotic surface is an asymptotic hyperboloid $\hyp^-$ rather than $\scri^-$. To see this, we introduce a coordinate system as follows \cite{Campiglia:2015qka} (see also \cite{Porrill,Cutler}). Let
\be
    \tau^2 \defn t^2 - r^2 \eqsp \rho \defn \tanh^{-1} r/t. 
    \label{taurho}
\ee
where $t,r$ are the standard Minkowski time and radial coordinates.
These coordinates foliate the interiors of the future/past light cone of an arbitrary choice of origin in Minkowski spacetime by a family of Riemannian hyperboloids with \(\tau = \text{constant}\) (see \cref{fig:hyperboloids}). The induced metric on the 3-dimensional unit-hyperboloid \(\hyp\) (with \(\tau^2 = 1\)) is given by
\be
    ds_\hyp^2 = \frac{d\rho^2}{1+\rho^2} + \rho^2 s_{AB} dx^A dx^B
\ee
where \(s_{AB}\) is the metric on the unit \(2\)-sphere and $x^A$ are coordinates on the sphere. The metric on a hyperboloid with \(\tau^2 \neq 1\) is just \(\tau^2 ds_\hyp^2\). Note that any point on \(\hyp\) can also be thought of as a unit-normalized timelike vector \(p\) in Minkowski spacetime, and we will use the notation \(p = (\rho , x_p^{A})\) to denote points on \(\hyp\). The induced volume element on \(\hyp\) is then
\be\label{eq:hypmeasure}
    d^3p \defn \frac{\rho^2}{\sqrt{1+\rho^2}} d\rho \dS.
\ee

The stationary phase method suggests that as $\tau \to - \infty$ at fixed $p$, there exists a gauge\footnote{In the Lorenz gauge for the electromagnetic vector potential, the scalar field in \cref{phiasymbeh} would have an additional overall phase \(e^{i q \log \tau}\) in its asymptotic behavior (see e.g. \cite{Campiglia:2015qka} or Ch. IV of \cite{Schiff_1949}). However this logarithmic ``Coulomb phase'' can be eliminated by a different choice of gauge. The vector potential in the Lorenz gauge is also badly behaved at null infinity (see \cref{fn:lorenz-A}).} such that (up to a constant phase factor) the leading order asymptotic behaviour of \(\varphi\) is given by \cite{Campiglia:2015qka} (see also \cite{Porrill}) \be
   \varphi \sim \frac{\sqrt{m}}{2 (2\pi \tau)^{3/2}} \lb[ c(p) e^{-im\tau} + i \bar b(p) e^{im\tau} \rb]
\label{phiasymbeh}
\ee
where \(p\) denotes a future-directed unit-normalized momentum and thus, a point \(p = (\rho, x_p^{A})\) on the unit-hyperboloid \(\hyp^-\) in the tangent space at past timelike infinity. Note that although each hyperboloid of constant $\tau$ extends to past null infinity, the hyperboloid \(\hyp^-\) corresponds to taking the limit $\tau \to - \infty$ at fixed $p = (\rho, x^A_p)$ and thus gives a representation of unit-timelike directions at past timelike infinity closely analogous to the description of spatial infinity given by Ashtekar and Hansen \cite{AH}.\footnote{A similar analysis at timelike infinity can be found in \cite{Porrill,Cutler}.} We will assume that the asymptotic behavior of $\varphi$ is given by \cref{phiasymbeh} and that $\hyp^-$ can be treated as a Cauchy surface. 

The initial data on $\hyp^{-}$ of a solution consists of the complex functions $b(p)$ and $c(p)$ appearing in \cref{phiasymbeh}. The symplectic form on this initial data can be written as \cite{Campiglia:2015qka}\footnote{Our convention for the symplectic form differs from that in \cite{Campiglia:2015qka} by a factor of \(-\half\).}
\be\label{eq:symp-massive}
    \Omega_{i^-}^\KG(\varphi_1,\varphi_2) = -\frac{im^2}{4(2\pi)^3} \int_{\hyp^-} d^3p~ \lb[ b_1(p) \bar b_2(p) + c_1(p) \bar c_2(p) - (1 \lra 2) \rb].
\ee
The symplectic form is a real-bilinear map on the real and imaginary parts of $b$ and $c$, but it is not complex-bilinear (or complex-bi-antilinear) in $b$ and $c$. On the complex plane, it is often convenient to treat $z=x + i y$ and $\bar{z} = x - iy$ as though they were independent quantities, imposing that they are conjugates only at the end of any calculation. For similar reasons, it is convenient to treat $\bar{b}$ and $\bar{c}$ as though they are quantities independent of $b$ and $c$ on phase space, imposing that they be conjugates of $b$ and $c$ at the end of any calculation. Thus, we will take a point in the asymptotic description of the classical phase space to be represented as the quadruple $(b(p),\bar{b}(p),c(p),\bar{c}(p))$. The symplectic form is then a complex-bilinear function of its variables.

Local observables are obtained by smearing the fields with a \emph{complex} test function $w(p)$ on $\hyp^{-}$.  The smeared fields $b(w)$ and $\bar{b}(w)$ are defined by
\be 
b(w)\defn \int_{\hyp^{-}}d^{3}p~b(p)\bar{w}(p) \, , \quad \quad \bar{b}(w)\defn \int_{\hyp^{-}}d^{3}p~\bar{b}(p)w(p).
\ee 
Note that we take $b(w)$ to be anti-linear in $w$ whereas $\bar{b}(w)$ is linear in $w$, so $\bar{b(w)} = \bar{b} (w)$. Note also that the Hamiltonian vector fields corresponding to these observables are given by $-\frac{4(2\pi)^{3}}{im^{2}}(0,\bar{w}(p),0,0)$ and $\frac{4(2\pi)^{3}}{im^{2}}(w(p),0,0,0)$ respectively. We similarly define the smeared local observables 
\be 
c(w)\defn \int_{\hyp^{-}}d^{3}p~c(p)w(p) \, , \quad \quad \bar{c}(w)\defn \int_{\hyp^{-}}d^{3}p~\bar{c}(p)\bar w(p)
\ee 
where we now take $c(w)$ to be linear and $\bar{c}(w)$ to be anti-linear in $w$. The nontrivial Poisson brackets are
    \begin{equation} 
    \pb{b (w_{1})}{\bar{b}(w_{2})} = -i \frac{4(2\pi) ^3}{m^2} \braket{w_{1},w_{2}}_{\hyp^-} 1 \eqsp \pb{{c}(w_{1})}{\bar{c}(w_{2})}= -i \frac{4(2\pi) ^3}{m^2} \braket{w_{2},w_{1}}_{\hyp^-} 1.
    \label{scalpbrel}
    \end{equation} 
    Here, the inner product $\braket{w_{1},w_{2}}_{\hyp^-}$ is the ordinary $L^{2}$ inner product on $\hyp^{-}$ with the volume element \cref{eq:hypmeasure}, which is antilinear in its first argument and is linear in its second argument. 

The asymptotic quantization algebra, $\Alg^{\KG}_{\inn}$, for the massive complex scalar field  is then defined by starting with the free algebra generated by the smeared fields $\op{b}(w)$, $\op{c}(w)$, their formal adjoints $\op{b}(w)^{\ast},\op{c}(w)^{\ast}$ and an identity $\op{1}$. We note that the adjoint operators $\op{b}(w)^{\ast}$ and $\op{c}(w)^{\ast}$ correspond to the complex conjugate observables $\bar{b}(w)$ and $\bar{c}(w)$ respectively. We then factor this algebra by the analog of the linearity condition \ref{A31}, the commutation relations
    \begin{equation} 
    [\op{b}(w_{1}),\op{b}(w_{2})^{\ast}] = \frac{4(2\pi) ^3}{m^2}\braket{w_{1},w_{2}}_{\hyp^-}\op{1} \eqsp [\op{c}(w_{1}),\op{c}(w_{2})^{\ast}]= \frac{4(2\pi) ^3}{m^2} \braket{w_{2},w_{1}}_{\hyp^-} \op{1}
    \label{scalcommrel}
    \end{equation} 
and vanishing commutators for all other fields. This completes the specification of the algebra $\Alg^{\KG}_{\inn}$ of local observables of the massive, charged Klein-Gordon field.

The Hadamard condition for states $\s^{\KG}$ on $\Alg^{\rm KG}_{\inn}$, is that the $2$-point functions $\s^{\KG}(\op{b}(p_{1})\op{b}(p_{2}))$, $\s^{\KG}(\op{c}(q_{1})\op{c}(q_{2}))$, $\s^{\KG}(\op{b}(p_{1})\op{c}(q_{2}))$, and $\s^{\KG}(\op{b}(p_{1})^{\ast}\op{c}(q_{1}))$ are smooth, whereas the remaining $2$-point functions have the form
\begin{subequations}\label{bchad}\begin{align}
    \s^{\KG}(\op{b}(p_{1})\op{b}(p_{2})^{\ast}) &=  \frac{4(2\pi) ^3}{m^2} \delta_{\hyp}(p_{1},p_{2})+B(p_{1},p_{2})
    \label{bhad} \\
    \s^{\KG}(\op{c}(q_{1})\op{c}(q_{2})^{\ast}) &=  \frac{4(2\pi) ^3}{m^2} \delta_\hyp(q_{2},q_{1})+C(q_{2},q_{1})
    \label{chad}
\end{align}\end{subequations}
where it is understood that $\delta_{\hyp}$ is to be smeared with a complex conjugate test function $\bar{w}$ in its first argument and a test function $w$ in its second argument, and the  functions $B$ and $C$ are (state-dependent) smooth functions on $\mc{H}\times \mc{H}$. Furthermore, the connected $n$-point functions for $n \neq 2$ of $\s^{\KG}$ are required to be smooth. Note that the $2$-point function of the Poincar\'e invariant vacuum state $\s^{\KG}_{0}$ is given by \cref{bchad} with $B=C=0$. 

In addition, we impose the following decay condition on states: We require that $B$, $C$ and all the connected $n$-point functions for $n\neq 2$ of $\s^{\KG}$ decay for any set of $|p_{i}|\to \infty$ as $O((\sum_{i}p_{i}^{2})^{-\half-\epsilon})$ for some $\epsilon>0$. This completes our specification of the regularity conditions on states.

\subsection{Extension of the asymptotic quantization algebra to include charges and Poincar\'e generators}
\label{subsec:chmem}

The algebra $\Alg_{\inn}$ that we have defined in the previous subsection was generated by the local field observables of the asymptotic ``in'' fields. Thus, the only observables represented in $\Alg_{\inn}$ are the local fields. However, there are additional observables of interest, where, here and elsewhere in this paper, we use the term ``observables'' in the precise sense explained in \cref{sec:classphase}. In this section, we will extend $\Alg_{\inn}$ to the algebra $\Algex$ by the addition of generators of large gauge transformations (i.e. ``charges''). We will then further extend this algebra to an algebra $\Algexqp$ that includes the generators of Poincar\'e symmetries. We will construct these algebras by obtaining observables on the classical phase space that generate large gauge transformations and Poincar\'e symmetries. These observables automatically have well-defined Poisson brackets with themselves and with the local fields. We then will obtain $\Algex$ by starting with the free algebra generated by $\Alg_{\inn}$ together with the observables that generate large gauge transformations and then factoring by the commutation relations obtained from the Poisson brackets. We will then further enlarge this algebra to $\Algexqp$ by including the observables that generate Poincar\'e symmetries.

We first consider the large gauge charges. As stated above, QED has an invariance under \cref{gaugetrans}. The transformations \cref{gaugetrans} with $\LGT$ vanishing at infinity are genuine gauge transformations in the sense that the infinitesimal versions of these transformations are degeneracies of the symplectic form. In order to construct a phase space with a nondegenerate symplectic form, one must pass to the space of gauge orbits \cite{Lee_1990}, so fields that differ by a gauge transformation correspond to the same point of phase space. However, as previously noted in \cref{sec:intro}, the transformations \cref{gaugetrans} with $\LGT = \LGT(x^A)$ are not degeneracies of the symplectic form. Such ``large gauge transformations'' must be treated as symmetries and they act nontrivially on the classical phase space. The infinitesimal version of these symmetries defines a vector field on phase space. We will show that this vector field on phase space is generated by a classical observable, which will be referred to as a ``charge.''	 Consequently, we can expect that the quantum algebra $\Alg_{\inn}$ can be extended to include quantum representatives of the charges.

Since the asymptotic description of phase space is the Cartesian product of the Klein-Gordon and Maxwell phase spaces, we can separately consider the action of large gauge transformations on the Klein-Gordon and Maxwell fields separately. We will thereby obtain two ``charges'': (i) a charge $\mc Q_{i^-}$ that generates large gauge transformations on the Klein-Gordon field and (ii) a ``memory'' quantity that generates large gauge transformations on the Maxwell field. The sum of these two, denoted $\mc Q_{i^0}$, generates large gauge transformations on the full phase space. The reason for the use of ``${i^0}$'' in the notation for the total charge will be explained below.

We first consider the action of the large gauge transformations on the classical Klein-Gordon phase space, i.e., on the asymptotic fields on $\hyp^-$. The large gauge transformations are parametrized by a smooth function \(\LGT(x^{A})\) on \(\bb S^2\), which describes the asymptotic behavior of the transformation \cref{gaugetrans} on the scalar field as $\rho \to \infty$. It is useful to pick a unique representative of this transformation throughout $\hyp^-$ as follows.  Let \(\LGT_\hyp(p)\) be the unique function on \(\hyp^-\) which satisfies
\be\label{eq:lgt-hyp-eqn}
    \triangle_\hyp \LGT_\hyp(p) = 0 \eqsp \lim_{\rho \to \infty} \LGT_\hyp(p) = \LGT(x^{A})
\ee
where \(\triangle_\hyp\) is the Laplace operator on \(\hyp^-\). The solution \(\LGT_\hyp(p)\) can be expressed in terms of the boundary value \(\LGT(x^{A})\) using a Green's function as \cite{C-Green}
\be\label{eq:lgt-hyp-soln}
    \LGT_\hyp(p) = \int_{\bb{S}^{2}} \dS~ G_\hyp(p, x^{A}) \LGT(x^{A}) \eqsp G_\hyp(p,x^{A}) = \frac{1}{4\pi (\sqrt{1+\rho^2} - \rho \hat{p} \cdot \hat r )^2 } .
\ee
Here, $\hat{r}$ is the unit vector in $\bb{R}^{3}$ corresponding to the point $x^{A}$ on the unit $2$-sphere, and $\hat{p}$ denotes the projection of the point \(p \in \hyp^-\) onto the unit $2$-sphere, also represented as a unit vector in $\bb{R}^{3}$. The Euclidean dot product $\hat{p}\cdot \hat{r}$ of these unit vectors is the cosine of the geodesic distance between two points on $\bb{S}^{2}$ with respect to the unit $2$-sphere metric. Note that this Green's function satisfies
\be
\label{eq:S2intGH}
    \int_{\bb S^2} \dS~ G_\hyp(p,x^A) = 1.
\ee
In terms of $\LGT_{\hyp}$, the action of the large gauge transformations on the asymptotic scalar field is given by
\be
    b(p) \mapsto b(p) e^{-iq\LGT_\hyp(p)} \eqsp \bar b(p) \mapsto \bar b(p) e^{iq\LGT_\hyp(p)} \eqsp c(p) \mapsto c(p) e^{iq\LGT_\hyp(p)} \eqsp \bar c(p) \mapsto \bar c(p) e^{-iq\LGT_\hyp(p)} .
    \label{gaugemassive0}
\ee
The infinitesimal action of large gauge transformations on phase space is given by
\be 
(b(p),\bar{b}(p),c(p),\bar{c}(p))\to (b(p),\bar{b}(p),c(p),\bar{c}(p)) + i q \LGT_{\hyp}(p) \epsilon~ (-b(p),\bar{b}(p),c(p),-\bar{c}(p)).
\label{lgvf}
\ee 
This transformation is of the form \cref{Xaffine} with $\chi_0 = 0$ and
\be
L (b,\bar{b}, c,\bar{c}) = (b',\bar{b^{\prime}}, c',\bar{c^{\prime}})
\ee
with $b'(p) = - i q \LGT_\hyp(p) b(p)$, $c'(p) =  i q \LGT_\hyp(p) c(p)$. 
The linear map $L$ satisfies \cref{Bsym} so we obtain the observable
\be
\label{classicchargei-}
    \mc Q_{i^-}(\LGT) \defn \frac{1}{2} \Omega_{i^-}^\KG( (b,\bar{b},c,\bar{c}), L(b,\bar{b},c,\bar{c})) =  \frac{q m^2}{4(2\pi)^3}  \int_{\hyp^-} d^3p~ \LGT_\hyp(p) \lb[ b(p) \bar b(p) - c(p) \bar c(p)  \rb] .
\ee
Note that the integrand on the right-hand-side of \cref{classicchargei-} corresponds to the asymptotic limit to $\hyp^-$ of $J_\mu \tau^\mu$, where $\tau^\mu$ is the unit normal to the surfaces of constant $\tau$ and the charge-current vector $J^\mu$ of the scalar field is given by
\be
J^\mu = -\frac{iq}{2} [\bar\varphi D^\mu \varphi - \varphi D^\mu \bar\varphi] \, .
\label{JKG}
\ee
Thus, for \(\LGT(x^{A}) = \text{constant}\), $\mc Q_{i^-}$ is the total ordinary electric charge of the massive scalar field \cite{Campiglia:2015qka}.

Since the observables $\mc Q_{i^-}(\LGT)$ generate the large gauge transformations \cref{lgvf}, it is straightforward to compute their Poisson brackets. The Poisson brackets of the charges with themselves vanish
\be
\pb{\mc Q_{i^-}(\LGT_1)}{\mc Q_{i^-}(\LGT_2)} = 0
\label{pbchch}
\ee
The Poisson brackets of the charges with the smeared fields are
\be
    \pb{\mc Q_{i^-}(\LGT)}{b(w)} = -iqb( \LGT_\hyp w) &\eqsp \pb{\mc Q_{i^-}(\LGT)}{\bar b(w)} = iq\bar b( \LGT_\hyp w) \\
    \pb{\mc Q_{i^-}(\LGT)}{c(w)} = iqc( \LGT_\hyp w) &\eqsp \pb{\mc Q_{i^-}(\LGT)}{\bar c(w)} = -iq\bar c(\LGT_\hyp w) \, .
\label{pbchbc}
\ee
Since $\mc Q_{i^-}(\LGT)$ is an observable on the Klein-Gordon phase space, it has vanishing Poisson bracket with all electromagnetic observables.

We now consider the action of large gauge transformations on the Maxwell phase space. The large gauge transformation \(\LGT = \LGT(x^A)\) acts on the Maxwell phase space by
\be
    A_A \mapsto A_A + \ms D_A \LGT \eqsp E_A \mapsto E_A.
\label{lgauge}
\ee
This affine transformation is generated by $\tfrac{1}{4\pi}\Delta(\LGT)$ where \(\Delta(\LGT)\) is defined by
\be
    \Delta (\LGT) \defn -\int_{\scri^-} dv\dS~ E_A(v,x^{B}){\ms D}^A \LGT(x^{B}).
\label{memobs}
\ee
Thus, $ \Delta(\lambda)$ is an observable on the Maxwell phase space. We refer to $\Delta (\lambda)$ as the {\em memory} of the Maxwell field associated with the large gauge transformation $\lambda$. Since the local electromagnetic field observables $E(s)$ (see \cref{esmear}) are invariant under \cref{lgauge}, $ \Delta(\lambda)$ has vanishing Poisson bracket with all local electromagnetic field observables. Since $ \Delta(\lambda)$ is an observable on the Maxwell phase space, it also has vanishing Poisson bracket with all Klein-Gordon observables.

The generator of gauge transformations on the full phase space of the Klein-Gordon and Maxwell fields is given by the sum of \cref{classicchargei-} and \cref{memobs}
\be
\mc Q_{i^0}(\lambda) \defn \mc Q_{i^-}(\lambda) + \frac{1}{4\pi}\Delta (\lambda) \, .
\label{Q0def}
\ee
The Poisson brackets of $\mc Q_{i^0}(\lambda)$ with all local field observables are the same as those of $ \mc Q_{i^-}(\lambda)$.
The subscript ``$i^0$'' has been placed on $\mc Q_{i^0}(\lambda)$ because its value can be computed by taking limits of surface integrals of the electric field as one approaches spatial infinity, $i^0$, along $\scri^-$. This can be shown by the following lengthy argument.

First, we show that the charge $\mc Q_{i^-}(\lambda)$ can be computed as a bulk limit of the electric field. In the bulk spacetime, the massive scalar is coupled to the electromagnetic field via the Maxwell equation
\be\label{eq:Max-eqn}
    \frac{1}{4\pi} \nabla_\nu F^{\nu \mu} = J^\mu
\ee
with $J^\mu$ given by \cref{JKG}. We assume that the limit to $\hyp^-$ of the ``electric field''
\be
    \mc E_\mu (p) = \lim_{\tau \to - \infty} \tau^2 {h_\mu}^\nu F_{\nu \sigma} \tau^\sigma \,
    \label{Ehyp}
\ee
exists and defines a smooth tensor field ${\mc E}_a$ on \(\hyp^-\), where \(h_{\mu \nu} = g_{\mu \nu} + \tau_\mu \tau_\nu\) is the induced metric on the hyperboloids of constant $\tau$. From the Maxwell equation (\ref{eq:Max-eqn}) and the falloff of the scalar field, it follows that there exists an electric potential \(V(p)\) on $\hyp^-$ so that \(\mc E_a(p) = \mc{D}_a V(p)\), which satisfies
\be\label{eq:Lap-V}
    \frac{1}{4\pi}\triangle_\hyp V(p) = \frac{qm^2}{4(2\pi)^3}  \lb[ b(p) \bar b(p) - c(p) \bar c(p)  \rb]\,
\ee
where ${\mathcal D}_a$ denotes the derivative operator on $\hyp^-$ and \(\triangle_\hyp\) again denotes the Laplacian on $\hyp^-$. By Green's identity, for any large gauge transformation $\lambda$, we have
\be
\frac{1}{4\pi}{\mathcal D}^a \left( \lambda_\hyp {\mathcal D}_a V - V {\mathcal D}_a \lambda_\hyp \right) &= \frac{1}{4\pi} ( \lambda_\hyp \Delta_\hyp V - V  \Delta_\hyp \lambda_\hyp) \\
&=  \frac{qm^2}{4(2\pi)^3}  \lb[ b(p) \bar b(p) - c(p) \bar c(p)  \rb]
\ee
Integrating this equation over $\hyp^-$ and applying Gauss' theorem to the left side, we obtain
\be
\lim_{\rho \to \infty} \frac{1}{4\pi} \int_{\bb S^2}\dS~ \lambda(x^{A}) (\cosh\rho)^2 \rho^a \mc E_a & = \frac{qm^2}{4(2\pi)^3}  \int_{\hyp^-} d^3p~ \lambda_\hyp(p) \lb[ b(p) \bar b(p) - c(p) \bar c(p)  \rb] \\ 
&= \mc Q_{i^-}(\lambda)
\label{gaussform}
\ee
where \(\rho^a\) is the unit-spacelike-normal to the \(\rho = \text{constant}\) cross-sections of \(\hyp^-\). 
Thus, as we desired to show, the charge $\mc Q_{i^-}(\lambda)$ can be obtained as an asymptotic surface integral as $\rho \to \infty$ of the electric field ${\mc E}_a$ on $\hyp^-$, which itself is obtained as the bulk limit \cref{Ehyp} as $\tau \to -\infty$. For \(\lambda(x^{A}) = \text{constant}\), \cref{gaussform} corresponds to the usual Gauss law formula for charge.

Next, we assume that the analogue of the ``null regularity'' condition imposed at spatial infinity in \cite{KP-EM-match} holds at timelike infinity. This yields
\be\label{eq:timelike-NR}
    \lim_{\rho \to \infty} (\cosh\rho)^2 \rho^a \mc E_a \quad (\text{along } \hyp^-) = \lim_{v \to - \infty} F_{\mu\nu} l^\mu n^\nu \quad (\text{along } \scri^-)
\ee
where \(l^\mu\) is a vector field at \(\scri\) satisfying \(l_\mu l^\mu = 0\) and \(l^\mu n_\mu = -1\). This quantity is not a function on the electromagnetic phase space, i.e., it depends on non-radiative (Coulombic) information at $\scri^-$ that is obtained from bulk limits. It follows that
\be
\mc Q_{i^-}(\lambda) = \lim_{v \to - \infty} \frac{1}{4\pi} \int_{\bb{S}^{2}} \dS~\lambda(x^{A}) F_{\mu\nu} l^\mu n^\nu.
\ee

Now, Maxwell's equations imply that on $\scri^-$ we have
\be\label{eq:n_maxwell}
  \Lie_n (F_{\mu\nu} l^\mu n^\nu) =  \ms D^A E_A.
\ee
It follows immediately that
\be\label{eq:memphi1}
 \Delta (\lambda) &=  \frac{1}{4} \int_{\scri^-} dv \dS~\lambda(x^{A}) \frac{\partial (F_{\mu\nu} l^\mu n^\nu)}{\partial v} \\
 &=  \lim_{v \to + \infty} \frac{1}{4} \int_{\bb{S}^{2}} \dS~ \lambda(x^{A}) F_{\mu\nu} l^\mu n^\nu -  \lim_{v \to - \infty} \frac{1}{4} \int_{\bb{S}^{2}} \dS ~\lambda(x^{A}) F_{\mu\nu} l^\mu n^\nu.
\ee
Thus, we obtain our desired result
\be
\mc Q_{i^0}(\lambda) = \lim_{v \to + \infty} \frac{1}{4\pi}\int_{\bb{S}^{2}} \dS~\lambda(x^{A}) F_{\mu\nu} l^\mu n^\nu
\ee
which shows that $\mc Q_{i^0}(\lambda)$ can be computed in terms of a limit of $F_{\mu\nu} l^\mu n^\nu$ as one approaches $i^0$ along $\scri^-$. However, it should be kept in mind that $F_{\mu\nu} l^\mu n^\nu$ is not an observable on the Klein-Gordon and Maxwell phase spaces. The definition of $\mc Q_{i^0}(\lambda)$ as an observable is given by \cref{Q0def,classicchargei-,memobs}.

It is worth noting that on any cross-section \(S \cong \bb S^2\) of \(\scri\), we can define the Maxwell charge associated with the large gauge transformation \(\lambda\) by
\be\label{eq:EM-charge}
    \mc Q_S(\lambda) = \frac{1}{4\pi} \int_S \dS~ \lambda(x^{A})  F_{\mu\nu} l^\mu n^\nu .
\ee
This notion of ``large gauge charge at a finite advanced time'' may be useful for a number of purposes. However, unlike $\mc Q_{i^-}(\lambda)$ and $\mc Q_{i^0}(\lambda)$, the quantity $ \mc Q_S(\lambda)$ does {\em not} correspond to an observable on phase space. Thus, $ \mc Q_S(\lambda)$ does not have well-defined Poisson brackets with local field observables and we cannot expect $ \mc Q_S(\lambda)$ to have a well-defined counterpart in quantum field theory.\footnote{The nonexistence of operators corresponding to $ \mc Q_S(\lambda)$ in quantum field theory is in agreement with the arguments given in \cite{Bousso_2017}.}

We now turn to the extension of the algebra, $\Alg_{\inn}$, of local quantum observables to an algebra $\Algex$ that includes the large gauge charges $\mc Q_{i^-}(\lambda)$ and $\Delta(\lambda)$ (and thereby, $\mc Q_{i^0}(\lambda)$). 
We start with the free algebra generated by the smeared local field observables together with observables labeled as $\op{\mc Q}_{i^-}(\lambda)$ and $\op\Delta(\lambda)$ and their adjoints for all large gauge transformations \(\lambda(x^{A})\). We then factor this algebra by linearity in the test functions and $\lambda$ as well as Hermitian conditions for $\op{E}$, $\op{\mc Q}_{i^-}$, and $\op\Delta$. Finally, we factor the algebra by the commutation relations corresponding to all of the Poisson bracket relations we have obtained above.
This defines the desired extended algebra $\Algex$. 

However, since $\op\Delta(\lambda)$ commutes with all observables\footnote{Note, however, that this will not be the case after we further extend the algebra to $\Algexqp$ by including Poincar\'e generators, as we shall do below.} in $\Algex$, it follows that in any representation of $\Algex$, a shift of $\op\Delta(\lambda)$ by a multiple of the identity --- with no corresponding shift of $\op{E}$ --- would also yield a representation of the algebra. This implies that there are many states on the extended algebra where the value of the memory observable is not related to value of the electric field by \cref{memobs}. In order to exclude such states, we require as a further condition on states $\s$ (in addition to the Hadamard condition and the fall-off conditions of \cref{subsec:MaxMKGclass}) that\footnote{We could also impose conditions on higher $n$-point functions of memory, but we will not need these, so we shall not impose them here.}
\be
\omega(\op\Delta(\lambda)) =  -\int_{\scri^-}dv \dS~  \omega(\op E_A(v,x^{B})){\ms D}^A \lambda(x^{B}).
\label{mem1ptfun}
\ee
 
We have a similar multiple of the identity ambiguity for operator representatives of \(\op{\mc Q}_{i^-}(\lambda)\). We could similarly impose the additional requirement on states that
\be
\omega\!\left(\op{\mc Q}_{i^-}(\lambda) \right) = \frac{m^2}{4(2\pi)^3} q \int_{\hyp^-} d^3p~ \lambda_\hyp(p)~ \omega \!\left( \op b(p)^* \op b(p) - \op c(p)^* \op c(p)  \right)  
\label{Q1ptfun}
\ee
where the expected value of quadratic quantities was defined at the end of \cref{sec:revhad}.
However, for the massive scalar field, we will only be interested in states in the standard Fock space (see \cref{subsubsec:MaxMKGFockrepiminus}). For such states, instead of demanding \cref{Q1ptfun}, we can more simply demand that the charge \(\op{\mc Q}_{i^-}(\lambda)\) annihilates the Fock vacuum state, since this removes the multiple of the identity ambiguity in this representation and implies that \cref{Q1ptfun} holds for all states in this representation.

We now follow the same strategy to further extend the algebra $\Algex$ to an algebra $\Algexqp$ that also includes observables corresponding to the generators of Poincar\'e transformations. The first step is to write down the action of the Poincar\'e group on the classical phase space and show that its infinitesimal action is generated by an observable on the classical phase space. Again, we may consider the Poincar\'e action on the Maxwell phase space and the Klein-Gordon phase space separately. 

Poincar\'e transformations correspond to a particular class of diffeomorphisms of $\scri^-$, and these act naturally on the fields $A_A$ and $E_A$, so it is straightforward to determine their action on the Maxwell phase space. Lorentz transformations (with origin taken to be that used to define the hyperboloids of \cref{taurho}) similarly correspond to isometries of $\hyp^-$ and thus have a natural action on the asymptotic Klein-Gordon fields. Thus, we only need to explain the action of translations.
In terms of the asymptotic coordinates used above, it can be shown that, at leading order, a translation corresponds to the transformation \(\tau \mapsto \tau + f_\hyp(p)\) where the function \(f_{\hyp}(p)\) satisfies\footnote{One can also represent BMS supertranslations at timelike infinity by considering solutions \(f_\hyp(p)\) to \cref{eq:hyp-translations} on \(\hyp^-\) with the boundary value \(f(x^A)\) now being allowed to be any smooth function on \(\bb S^2\).}
\be\label{eq:hyp-translations}
    (\triangle_\hyp - 3) f_\hyp(p) = 0 \eqsp \lim_{\rho \to \infty} \rho^{-1} f_\hyp(p) = f(x^{A})
\ee
with \(f(x^{A})\) is a smooth function on \(\bb S^2\) supported only on the \(\ell = 0,1\) spherical harmonics. The solution can again be written in terms of a Green's function as \cite{C-Green}
\be\label{st-hyp-soln}
    f_\hyp(p) = \int_{\bb S^2} \dS~ \tilde G_\hyp(p , x^{A}) f(x^{A}) \eqsp \tilde G_\hyp(p,x^{A}) = \frac{1}{4\pi (\sqrt{1+\rho^2} - \rho \hat{p} \cdot \hat{r} )^3 } 
\ee
where the notation $\hat{p}$ and $\hat{r}$ is as explained below \cref{eq:lgt-hyp-soln}.
The action of these translations on the asymptotic fields is given by
\be
    b(p) \mapsto b(p) e^{-imf_\hyp(p)} \eqsp \bar b(p) \mapsto \bar b(p) e^{imf_\hyp(p)} \eqsp c(p) \mapsto c(p) e^{-imf_\hyp(p)} \eqsp \bar c(p) \mapsto \bar c(p) e^{imf_\hyp(p)} .
\ee

The infinitesimal action of an arbitrary Poincar\'e transformation on both the Maxwell and Klein-Gordon phase spaces is thus given by a linear transformation, $P$. It can be verified that $P$ satisfies \cref{Bsym}. We thereby obtain an observable, $F_P$, on phase space corresponding to an arbitrary infinitesimal Poincar\'e transformation $P$ by the formula,
\be
F_P (\phi) = \frac{1}{2} \Omega(\phi, P \phi)
\ee
where ``$\phi$'' stands for a point in the Cartesian product of the Maxwell and Klein-Gordon phase spaces. For a given choice of origin in the bulk spacetime, we can write an arbitrary infinitesimal Poincar\'e transformation as $P = T + X$ where $T$ is a translation and $X$ is a Lorentz transformation. We denote the corresponding observables as $F_T$ and $F_X$. The Poisson bracket of these observables with the local  asymptotic massive scalar field observables are given by
\be
\{F_T, b(w) \} = ib(f_{\hyp}w) \eqsp \{F_X, b(w) \} = b(\Lie_{X}w)
\label{transbc}
\ee
where \(\Lie_X w\) is the Lie derivative of the complex test function \(w(p)\) with respect to the Killing vector field on \(\hyp\) representing the Lorentz transformation \(X\). Analogous formulae also hold for the observable \(c(w)\). The Poisson bracket of the Poincar\'e obervables with the local electromagnetic field observables on \(\scri^-\) are given by
\be
\label{eq:PBFE}
\{F_T,  E(s)\} = E(\Lie_{T}s) \eqsp \{F_X,  E(s)\} = E(\Lie_{X}s)
\ee
where, now, the Poincar\'e translation \(T\) is represented by the vector field \(fn^\mu\) on \(\scri^-\) with \(f\) supported on the \(\ell=0,1\) spherical harmonics and the Lorentz transformation is a conformal Killing vector field \(X^A\) on \(\bb S^2\). Further, the Poisson brackets of the Poincar\'e observables with memory and charges are given by
\begin{subequations}\begin{align}
 \{F_T,\Delta(\LGT)\}=0 &\eqsp \{F_T,\mc{Q}_{i^{-}/i^{0}}(\LGT)\}=0
 \label{eq:PBFTmem} \\
\{ F_X,\Delta(\LGT)\}=\Delta(\Lie_{X}\LGT) &\eqsp \{F_X,\mc{Q}_{i^{-}/i^{0}}(\lambda)\}=\mc{Q}_{i^{-}/i^{0}}(\Lie_{X}\lambda).
\label{memlor}
\end{align}\end{subequations}
Finally the brackets of the Poincar\'e generators with themselves are 
\begin{subequations}\begin{align}
&\{F_{T_1}, F_{T_2}\}=0 \eqsp \{F_{X_1}, F_{X_2}\} = F_{[X_{1},X_{2}]} \label{eq:PBFTFX} \\
&\qquad \qquad   \{F_X, F_T \}= F_{T'}
\label{lortrans}
\end{align}\end{subequations} 
where in the above, if \(T\) is represented by a function \(f(x^A)\) then \(T'\) is represented by $f' = \Lie_{X}f - \half(\ms{D}_{A}X^{A}) f$. It should be noted that \cref{memlor} shows that memory is not Lorentz invariant unless it vanishes. Similarly, the charges at timelike and spatial infinity are not Lorentz invariant unless all of the charges (including the ordinary total electric charge) vanish. In addition, the nonvanishing of the Poisson brackets of $F_X$ with $\mc{Q}_{i^0}$ shows that the observables $F_X$ are not gauge invariant unless all of the charges vanish, as expected from the considerations given in \cite{Bonga:2019bim}.

The extended algebra $\Algexqp$ is now obtained by adding Hermitian elements $\op{F}_{\! P}$ for each Poincar\'e generator $P$ to $\Algex$ and factoring by commutation relations corresponding to all of the above Poisson bracket relations. 

\subsection{Fock representations}
\label{subsec:MaxMKGreps}

In the previous sections, we have constructed the asymptotic local field algebra $\Alg_{\inn}$ and we have extended it to the algebras $\Algex$ and $\Algexqp$ that include large gauge charges and Poincar\'e generators. The ordinary Minkowski vacuum state, $\omega_0\defn \s^{\KG}_{0}\otimes \s^{\EM}_{0}$, is the Gaussian state on $\Alg_{\inn}$ with vanishing $1$-point function and with $2$-point function given by \cref{hadformSR3,bchad} with $S_{AB} = B = C = 0$. We can extend its action to $\Algexqp$ such that for all $a \in \Algexqp$ we have $\omega_0(a\op\Delta) = \omega_0(\op\Delta a)= \omega_0(a\op{\mc Q}_{i^-}) = \omega_0(\op{\mc Q}_{i^-}a) = \omega_0 (a \op{F}_{\! P}) = \omega_0 ( \op{F}_{\! P} a) = 0$, i.e., such that $\omega_0$ is an eigenstate with eigenvalue zero of all the charges and Poincar\'e generators. The GNS representation of $\omega_0$ will yield the usual Fock space of incoming particle states. However, all states in this Fock space will be eigenstates of memory with vanishing eigenvalue. The corresponding construction of an ``out'' Hilbert space will similarly contain only states with vanishing memory. Consequently, as discussed at length in \cref{sec:intro}, this choice of Hilbert space is not adequate for scattering theory, since it does not contain the states with memory that arise from the scattering processes.

The purpose of this subsection is to construct a large supply of states --- including states with memory --- that later can be reassembled into a Hilbert space satisfying properties \ref{enum:S1}--\ref{enum:S5} of \cref{sec:intro}. We will do so by constructing the ordinary Fock representation of the asymptotic Klein-Gordon scalar field and the asymptotic electromagnetic field. We will then construct corresponding ``memory representations'' of the electromagnetic field by shifting the electromagnetic field by the identity multiplied by a classical electromagnetic field with the desired memory. We will thereby obtain states of the electromagnetic field with arbitrary memory.

\subsubsection{Fock representation of the massive field algebra}
\label{subsubsec:MaxMKGFockrepiminus}

In this subsection, we construct the standard Fock representation of the asymptotic Klein-Gordon scalar field. This will give us an ample supply of incoming states of the Klein-Gordon field.

The vacuum state on the extended asymptotic Klein-Gordon algebra $\Algexqp^{\rm KG}$ is the Gaussian algebraic state $\s^{\KG}_{0}$ with vanishing $1$-point function and $2$-point function given by (see \cref{bchad})
\be 
\label{2pti0}
\s^{\KG}_{0}(\op{b}(w_{1})\op{b}(w_{2})^{\ast}) = \frac{4(2\pi) ^3}{m^2} \braket{w_1, w_2}_\hyp \eqsp \s^{\KG}_{0}(\op{c}(w_{1})\op{c}(w_{2})^{\ast}) = \frac{4(2\pi) ^3}{m^2} \braket{w_2,w_1}_\hyp
\ee 
for all test functions $w_{1}(p),w_{2}(p)$, where $\braket{\,, \,}_\hyp$ is the $L^2$ inner product on $\hyp^{-}$. Furthermore, $\s^{\KG}_{0}$ is an eigenstate of eigenvalue zero of $\cop_{i^{-}}(\lambda)$ and all of the Poincar\'e generators.

The GNS construction for $\s^{\KG}_{0}$ yields a Hilbert space $\Fock^{\KG}$ with a natural Fock space structure. Concretely this construction is obtained as follows. On the space of complex test functions on \(\hyp^{-}\) we define the inner product 
\be
\braket{w_1 | w_2 } \defn \s^{\KG}_{0}(\op{b}(w_{1})\op{b}(w_{2})^*) = \frac{4(2\pi) ^3}{m^2} \braket{w_1, w_2}_\hyp
\ee
using the \(2\)-point function in \cref{2pti0}. This is a non-degenerate, positive, Hermitian inner product. Let \(\Hilb^{\KG}\) be the completion of the space of test functions in this inner product, i.e., \(\Hilb^{\KG}\) is the Hilbert space of square-integrable functions (i.e. wave packets) of the timelike momentum \(p\) represented as points on \(\hyp^{-}\). This Hilbert space serves as the ``one particle'' Hilbert space for particles in the Fock space. The Fock space of particles is given by 
\be 
\label{Focki0}
\Fock^{\KG}_{\rm particles} = \mathbb{C}\oplus \bigoplus_{n\geq 1}\underbrace{\big(\Hilb^{\KG} \otimes_{S}\dots \otimes_{S}\Hilb^{\KG}\big)}_{n \text{ times}}.
\ee 
where $\otimes_{S}$ is the symmetrized tensor product. On this Fock space the \(\op b(w)^*\) acts as a creation operator and \(\op b(w)\) acts as the annihilation operator for a particle wave packet \(w(p)\). 
An identical construction with the inner product 
\be
\s^{\KG}_{0}(\op{c}(w_{1})\op{c}(w_{2})^*) = \frac{4(2\pi) ^3}{m^2} \braket{w_2,w_1}_\hyp
\ee
gives the ``one antiparticle'' Hilbert space $\antiHilb^{\KG}$, and the corresponding Fock space \(\Fock^{\KG}_{\rm antiparticles}\) on which \(\op c(w)^*\) acts as a creation operator and \(\op c(w)\) acts as the annihilation operator for an antiparticle wave packet \(\bar w(p)\). Since the operators \(\op b(w)\) and \(\op b(w)^*\) commute with \(\op c(w)\) and \(\op c(w)^*\), the full Fock space representation is given by the tensor product of the particle and antiparticle Fock spaces
\be
    \Fock^{\KG} = \Fock^{\KG}_{\rm particles} \otimes \Fock^{\KG}_{\rm antiparticles}\, .
\ee
The algebraic state $\s^{\KG}_{0}$ corresponds to the vacuum state of the Fock space, which we denote as $\ket{\s^{\KG}_{0}}\in \Fock^{\KG}$. We have
\be 
\op{b}(w) \ket{\s^{\KG}_{0}} = \op{c}(w) \ket{\s^{\KG}_{0}} = \cop_{i^{-}}(\lambda) \ket{\s^{\KG}_{0}}=0 \eqsp \text{for all } w(p), \lambda(x^{A})\, .
\ee 
A dense set of Hadamard states in this Fock space is generated by the linear span of the vacuum \(\ket{\s^{\KG}_0}\) and symmetric tensor products of the particle and antiparticle wave packet states with test functions \(w(p)\). 

The large gauge transformations and Poincar\'e transformations have a strongly continuous unitary action on \(\Fock^{\KG}\). The vacuum state \(\ket{\s^{\KG}_0}\) is invariant under these transformations. The large gauge transformations and translations act on the one-particle/antiparticle spaces as multiplication by a phase. The Lorentz group acts on the one-particle/antiparticle spaces by its natural action on $L^2(\hyp^{-})$. The action on the full Fock space is immediately obtained by extending the action to symmetric tensor products of the one-particle/antiparticle spaces \cite{Araki:1999ar}.

In the construction of the Faddeev-Kulish representations (see \cref{subsec:MaxMKGKFreps}) it will be useful to work with ``improper states'' of definite particle/antiparticle momenta. Formally, these states correspond to applying the point-wise creation operators $\op{b}(p)^*$ and $\op{c}(q)^*$ to the vacuum
\be 
\label{nparticleimp}
\ket{p_{1},\dots , p_{n}} = \frac{1}{\sqrt{n!}}\op{b}(p_{1})^*\dots \op{b}(p_{n})^*\ket{\s^{\KG}_{0}} \textrm{ and }\ket{q_{1},\dots , q_{m}} = \frac{1}{\sqrt{m!}}\op{c}(q_{1})^*\dots \op{c}(q_{m})^*\ket{\s^{\KG}_{0}}
\ee 
where $p,q\in \hyp^{-}$ correspond to the momenta of particles and antiparticles respectively. Although \cref{nparticleimp} is well-defined if we smear with test functions in all variables, the definite momentum states $\ket{p_{1},\dots , p_{n}} $ and $\ket{q_{1},\dots , q_{m}}$ themselves have infinite norm and are not genuine states in $\Fock^{\KG}$. However, we can make mathematical sense of these improper states and their relationship to $\Fock^{\KG}$ in a precise way as follows. Let $\Hilb^{\KG}_{p}\cong \mathbb{C}$, be the one-dimensional complex Hilbert spaces spanned by the symbol \(\ket{p}\). Thus, \(\ket{p}\) is a genuine state in $\Hilb^{\KG}_{p}$. Similarly, for all $p_1, \dots, p_n$ we define the one-dimensional complex Hilbert space $\Hilb^{\KG}_{p_{1}\dots p_{n}}$ by
\be 
\Hilb^{\KG}_{p_{1}\dots p_{n}}= \Hilb^{\KG}_{p_{1}}\otimes_{S}\dots \otimes_{S}\Hilb^{\KG}_{p_{n}} 
\label{hilmomeig}
\ee 
where the symmetric tensor product symbol indicates here that we identify the Hilbert spaces that differ by a permutation of $p_1, \dots, p_n$. Then $\Hilb^{\KG}_{p_{1}\dots p_{n}}$ is spanned by $\ket{p_{1},\dots,p_{n}} \defn \ket{p_1} \otimes_{S}\dots \otimes_{S} \ket{p_n}$. We similarly define the $n$-antiparticle Hilbert spaces of definite momenta. The Fock space $\Fock^{\KG}$ is then given by the direct sum of the direct integral of these Hilbert spaces
\be
\label{Fockmomentum}
\Fock^{\KG} \cong \bigoplus_{n,m\geq 0}~\int_{\hyp^{n+m}}d^{3}p_{1}\dots d^{3}p_{n}d^{3}q_{1}\dots d^{3}q_{m}~\Hilb^{\KG}_{p_{1}\dots p_{n}}\otimes \antiHilb^\KG_{q_{1}\dots q_{m}}
\ee 
where $d^3 p$ and $d^3q$ denote the Lorentz invariant measure \cref{eq:hypmeasure} on the hyperboloid. States in the Fock space are thus given by expressions of the form
\be\label{eq:state-momentum-decomp}
\ket{\Psi} = \sum_{n,m}\int_{\hyp^{n+m}}d^{3}p_{1}\dots d^{3}p_{n}d^{3}q_{1}\dots d^{3}q_{m}~\psi(p_{1},\dots p_{n},q_{1},\dots,q_{m})\ket{p_{1}\dots p_{n}}\otimes \ket{q_{1}\dots q_{m}}
\ee 
where $\psi$ is a complex \(L^2\)-function of $p_{i}$ and $q_{i}$ that is invariant under permutations of the $p_i$ and permutations of the $q_i$. The quantities $\ket{p_{1}\dots p_{n}}$ and  $\ket{q_{1}\dots q_{m}}$ appearing in this equation are the mathematically well-defined basis elements of $\Hilb^{\KG}_{p_{1}\dots p_{n}}$ and $\antiHilb^{\KG}_{q_{1}\dots q_{m}}$.

The action of the charge operator $\cop_{i^{-}}(\lambda)$ on $\Fock^{\KG}$ can be expressed very conveniently in this representation of the Fock space, since the direct integral decomposition \cref{Fockmomentum} corresponds to the spectral decomposition of the operator $\cop_{i^{-}}(\lambda)$. Formally, the states $\ket{p_{1},\dots , p_{n}} $ and $\ket{q_{1},\dots , q_{m}}$ are eigenstates of $\cop_{i^{-}}(\lambda)$ with eigenvalues given by
\be 
\label{Qiminuseigen}
&\op{\mc{Q}}_{i^{-}}(\lambda) \ket{p_{1},\ldots, p_{n}} = q\bigg(\sum_{i=1}^{n}\lambda_{\hyp}(p_{i})\bigg)\ket{p_{1},\ldots, p_{n}} \\
&\op{\mc{Q}}_{i^{-}}(\lambda) \ket{q_{1},\ldots, q_{n}} = -q\bigg(\sum_{i=1}^{n}\lambda_{\hyp}(q_{i})\bigg)\ket{q_{1},\ldots, q_{n}}
\ee 
where $\lambda_{\hyp}$ is given by \cref{eq:lgt-hyp-soln}. Note that the unsmeared version of \cref{Qiminuseigen} is
\be 
&\op{\mc{Q}}_{i^{-}}(x^{A}) \ket{p_{1},\ldots, p_{n}} = q\bigg(\sum_{i=1}^{n}G_{\hyp}(p_{i},x^{A})\bigg)\ket{p_{1},\ldots, p_{n}} \\
&\op{\mc{Q}}_{i^{-}}(x^{A}) \ket{q_{1},\ldots, q_{n}} = -q\bigg(\sum_{i=1}^{n}G_{\hyp}(q_{i},x^{A})\bigg)\ket{q_{1},\ldots, q_{n}}.
\ee 
where $G_{\hyp}(p,x^{A}) = (4\pi)^{-1}(\sqrt{1+\rho^{2}}-\rho \hat{p}\cdot \hat{r})^{-2}$ (see \cref{eq:lgt-hyp-soln}). 
 For a state $\ket{\Psi}$ in the Fock space lying in the subspace of $n$ particles and $m$ antiparticles, the action of the charge operator is given by
\be
\cop_{i^-}(\lambda) \ket{\Psi} = q\int_{\hyp^{n+m}}d^{3}p_{1}\dots d^{3}p_{n}d^{3}q_{1}\dots d^{3}q_{m}~\psi(p_{1},\dots p_{n},q_{1},\dots,q_{m}) \times \\
\bigg(\sum_{i=1}^{n}\lambda_{\hyp}(p_{i}) - \sum_{i=1}^{m}\lambda_{\hyp}(q_{i})\bigg)
\ket{p_{1}\dots p_{n}}\otimes \ket{q_{1}\dots q_{m}}.
\ee
All states in this subspace are eigenstates of the total charge operator $\cop_{i^{-}}(1)$
 \be 
\cop_{i^-}(1)\ket{\Psi} = q(n-m) \ket{\Psi}\, .
\ee
However, for non-constant $\lambda$, there are no proper eigenstates of $\cop_{i^-}(\lambda)$ apart from the vacuum state.

\subsubsection{Fock representations of the Maxwell field algebra}
\label{subsubsec:MaxMKGFockrepscriminus}

The vacuum state on the extended asymptotic algebra $\Algexqp^{\rm EM}$ of the electromagnetic field is the Gaussian algebraic state \(\s_0^{\EM}\) with vanishing $1$-point function and $2$-point function given by (see \cref{hadformSR3})
\be 
\label{2ptscri0}
    \s_{0}^{\EM}(\op{E}(s_{1})\op{E}(s_{2}))=- \int_{\mathbb{R}^{2}\times \mathbb{S}^{2}}dv_{1}dv_{2}\dS~\frac{q_{AB}s^{A}_{1}(v_{1},x^{A})s^{B}_{2}(v_{2},x^{A})}{(v_{1}-v_{2}-i0^{+})^{2}}. 
\ee
Furthermore, \(\s_0^{\EM}\) is an eigenstate of eigenvalue zero of memory, $\op\Delta(\lambda)$, and all of the Poincar\'e generators.

The GNS construction for \(\s_0^{\EM}\) yields a Hilbert space $\Fock_{0}^{\EM}$ with a natural Fock space structure. Concretely this construction is obtained as follows. On the space of positive frequency Schwartz test functions on $\scri^-$ we define the inner product
\be 
\label{scriinnprod}
\braket{s_{1}|s_{2}}_{0} \defn  \s_{0}^{\EM}(\op{E}(s_{1})^* \op{E}(s_{2})) = 2\pi \int_0^\infty \omega d\omega \int_{\bb S^2} \dS~  \bar{\hat{s}^A_{1}(\omega, x^A)}\hat{s}_{A,2}(\omega,x^A)
\ee 
where the ``hat'' denotes the Fourier transform, and the final equality above is the Fourier space representation of \cref{2ptscri0}. We define the one-particle Hilbert space, $\Hilb_{0}^{\EM}$, to be the completion of the space of positive frequency test functions in this inner product.
The GNS Fock space associated with the vacuum state is then given by 
\be 
\Fock^{\EM}_{0}= \mathbb{C}\oplus \bigoplus_{n\geq 1}\underbrace{\big(\Hilb_{0}^{\EM} \otimes_{S}\dots \otimes_{S}\Hilb_{0}^{\EM}\big)}_{n\text{ times}}. 
\ee 
For real $s^A$, the smeared electric field operator $\op{E}_{0}(s)\defn \pi_{0}^{\EM}[\op{E}(s)]$ in this representation is given by
\be 
\label{eq:Eaadag}
\op{E}_{0}(s) = \op{a}_{0}(s^-) + \op{a}^\dagger_{0}(s^+)  
\ee 
where $\op a_0, \op a_0^\dagger$ are the usual annihilation and creation operators on the Fock space and superscripts ``$\pm$'' denote the positive/negative frequency parts, respectively. There is a dense subspace of Hadamard states given by the span of the vacuum \(\ket{\s_0^{\EM}}\) and any finite products of $\op{a}^\dagger_{0}(s_i^+)$ applied to the vacuum, with $s_i$ an arbitrary test function. The Poincar\'e transformations act on $\Hilb_{0}^{\EM}$ by their natural action on $\scri^-$, which gives rise to a strongly continuous action on $\Fock^{\EM}_{0}$. The vacuum state, \(\ket{\s_0^{\EM}}\), is invariant under these transformations.

The Fock space \(\Fock_0^{\EM}\) constructed above is the usual choice of Hilbert space for the ``in'' radiative states of the Maxwell fields. However, all states in this Hilbert space are eigenstates of memory, $\op\Delta(\lambda)$, with eigenvalue zero --- as follows immediately from the fact that the vacuum is an eigenstate of $\op\Delta(\lambda)$ with eigenvalue zero, and $\op\Delta(\lambda)$ commutes with $\op{E}(s)$ and hence with $\op{a}^\dagger_{0}(s^+)$. However, even in the classical theory, memory is not conserved between ``in'' and ``out'' states in generic scattering processes. Thus, even if we restrict to the zero-memory Fock space \(\Fock_0^{\EM}\) for the Maxwell in-states, the out states obtained will not have zero memory and hence will not live in the zero memory out-Fock space \(\Fock_0^{\EM}\). One is thus forced to consider states which have non-vanishing memory to describe scattering. Thus, the states in \(\Fock_0^{\EM}\) do {\em not} give us an ample supply of states to use in scattering theory.

However, we can construct different Fock representations containing states with nonvanishing memory as follows (see \cite{asymp-quant}). Choose a smooth classical
electric field \(e_A(x)\) on $\scri^-$ that satisfies our decay conditions but is such that the corresponding classical memory
\be
 \Delta (e, \LGT) = -  \int_{\scri^-} dv\dS~  e_A(v,x^{B}){\ms D}^A \LGT(x^{B})
\ee
is non-vanishing. Consider the algebra automorphism $\Aut_{e}: \Algex^{\EM} \to \Algex^{\EM}$ determined by
\be 
\label{memauto}
\Aut_{e}[\op{E}(s)] = \op{E}(s) + e(s)\op{1} \eqsp \Aut_{e} [\op{\Delta}(\lambda)] = \op{\Delta}(\lambda)+\Delta(e, \lambda)\op{1}.
\ee 
This is easily seen to define an automorphism, since the commutation relations are unaffected by shifting the operators by a multiple of \(\1\).
Using this automorphism we define a new algebraic state $\s_{e}^{\EM}$ by
\be 
\label{Lambdamem}
\s_{e}^{\EM}(\op{O})\defn  \s_{0}^{\EM}(\Aut_{e}[\op{O}]) \quad   \textrm{ for all $\op{O}\in \Algex^{\EM}$.}
\ee 
Then $\s_{e}^{\EM}$ is a Gaussian, Hadamard state on $\Algex^{\EM}$ that satisfies \cref{mem1ptfun} and, for each $\lambda$, is an eigenstate of $\op{\Delta}(\lambda)$ with eigenvalue $\Delta(e, \lambda)$. The GNS construction for $\s_{e}^{\EM}$ yields a Hilbert space \(\Fock_e^{\EM}\) with a Fock space structure and a vacuum state $\ket{\s_{e}^{\EM}}$ corresponding to $\s_{e}^{\EM}$. Every state in \(\Fock_e^{\EM}\) is an eigenstate of $\op{\Delta}(\lambda)$ with eigenvalue $\Delta(e, \lambda)$. Thus, this construction --- for the various different choices of classical electric field $e_A$ --- gives an ample supply of states with any desired memory.

It should be noted that if $e_A$ and $e'_A$ are smooth and satisfy our decay conditions, then the Fock representations obtained by the above GNS construction will be unitarily equivalent\footnote{However, if $e_A$ and $e'_A$ are not smooth, the norm (defined in \cref{scriinnprod}) of the positive frequency part of $e_A - e'_A$ need not be finite even when they have the same memory. In that case, the Fock space constructions will not be unitarily equivalent. This point will be relevant to the considerations of \cref{subsec:MaxZMKGKFrep}. \label{foot:ineqrep}} if and only if $ \Delta (e, \LGT) =  \Delta (e', \LGT)$ for all $\lambda$. Thus, there are as many unitarily inequivalent constructions as there are choices of memory one-form $\Delta_{A}(x^A)$ on \(\bb S^2\). In particular, there are uncountably many such constructions. If $e_A$ and $e'_A$ are such that $ \Delta (e, \LGT) =  \Delta (e', \LGT)$ (so that they give rise to unitarily equivalent representations), then the state $\s_{e'}^{\EM}$ --- which corresponds the the vacuum state in \(\Fock_{e'}^{\EM}\) ---  corresponds in \(\Fock_e^{\EM}\) to the coherent state associated with the classical solution $e'_A - e_A$. Thus, the representations with nonvanishing memory do not have a ``preferred'' vacuum state, i.e., the vacuum state of the Fock representation depends on the choice of representative classical electric field $e_A$. Nevertheless, the unitary equivalence class of the Fock representations $\Fock^{\EM}_{e}$ correspond to all smooth $e_{A}$ with memory $\Delta_{A}$. Thus, the ``memory representations'' can be labeled by the memory of the representative --- i.e., as $\Fock^{\EM}_{\Delta}$ rather than \(\Fock_e^{\EM}\) --- and we shall do so in the following.

It also should be noted that for any given choice of memory $\Delta_A(x^A)$ on \(\bb S^2\) and any given choice of frequency $\omega_0 > 0$ one can find a representative classical electric field $e_A(v, x^B)$ with memory equal to $\Delta_A(x^A)$ such that the Fourier transform of $e_A$ is nonvanishing only for frequencies $\omega < \omega_0$. Thus, the states in \(\Fock_{\Delta}^{\EM}\) can be viewed as differing from the states in \(\Fock_{0}^{\EM}\) only in the (arbitrarily) far infrared. However, for $\Delta_A(x^A) \neq 0$, if one tries to formally express a normalized state in \(\Fock_{\Delta}^{\EM}\) as a state in \(\Fock_{0}^{\EM}\), one will find that it has infinitely many ``soft photons'' and cannot be normalized. Thus, the states in \(\Fock_{\Delta}^{\EM}\) are genuinely different from states in \(\Fock_{0}^{\EM}\).

The above construction yields representations of $\Algex^{\EM}$ with any desired memory. We now consider whether these representations can be extended to representations of $\Algexqp^{\EM}$, i.e., whether one can define an action of Poincar\'e generators on \(\Fock_{\Delta}^{\EM}\) such that the commutation relations corresponding to \crefrange{transbc}{lortrans} hold. Consider, first, a translation. The natural action of a finite translation on the classical electric field $e_A$ maps it into an electric field $e'_A$ with the same memory. As noted above, the representation obtained from $e^{\prime}_A$ is therefore unitarily equivalent to the representation obtained from $e_A$. It follows that the natural action of finite translations can be represented by a unitary map on \(\Fock_{\Delta}^{\EM}\). This map is strongly continuous in the translation parameter, so we get a self-adjoint operator on \(\Fock_{\Delta}^{\EM}\) representing an arbitrary translation generator $T$, which satisfies all of the required commutation relations.\footnote{We emphasize that the spectrum of the energy operator corresponding to time translations is bounded below by zero but does not achieve the value zero for {\em any} state with memory.} Thus, the above Fock representations of $\Algex^{\EM}$ with nonvanishing memory can be extended to include Poincar\'e translations. 

However, the Fock representations \(\Fock_{\Delta}^{\EM}\) with nonvanishing memory {\em cannot} be extended to include the action of the Lorentz generators \cite{Roepstorff:1970gg,asymp-quant}. As noted above, for all $\lambda$ all states in \(\Fock_{\Delta}^{\EM}\) are eigenstates of $\op{\Delta}(\lambda)$ with eigenvalue $\Delta(e, \lambda)$. Thus, for all $\lambda$, the memory operator commutes with all operators on \(\Fock_{\Delta}^{\EM}\). However, if $\Delta(e, \lambda)$ is nonvanishing for some $\lambda$, then by \cref{memlor} some Lorentz generator $X$ must have a nonvanishing commutator with $\op{\Delta}(\lambda)$. Thus, the above Fock representations of $\Algex^{\EM}$ with nonvanishing memory \emph{cannot} be extended to representations of $\Algexqp^{\EM}$.

\subsection{Faddeev-Kulish representation}
\label{subsec:MaxMKGKFreps}

We turn now to the issue of whether we can find Hilbert spaces of incoming and outgoing states that satisfy properties \ref{enum:S1}--\ref{enum:S5} listed in \cref{sec:intro}. The standard choice of ``in'' Hilbert space $\Fock_{\inn} = \Fock^{\KG} \otimes \Fock^{\EM}_{0}$ and correspondingly constructed standard ``out'' Hilbert space $\Fock_{\rm out}$ of the ``out'' algebra does not work, since all states in $\Fock_{\inn}$ and $\Fock_{\rm out}$ have vanishing memory, but scattering takes states with vanishing memory to states with nonvanishing memory. As we shall discuss further in \cref{sec:NKFreps}, one could attempt to allow memory by replacing $\Fock^{\EM}_{0}$ with a direct sum, $\oplus_\Delta \Fock_{\Delta}^{\EM}$, over all unitarily inequivalent memory Fock spaces. However, since there are uncountably many such memory Fock spaces, this would give a non-separable Hilbert space, in violation of property \ref{enum:S5}. Furthermore, each state in such a direct sum would have a nonvanishing probability for only a countable number of discrete values of memory.  However, scattering with an ``in'' state of this sort surely does not produce an ``out'' state of this sort, so property \ref{enum:S4} also will not be satisfied by this choice of the ``in'' and ``out'' Hilbert spaces. Finally, by \cref{memlor}, since the memory is not Lorentz invariant there cannot be continuous action of Lorentz on the direct sum --- in violation of property \ref{enum:S3} --- so the angular momentum cannot be defined. A more promising possibility would be to take some sort of direct integral of memory representation Hilbert spaces. However, as we shall discuss further in \cref{sec:NKFreps}, the natural Lorentz invariant Gaussian measure on memory has support on memories that are too singular to be admissible, and there does not appear to any other choices of measure for a direct integral construction that have the prospect of satisfying properties \ref{enum:S3} or \ref{enum:S4}.

Nevertheless, it is possible to give a construction, due to Faddeev and Kulish \cite{Kulish:1970ut}, of ``in'' and ``out'' Hilbert spaces that satisfy \ref{enum:S1}--\ref{enum:S5}. The construction involves taking a direct integral over the memory Fock spaces of the electromagnetic field but correlating these Fock spaces with (improper) momentum eigenstates of the massive Klein-Gordon field so as to produce states with vanishing charges \(\mc Q_{i^0}(\lambda)\) at spatial infinity. This is a useful construction because of the fact that, as shown in \cite{CE,Henneaux:2018gfi,KP-EM-match,Mohamed_2021}, for solutions to the Maxwell equations that are suitably regular at spatial infinity, the charges \(\mc Q_{i^0}^\inn(\lambda)\) obtained from the limit along past null infinity are matched antipodally to the similarly defined charges \(\mc Q_{i^0}^\out(\lambda)\) obtained from the limit along future null infinity,
\be
\label{eq:chargecons}
    \mc Q_{i^0}^\inn(\lambda) = \mc Q_{i^0}^\out(\lambda \circ \Upsilon)
\ee
where $\Upsilon$ is the antipodal map on ${\bb S}^2$. Thus, any ``in'' state that is an eigenstate of $ \mc Q_{i^0}^\inn(\lambda)$ for all $\lambda$ should evolve to an ``out'' state that is an eigenstate of $\mc Q_{i^0}^\out(\lambda\circ \Upsilon )$ of the same eigenvalue. However, since by \cref{memlor} the Lorentz group generators have nontrivial commutators with the charges at spatial infinity, the Lorentz group generators cannot act on a Hilbert space of states of definite charges {\em except} in the case where all of the charges vanish, $ \mc Q_{i^0}(\lambda) = 0$ for all $\lambda$ \cite{Gervais_1980,Buchholz86,Frohlich:1978bf,Frohlich:1979uu}. Therefore, we seek to construct ``in'' and ``out'' Hilbert spaces composed of states that are eigenstates of eigenvalue zero of all of the large gauge charges (including the total electric charge) at spatial infinity.

To construct an ``in'' Hilbert space with $ \mc Q_{i^0}(\lambda) = 0$ for all $\lambda$, we make use of the relation
\be 
\label{sptltimemem}
\cop_{i^{0}}(\lambda) = \cop_{i^{-}}(\lambda)+\frac{1}{4\pi} \op{\Delta}(\lambda)
\ee 
(see \cref{Q0def}). We start with the one-dimensional Hilbert space $\Hilb^{\KG}_{p_{1}\dots p_{n}}\otimes \antiHilb^{\KG}_{q_{1}\dots q_{n}}$ of $n$ incoming particles and $n$ incoming antiparticles in momentum states $p_1, \dots, p_n$ and $q_1, \dots, q_n$, respectively (see \cref{hilmomeig}). This state has vanishing total electric charge and large gauge charges
\be
{\mc Q}_{i^{-}}(\lambda) = q\sum_{i=1}^{n} \left(\lambda_{\hyp}(p_{i}) -  \lambda_{\hyp}(q_{i}) \right)
\ee
(see \cref{Qiminuseigen}), with $\lambda_{\hyp}(p)$ given by \cref{eq:lgt-hyp-soln}. Therefore, we can obtain a state with $\cop_{i^{0}}(\lambda) = 0$ for all $\lambda$ if we can find a memory representation $\ms{F}_{\Delta}^{\EM}$ such that for all $\lambda$ we have
\be
\Delta(\lambda;p_{1},\dots,q_{n}) =- 4 \pi q \sum_{i=1}^{n} \left(\lambda_{\hyp}(p_{i}) -  \lambda_{\hyp}(q_{i}) \right).
\label{dress1}
\ee
This will be the case if for any $p$ we can solve
\be 
\label{memn1m0}
\ms{D}^{A}\Delta_{A}(x^A;p) =  4 \pi q (G_{\hyp}(p, x^A) - 1)
\ee
with $G_{\hyp}$ given by \cref{eq:lgt-hyp-soln}. If we can solve \cref{memn1m0}, then the memory representation \(\Fock_{\Delta}^{\EM}\) obtained from any $e_A$ such that
\be
-\int_{-\infty}^{\infty}dv~ e_A(v, x^A)   = \sum_{i=1}^{n} \left( \Delta_A (x^A; p_{i}) - \Delta_A (x^A; q_{i}) \right)
\label{dress2}
\ee
will have memory satisfying \cref{dress1}. 

We can decompose any one-form on $\bb{S}^2$ such as $\Delta_A$ into its electric and magnetic parts as
\be
\label{eq:memelmag}
\Delta_A = \ms{D}_A \alpha + {\epsilon_A}^B \ms{D}_B \beta
\ee
The magnetic part\footnote{As previously explained (see \cref{magpar}), we restrict consideration in any case to purely electric parity memory.} will not contribute to $\ms{D}^A \Delta_A$, so \cref{memn1m0} becomes
\be
\ms{D}^{2} \alpha =  4 \pi q(G_{\hyp}(p, x^A) -1)
\label{lapalpha}
\ee
By \cref{eq:S2intGH}, the right side is orthogonal to the $\ell=0$ spherical harmonic, so this equation can be uniquely solved. Since $G_{\hyp}(p, x^A)$ is smooth, it follows that $\Delta_{A}(x^A;p)$ is smooth and, hence, $e_A(v, x^A)$ can be chosen to be smooth. 

We now have all of the ingredients needed for the Faddeev-Kulish construction. As described in the previous subsection, the standard Fock space $\Fock^{\KG}$ for the Klein-Gordon field can be obtained by taking a direct sum of direct integrals of the one-dimensional Hilbert spaces $\Hilb^{\KG}_{p_{1}\dots p_{n}}\otimes \antiHilb^{\KG}_{q_{1}\dots q_{m}}$ of momentum eigenstates (see \cref{Fockmomentum}). As stated above, the standard ``in'' Hilbert space is then obtained by taking the tensor product of this Klein-Gordon Fock space with the standard (zero memory) Fock space $\Fock^{\EM}_{0}$ for the electromagnetic field. The Faddeev-Kulish construction modifies this procedure as follows. Prior to taking the direct integral, we pair the state $\ket{p_{1}\dots p_{n}}\otimes \ket{q_{1}\dots q_{n}}$ with the Fock representation \(\Fock_{\Delta(p_1, \dots, q_n)}^{\EM}\) obtained from an electric field $e_A$ on $\scri^-$ satisfying \cref{dress2}. All electromagnetic states in this representation have memory given by \cref{dress1}, so the states in $\Hilb^{\KG}_{p_{1}\dots p_{n}}\otimes \antiHilb^{\KG}_{q_{1}\dots q_{n}} \otimes \Fock_{\Delta(p_1, \dots, q_n)}^{\EM}$ have $ \mc Q_{i^0}(\lambda) = 0$ for all $\lambda$. We now take the direct integral of these Hilbert spaces over $p_1, \dots, q_n$ and the direct sum over $n$ to obtain the Faddeev-Kulish ``in'' Hilbert space
\be 
\label{mQEDKFQ}
\Hilb_{\inn}^{\rm FK} \defn \bigoplus_{n=0}^\infty ~ \int_{\hyp^{2n}}d^{3}p_{1}\dots d^{3}p_{n}d^{3}q_{1}\dots d^{3}q_{n}~\Hilb^{\KG}_{p_{1}\dots p_{n}}\otimes \antiHilb^{\KG}_{q_{1}\dots q_{n}}\otimes \Fock^{\EM}_{\Delta(p_{1},\dots,q_{n})}
\ee 
All states in $\Hilb_{\inn}^{\rm FK}$ are eigenstates of $\cop^{\inn}_{i^{0}}(\lambda)$ with eigenvalue zero for all $\lambda$.
The ``out'' Hilbert space $\Hilb_{\rm out}^{\rm FK}$ is constructed similarly.

It should be noted that $\Hilb_{\inn}^{\rm FK}$ does not carry a representation of the algebra $\Algexqp$ or even of the unextended algebra $\Alg_{\inn}$. The massive field operators $\op{b}(w)$, $\op{c}(w)$ have nontrivial commutators with 
with $\cop_{i^{0}}(\lambda)$ (see \cref{pbchbc}) and cannot be made to act on $\Hilb_{\inn}^{\rm FK}$. However, all gauge invariant observables in $\Algexqp$ commute with $\cop_{i^{0}}(\lambda)$, and therefore $\Hilb_{\inn}^{\rm FK}$ carries a representation of the subalgebra of gauge invariant observables.

The Faddeev-Kulish ``in'' and ``out'' Hilbert spaces can be seen to satisfy requirements \ref{enum:S1}--\ref{enum:S5} of \cref{sec:intro} as follows. Requirement \ref{enum:S1} is automatically satisfied, since $\Hilb_{\rm out}^{\rm FK}$ is obtained by the same construction as $\Hilb_{\inn}^{\rm FK}$. Satisfaction of requirement \ref{enum:S2} follows from conservation of the large gauge charges, \cref{eq:chargecons}, which implies that any state in $\Hilb_{\inn}^{\rm FK}$ must evolve to an eigenstate of eigenvalue zero of $\cop^{\rm out}_{i^{0}}(\lambda)$ for all $\lambda$, which, presumably, must lie in $\Hilb_{\rm out}^{\rm FK}$. With regard to requirement \ref{enum:S3}, the translation group acts naturally on both $\Hilb^{\KG}_{p_{1}\dots p_{n}}\otimes \antiHilb^{\KG}_{q_{1}\dots q_{n}}$ and $\Fock^{\EM}_{\Delta}$ so there is no problem obtaining its action on $\Hilb_{\inn}^{\rm FK}$ \cite{Frohlich:1978bf}. A Lorentz transformation $\Lambda$ maps $\Hilb^{\KG}_{p_{1}\dots p_{n}}\otimes \antiHilb^{\KG}_{q_{1}\dots q_{n}}$ to $\Hilb^{\KG}_{\Lambda p_{1}\dots \Lambda p_{n}}\otimes \antiHilb^{\KG}_{\Lambda q_{1}\dots \Lambda q_{n}}$ and maps $\Fock^{\EM}_{\Delta}$ to $\Fock^{\EM}_{\Lambda \Delta}$. However, since $\Lambda [e_A (p_{1}\dots q_{n})]$ defines the same memory Fock space as $e_A (\Lambda p_{1}\dots \Lambda q_{n})$, there is no problem obtaining an action of the Lorentz group on $\Hilb_{\inn}^{\rm FK}$, so requirement \ref{enum:S3} is satisfied \cite{Frohlich:1978bf}. Note that there would be a problem with obtaining Lorentz group action if we had similarly constructed a Hilbert space of eigenstates of $\cop_{i^{0}}(\lambda)$ with nonvanishing eigenvalues \cite{Gervais_1980,Buchholz86}. With regard to requirement \ref{enum:S4} since, as discussed above, $e_A$ can be chosen to be of arbitrarily low frequency, each $\Fock^{\EM}_{\Delta}$ contains representatives of any desired ``hard'' photon state. It is clear that $\Hilb_{\inn}^{\rm FK}$ contains states of arbitrary momenta of the charged particles and antiparticles provided that the number of particles and antiparticles are equal. As discussed in \cref{sec:intro}, although this equality of particle and antiparticle number yields a genuine restriction on the allowed states, one can deal with this in the consideration of scattering by putting any extra/unwanted particles ``behind the moon.'' Thus, arguably, requirement \ref{enum:S4} is satisfied. Finally, it is straightforward to show that requirement \ref{enum:S5} is satisfied.

The states in $\Hilb_{\inn}^{\rm FK}$ correspond to incoming particles/antiparticles together with incoming photons in states whose memory is highly correlated with the momenta of the particles and antiparticles. As mentioned above, we may view the states in any memory Fock space $\Fock^{\EM}_{\Delta}$ as corresponding to a state in $\Fock^{\EM}_{0}$ together with infinitely many ``soft photons.'' Thus, we may view the memory associated with $e_A (p_{1}\dots q_{n})$ as ``dressing'' the incoming charged particle state $\ket{p_{1}\dots p_{n}}\otimes \ket{q_{1}\dots q_{n}}$ with a ``soft photon cloud.'' In $\Hilb_{\inn}^{\rm FK}$, all charged particle states must be ``dressed'' in this manner.

The ``dress requirements'' imposed by $\Hilb_{\inn}^{\rm FK}$ have a number of unpleasant consequences. Most notably, one cannot consider a coherent superposition of charged particle states of different momenta, since charged particle states with different momenta are required to be dressed with soft photon clouds corresponding to different representations of the electromagnetic field. These orthogonal soft photon clouds will preclude any interference effects arising from superposing charged particle states of different momenta. Nevertheless, as we have argued above, $\Hilb_{\inn}^{\rm FK}$ contains a supply of states that is adequate for analyzing many scattering processes of interest.

\section{QED with a massless, charged Klein-Gordon field}
\label{sec:MaxZMKG}

In this section, we consider QED with the massive charged Klein-Gordon field of \cref{sec:MaxMKG} replaced by a massless charged Klein-Gordon field. Thus, we consider the theory defined by the Lagrangian \cref{lqed} with $m=0$. Most of the analysis carries through in close parallel with the massive case. However, as we shall see, a significant difference arises in the construction of the Faddeev-Kulish Hilbert space due to the fact that the memory representations of the electromagnetic field needed in the construction are singular. In \cref{subsec:YM}, we shall consider the source-free Yang-Mills case. In addition to the problems of massless QED, a new problem arises from the fact that the ``soft dressing'' contributes to the Yang-Mills charge-current flux, thereby invalidating the construction of eigenstates of large gauge charges via ``dressing.'' Although one can obtain charge eigenstates by other means, there are insufficiently many eigenstates to obtain Hilbert spaces for scattering.

\subsection{Asymptotic quantization algebra}
\label{subsec:MaxZMKGclassquant}

The asymptotic quantization of the electromagnetic field was already given in \cref{subsec:MaxMKGclass}, so we need only give the asymptotic quantization of the massless charged Klein-Gordon field. As discussed in \cref{sec:classphase}, the asymptotic behavior of a massless scalar field in the asymptotic past is described by
\be\label{eq:massless-falloff}
    \Phi(x) \defn \lim_{\scri^-} \Omega^{-1} \varphi
\ee
(see \cref{phitil}). The symplectic form is given by
\be
    \Omega_{\scri}^{\KG0}((\Phi_1,\bar\Phi_1), (\Phi_2,\bar\Phi_2) ) = - \frac{1}{2} \int_{\scri^-} d^3x \lb[ \Phi_1 \partial_v \bar\Phi_2 + \bar \Phi_1 \partial_v \Phi_2 - (1 \lra 2) \rb] \, ,
\ee
where the superscript ``\(\KG0\)'' denotes that this is the symplectic form of a massless scalar field. The above symplectic form differs from \cref{symprodscri} only in that we are now considering a complex, rather than real, scalar field. For the same reasons as indicated below \cref{eq:symp-massive}, it is convenient to treat \(\Phi \) and \( \bar\Phi\) as though they were independent quantities and to take the asymptotic phase space to consist of the pairs \((\Phi, \bar\Phi)\). Then $\Omega_{\scri}^{\KG 0}$ is a complex-bilinear function of its variables. It is convenient, as in \cref{scriobs} of \cref{sec:classphase},  to write $\Pi = \partial_v \Phi$ and $\bar{\Pi} = \partial_v \bar{\Phi}$.

In parallel with \cref{scriobs}, the local scalar field observables on $\scri^-$ are 
\be\label{eq:Pi-smeared-defn}
    \Pi(s) \defn \int_\scri d^3x~ \Pi(x) s(x) \eqsp \bar\Pi(s) \defn \int_\scri d^3x~ \bar\Pi(x) \bar s(x)
\ee
where \(s(x)\) is a smooth complex function on \(\scri^-\) with conformal weight \(-1\). Note that we take \(\Pi(s)\) to be linear in \(s\) while \(\bar\Pi(s)\) is antilinear in the test function \(s(x)\). The Hamiltonian vector fields for these observables are given by the pairs \((0,s)\) and \((\bar s,0)\), respectively. The only nonvanishing Poisson brackets are
\be
\lb\{ \bar\Pi(s_1), \Pi(s_2) \rb\} = - \Omega_{\scri}^{\KG 0}((\bar s_1,0), (0,s_2))1 =  \frac{1}{2} \int_{\scri^-} d^3x \lb[ \bar s_1 \partial_v s_2 - s_2 \partial_v \bar s_1 \rb]\,.
\ee
The additional factor of \(2\) in the above formula relative to \cref{scricomm0} arises because we are now working with a complex scalar field.

The asymptotic quantization algebra, $\Alg_{\inn}^{\rm KG0}$, for the massless charged Klein-Gordon field is defined by starting with the free, unital $\ast$-algebra generated by $\op{\Pi}(s)$, its formal adjoint $\op{\Pi}(s)^*$, and the identity $\op{1}$. We then factor this algebra by the linearity condition\footnote{Since $\op{\Pi}(s)$ is a complex scalar field, the scalar multiplication in the linearity condition \ref{A21} must be extended to $\bb{C}$.} \ref{A21} and the commutation relation
\begin{equation} 
\label{commPiPiJPi}
[\op{\Pi}(s_{1})^*,\op{\Pi}(s_{2})]=-i \Omega_{\scri}^{\KG  0}((\bar s_1,0), (0,s_2)) \op{1}
\end{equation}
together with vanishing commutators for $\op{\Pi}(s_{1})$ and $\op{\Pi}(s_{2})$.

The Hadamard condition on states $\s^{\KG 0}$ on $\Alg_{\inn}^{\rm KG0}$ is that the $2$-point function $\s^{\KG 0}(\op{\Pi}(x_{1})\op{\Pi}(x_{2}))$ is smooth, whereas
    \be
    \s^{\KG 0}(\op{\Pi}(x_{1})^{\ast}\op{\Pi}(x_{2}))=-\frac{1}{\pi}\frac{\delta_{\bb{S}^{2}}(x^{A}_{1},x^{A}_{2})}{(v_{1}-v_{2}-i0^{+})^{2}}+P(x_{1},x_{2})
    \label{kgmasslesshad}
    \ee 
where $P$ is a (state dependent) smooth function on $\scri^- \times \scri^-$ with $P(x_1, x_2) = \bar{P}(x_2, x_1)$. In addition, the connected $n$-point functions for $n\neq  2$ of $\s^{\KG 0}$ are required to be smooth. The $2$-point function of the Poincar\'e invariant vacuum state $\s^{\KG 0}_{0}$ is given by \cref{kgmasslesshad} with $P=0$.

Finally, we require that $P$ and all connected $n$-point functions of $\s^{\KG 0}$ for $n\neq 2$ decay for any set of $|v_{i}|\to \infty$ as $O((\sum_{i}v_{i}^{2})^{-\half-\epsilon})$ for some $\epsilon>0$.

\subsection{Extension to include charges and Poincar\'e generators}
\label{subsec:MaxMlessKGext}

We have already given the extension of $\Alg_{\inn}^{\EM}$ to $\Algex^{\EM}$ and $\Algexqp^{\EM}$ in \cref{subsec:chmem}, so we need only obtain the charge and Poincar\'e observables for the massless Klein-Gordon field to obtain the desired extensions of the algebra of observables for massless QED.

Classically, the action of the large gauge transformations parametrized by the smooth function \(\lambda(x^A)\) on \(\bb S^2\) is given by 
\be
\Phi(x) \to e^{i q \lambda} \Phi(x)
\label{kggauge}
\ee
where $q$ is the charge of the Klein-Gordon field. The observables $\Pi$ and $\bar{\Pi}$ transform as
\be
    \Pi(s) \mapsto \Pi(e^{iq\lambda}s) \eqsp \bar\Pi(s) \mapsto \bar\Pi(e^{iq\lambda} s).
\ee
The vector field field on the asymptotic Klein-Gordon phase space associated with infinitesimal gauge transformation is thus \( i q\lambda (- \Phi, \bar \Phi)\). This is the Hamiltonian vector field of the observable
\be
    \mc J(\lambda) = - \frac{iq}{2} \int_{\scri^-} d^3x~  \lambda(x^A) \lb[ \Phi(x) \bar\Pi(x) - \bar\Phi(x) \Pi(x) \rb].
\label{kgflux}
\ee
    Thus, $\mc J(\lambda)$ is the infinitesimal generator of the gauge transformations \cref{kggauge}, i.e., it is the contribution of the KG field to the ``charge''. (However, we use the letter ``$\mc J$'' rather than ``$\mc Q$'' since the right side of \cref{kgflux} corresponds to the integrated Klein-Gordon charge-current flux $J_{\mu}n^{\mu}$ through $\scri^-$.) The Poisson brackets $\mc J(\lambda)$ with \(\Pi(s)\) and \(\bar\Pi(s)\) are
\be
    \big\{ \mc J(\lambda), \Pi(s) \big\} = q\Pi(i\lambda s) \eqsp \big\{ \mc J(\lambda), \bar\Pi(s) \big\} = q\bar\Pi(i\lambda s)
\ee
whereas $\{ \mc J(\lambda), \mc J(\lambda') \} =0$. Of course, $ \mc J(\lambda)$ has vanishing Poisson brackets with the electromagnetic field observables.

As previously found in \cref{subsec:chmem}, the generator of large gauge transformations on the asymptotic Maxwell phase space is the memory, \cref{memobs}. Thus, the observable that generates large gauge transformations on the full Klein-Gordon-Maxwell phase space is\footnote{If a massive charged Klein-Gordon also is present, then the additional term $\mc Q_{i^-}(\lambda)$ would also be present on the right side of \cref{qi0massless}.}
\be
    \mc Q_{i^0}(\lambda) = \mc J (\lambda) + \frac{1}{4\pi} \Delta(\lambda) .
\label{qi0massless}
\ee
By arguments similar to those given in the massive case in \cref{subsec:chmem}, it can be seen that $\mc Q_{i^0}(\lambda)$ can be obtained by taking limits of surface integrals of the electric field as one approaches $i^0$ along $\scri^-$, so the subscript ``$i^0$'' is appropriate.

In parallel with the massive case, the algebra $\Alg_{\inn}^0 \defn \Alg_{\inn}^{\rm KG0} \otimes \Alg_{\inn}^{\rm EM}$ can now be extended to an algebra $\Algex^0$ by including the algebra elements $\currop(\lambda)$ and $\op{\Delta}(\lambda)$ (and, hence, 
$\cop_{i^{0}}(\lambda)$) satisfying commutation relations corresponding to the above Poisson bracket relations.

The Poincar\'e transformations act naturally on $\scri^-$. As in the massive case, each infinitesimal Poincar\'e transformation $P$ is generated by an observable $F_P$ on phase space. Writing $P= T + X$ where $T$ is a translation and $X$ is a Lorentz transformation, the Poisson brackets of the Poincar\'e generators with the Klein-Gordon observables are
\begin{subequations}\begin{align}
    \{F_{T},\Pi(s)\}=\Pi(\Lie_{T}s) &\eqsp \{F_{X},\Pi(s)\}=\Pi(\Lie_{X}s+\tfrac{1}{2}s \ms{D}_{A}X^{A}) \\
\{F_{T},\mc{J}(\lambda)\}=0 &\eqsp \{F_{X},\mc{J}(\lambda)\}=\mc{J}(\Lie_{X}\lambda)
\end{align}\end{subequations}
and the Poisson brackets of the Poincaré generators with memory, charges at spatial infinity and themselves are given by \crefrange{eq:PBFTmem}{lortrans}. 

As in the massive case, the algebra $\Algex^0$ can be further extended to an algebra $\Algexqp^0$ by including algebra elements associated with these observables satisfying commutation relations corresponding to these Poisson bracket relations.

\subsection{Fock representations}
\label{subsubsec:HadstatesFockExtAlg}

 In analogy with the massive case, we now construct the Fock representation of $\Alg_{\inn}^{\rm KG0}$ based upon the Poincar\'e invariant vacuum state.\footnote{In a similar manner to the electromagnetic case, there exists a memory effect for the massless Klein-Gordon field as well as a ``scalar charge'' at spatial infinity relating the ``in'' and ``out'' memories (see e.g. \cite{Ferko:2021bym} and sec.~F.2 of \cite{Satishchandran_2019}). In the absence of a ``source'' for the massless scalar field, the scalar memory is conserved in scattering. Therefore in massless QED, we may restrict attention to ``in'' states with zero ``scalar memory'' for simplicity.} The Fock representations of $\Alg_{\inn}^{\rm EM}$ of interest were already constructed in \cref{subsubsec:MaxMKGFockrepscriminus}.
 
The Poincar\'e invariant vacuum state is the Gaussian state determined by a vanishing \(1\)-point function and \(2\)-point function given by (see \cref{kgmasslesshad})
 \be 
 \label{eq:Pi1pt2pt}
 \s_{0}^{\KG 0}(\op{\Pi}(s_{1})^{\ast}\op{\Pi}(s_{2}))=-\frac{1}{\pi}\int_{\bb{R}^2\times \bb{S}^{2}}dv_{1}dv_{2}d \Omega~\frac{\bar{s_{1}(v_{1}, x^A)}s_{2}(v_{2}, x^A)}{(v_{1}-v_{2}-i0^{+})^{2}}
 \ee 
 for all test functions $s_{1}(x)$ and $s_{2}(x)$. The $2$-point function gives rise to the inner product
\be\label{innprodpart}
\braket{s_{1}|s_{2}}\defn  \s_{0}^{\KG 0}(\op{\Pi}(s_{1})^{\ast}\op{\Pi}(s_{2})) = 2\int_0^\infty  \omega d\omega \int_{\bb{S}^{2}} \dS~\bar{\hat{s}_{1}(\omega, x^A)} \hat{s}_{2}(\omega, x^A).
\ee
on complex-valued test functions on $\scri^-$, where in the last equality we have rewritten \cref{eq:Pi1pt2pt} in terms of the positive frequency parts of the Fourier transform of the test functions. The completion of the space of test functions yields the ``one-particle'' Hilbert space $\Hilb^{\KG 0}$. The corresponding Fock space for particles is 
 \be 
 \Fock^{\KG 0}_{\text{particles}} = \bb{C}\oplus \bigoplus_{n\geq 1}\underbrace{\big(\Hilb^{\KG 0}\otimes_{S}\dots \otimes_{S}\Hilb^{\KG 0}\big)}_{n\text{ times}}.
 \ee 
 An identical construction with the inner product $\bar{\braket{s_{1}|s_{2}}}$ yields the ``one antiparticle'' Hilbert space $\antiHilb^{\KG 0}$ and corresponding Fock space $\Fock^{\KG 0}_{\text{antiparticles}}$. One can decompose $\op{\Pi}(s)$ and $\op{\Pi}(s)^{\ast}$ into creation and annihilation operators for particles and antiparticles as in \cref{eq:Eaadag} but we shall not need to do so here. The full Fock space of particles and antiparticles is then 
 \be 
 \Fock^{\KG 0}=\Fock^{\KG 0}_{\text{particles}}\otimes \Fock^{\KG 0}_{\text{antiparticles}}.
 \ee 
 The algebraic state $\s_{0}^{\KG 0}$ is the vacuum state of the Fock space which we denote as $\ket{\s_{0}^{\KG 0}}\in \Fock^{\KG 0}$. A dense set of Hadamard states in this Fock space is generated by the linear span of the vacuum $\ket{\s_{0}^{\KG 0}}$ and symmetric tensor products of particle and antiparticle wave packet states. 
 
 As in the massive case, the large gauge transformations and Poincar\'e transformations have a strongly continuous unitary action on $ \Fock^{\KG 0}$, so $ \Fock^{\KG 0}$ carries a representation of $\Algexqp^{\rm KG0}$. The vacuum state $\ket{\s_{0}^{\KG 0}}$ is invariant under these transformations. The large gauge transformations act on the one-particle/antiparticle spaces as multiplication by a phase. The Poincar\'e group acts on the one-particle/antiparticle spaces by its natural action on $\scri^-$. 
 
As in the massive case, it will be useful to express $ \Fock^{\KG 0}$ as a direct integral over improper momentum eigenstates. It is useful to parametrize the plane wave solution of $4$-momentum $p$ by $p = (\omega, x_p^A)$, where $\omega$ is the frequency of the wave and $x_p^A \in \bb{S}^2$ is the direction of the plane wave. As in the massive case, we define $\Hilb^{\KG 0}_{p}$ to be the one-complex-dimensional Hilbert space for particles spanned by $\ket{p}$ and we define 
 \be 
 \Hilb^{\KG 0}_{p_{1}\dots p_{n}}=\Hilb^{\KG 0}_{p_{1}}\otimes_{S}\dots\otimes_{S} \Hilb^{\KG 0}_{p_{n}} ,
 \ee 
which is spanned by $\ket{p_1, \dots, p_n}$. We similarly define $\antiHilb^{\KG 0}_{q}$ and $\antiHilb^{\KG 0}_{q_{1}\dots q_{n}}$ for antiparticles. The Fock space can then be written as
 \be 
 \label{eq:Fockscrimom}
 \Fock^{\KG 0} \cong \bigoplus_{n,m\geq 0}\int_{(C^{+})^{n+m}}d^{3}p_{1}\dots d^{3}p_{n}d^{3}q_{1}\dots d^{3}q_{m}~\Hilb^{\KG 0}_{p_{1}\dots p_{n}}\otimes \antiHilb^{\KG 0}_{q_{1}\dots q_{m}}
 \ee 
where $d^3 p$ denotes the Lorentz invariant measure $d^{3}p=\omega d\omega\dS$ on the positive frequency ``cone'' $C^{+}\defn \{(\omega,x_p^{A}) ~|~ \omega>0\}$. An arbitrary state $\ket{\Psi} \in  \Fock^{\KG 0}$ can be expressed as 
 \be 
\ket{\Psi}=\sum_{n,m}\int_{(C^{+})^{n+m}} d^{3}p_{1}\dots d^{3}p_{n}d^{3}q_{1}\dots d^{3}q_{m}~\psi_{nm} (p_{1},\dots,p_{n},q_{1},\dots,q_{m})\ket{p_{1}\dots p_{n}}\otimes \ket{q_{1}\dots q_{m}}
 \ee 
where $\psi_{nm}$ is a complex, square-integrable function invariant under permutations of $p_{i}$ and permutations of $q_{i}$ and supported on non-negative frequencies.

Again, the Fock space decomposition \cref{eq:Fockscrimom} corresponds to the spectral decomposition of the charge-current flux operator $\currop (\lambda)$. Formally, we have
 \be 
 \currop (\lambda) \ket{p_{1}\dots p_{n}}\otimes \ket{q_{1}\dots q_{m}} =q\bigg(\sum_{i=1}^{n}\lambda(x^A_{p_i})-\sum_{i=1}^{m}\lambda(x^A_{q_i})\bigg) \ket{p_{1}\dots p_{n}}\otimes \ket{q_{1}\dots q_{m}} \, .
 \ee 
The formal ``unsmeared'' action of $\currop(x^{A})$ on plane wave states is the sum of $\delta$-functions on $\bb{S}^{2}$ whose support is determined by the momenta of the plane waves 
\begin{align}
     &\currop (x^{A})\ket{p_{1}\dots p_{n}}=q\bigg(\sum_{i=1}^{n}\delta_{\bb{S}^{2}}(x_{p_i}^A,x^{A})\bigg)\ket{p_{1}\dots p_{n}} \\ 
     &\currop (x^{A})\ket{q_{1}\dots q_{n}}=-q\bigg(\sum_{i=1}^{n}\delta_{\bb{S}^{2}}(x_{q_i}^A,x^{A})\bigg)\ket{q_{1}\dots q_{n}}.
\end{align}
The action of $ \currop (\lambda)$ on a proper state $\ket{\Psi}\in \Fock^{\KG 0}$ lying in the subspace of $n$ particles and $m$ antiparticles is given by
\be
 \currop(\lambda)\ket{\Psi}=\int&d^{3}p_{1}\dots d^{3}p_{n}d^{3}q_{1}\dots d^{3}q_{m}\psi(p_{1},\dots,p_{n},q_{1},\dots,q_{m})\times \\
 &q\bigg(\sum_{i=1}^{n}\lambda(x^A_{p_i})-\sum_{i=1}^{m}\lambda(x^A_{q_i})\bigg)\ket{p_{1}\dots p_{n}}\otimes \ket{q_{1}\dots q_{m}}.
 \ee
All states in this subspace are eigenstates of the total charge operator $\currop (1)$
\be 
\currop(1)\ket{\Psi} = q(n-m)\ket{\Psi}.
\ee
However, for non-constant $\lambda$, there are no proper eigenstates of $\currop(\lambda)$ apart from the vacuum state. 

\subsection{Faddeev-Kulish representation}
\label{subsec:MaxZMKGKFrep}

We now turn to the construction of the analog for massless QED of the Faddeev-Kulish Hilbert space for massive QED given in \cref{subsec:MaxMKGKFreps}. Again, the key idea is to make use of conservation of large gauge charge at spatial infinity, \cref{eq:chargecons}, and construct ``in'' and ``out'' Hilbert spaces composed of states that are eigenstates of all the large gauge charges at spatial infinity (including total electric charge). 

The large gauge charges at spatial infinity are now given by
\be 
\label{chargecurrmem}
\cop_{i^{0}}(\lambda)= \currop(\lambda) +\frac{1}{4\pi}\op{\Delta}(\lambda).
\ee 
In parallel with the massive case, to obtain states with vanishing charge, we start with the one-dimensional Hilbert space  $\Hilb^{\KG 0}_{p_{1}\dots p_{n}} \otimes  \antiHilb^{\KG 0}_{q_{1}\dots q_{n}}$, which has vanishing total electric charge and charge-current flux given by
\be
\mc{J}(\lambda) = q\sum_{i=1}^{n} \left( \lambda(x^A_{p_i})-\lambda(x^A_{q_i}) \right).
\ee
We wish to pair this state with the memory Fock space of the electromagnetic field, with memory given by
\be 
\label{impcharge}
\Delta(\lambda;p_{1},\dots,q_{n}) =  -4\pi q \sum_{i=1}^{n} \left( \lambda(x^A_{p_i})-\lambda(x^A_{q_i}) \right)
\ee 
for all $\lambda$. If, for all $p_1, \dots, q_n$ we can find a classical, smooth electromagnetic field $e_A (p_1, \dots, q_n)$ that has this memory satisfying \cref{impcharge}, then the Faddeev-Kulish Hilbert space
\be 
\label{mQEDKFQ2}
\Hilb_{\inn}^{\rm FK0} \defn \bigoplus_{n=0}^\infty ~ \int d^{3}p_{1}\dots d^{3}p_{n}d^{3}q_{1}\dots d^{3}q_{n}~\Hilb^{\KG 0}_{p_{1}\dots p_{n}}\otimes \antiHilb^{\KG 0}_{q_{1}\dots q_{n}}\otimes \Fock^{\EM }_{\Delta(p_{1},\dots,q_{n})}
\ee 
should satisfy the desired conditions \ref{enum:S1}--\ref{enum:S5} of \cref{sec:intro}. Thus, the key issue is whether we can obtain an acceptable solution to \cref{impcharge}.

In parallel with the massive case (see \cref{lapalpha}), we will be able to solve \cref{impcharge} if and only if we can solve 
\be
\ms{D}^A \Delta_A = \ms{D}^{2} \alpha = q [ 4\pi \delta_{\bb{S}^2}(x^A, x^A_{p_i})  -1].
\label{lapalpha2} 
\ee
In contrast to \cref{lapalpha}, the right side of \cref{lapalpha2} is not smooth. The general solution to \cref{lapalpha2} is
\be
\alpha =  q \log (1- \hat{r} \cdot \hat{p}_{i})+ \text{const.}
\ee
where the dot product is defined by viewing $x^A, x^A_{p_i} \in \bb{S}^2$ as unit vectors $\hat{r},\hat{p}_{i}$ in $\bb{R}^3$, respectively, and taking their Euclidean inner product. Thus,
\be
\Delta_A(x^{A};p) =  q \ms{D}_A \log (1- \hat{r} \cdot \hat{p}_{i})
\label{memln}
\ee
is the unique solution to \cref{lapalpha2}, and 
\be
\label{eq:memp1pnq1qn}
\Delta_A (x^{A}; p_{1},\dots,p_{n},q_{1},\dots,q_{n}) =  \sum_{i=1}^{n} \left( \Delta_A (x^{A}, {p_i})-\Delta_A (x^{A}, {q_i}) \right)
\ee
will yield a solution to \cref{impcharge} via
\be
\Delta(\lambda;p_{1},\dots,p_{n},q_{1},\dots,q_{n}) = \int_{\bb{S}^2} \dS~ \ms{D}^A \lambda~\Delta_A (x^{A}; p_{1},\dots,p_{n},q_{1},\dots,q_{n}). 
\ee

Thus, for massless QED we can solve \cref{impcharge} and perform the Faddeev-Kulish Hilbert space construction \cref{mQEDKFQ2}. However, there is now a very significant difference from the massive case. As can be seen from \cref{memln}, the required memory diverges as $1/|x^{A}- x^{A}_{p_i}|$ at each particle and antiparticle momentum and, hence, is not square integrable on $\bb{S}^2$. Consequently, any classical electric field $e_A (v, x^A; p_{1},\dots,q_{n})$ that gives rise to the required memory and satisfies our required fall-off conditions in $v$ cannot be smooth and, indeed, cannot be square integrable on all spheres. It follows that the states in $\Fock^{\EM}_{e(p_{1},\dots,q_{n})}$ cannot be Hadamard.\footnote{The failure of $e_A (v, x^A; p_{1},\dots,q_{n})$ to be square integrable implies that two different choices of ``dressing'' $e$ and $e^{\prime}$ with memory \cref{eq:memp1pnq1qn} will, in general, yield unitarily {\em inequivalent} Fock representations (see \cref{foot:ineqrep}). Therefore we must label these singular representations by the choice of dressing, $e$, rather than the memory, $\Delta$.}  Furthermore, the failure of $e_A (v, x^A; p_{1},\dots,q_{n})$ to be square integrable on spheres implies that its classical energy flux through $\scri^-$ diverges
\be
\int_{\scri^-} d^{3}x~ T_{vv} = \infty \, .
\ee
It follows that all states in $\Fock^{\EM}_{e(p_{1},\dots,q_{n})}$ have infinite expected energy flux through $\scri^-$. 

Thus, although each charged particle and antiparticle can be ``dressed with soft photons'' in a manner similar to the massive case, we find that in massless QED this ``dressing'' has nontrivial angular singularities. These angular singularities correspond to the ``collinear divergences'' that arise in perturbative scattering calculations in massless QED when working with momentum eigenstates. If one chooses to ignore the physical effects of the soft photons and calculate only probabilities for inclusive ``hard'' processes, the collinear divergences can be dealt with by imposing an angular cutoff. Indeed the ``Kinoshita-Lee-Nauenberg (KLN) theorem'' states that, in addition to imposing a frequency cutoff,  if one imposes an angular cut-off one can again obtain (inclusive) cross-sections by summing over all low frequency and small angle quanta in the cutoff state and then removing the cutoffs \cite{K_1962,LN_1964}. However, the whole point of the Faddeev-Kulish construction is to take the states in the Faddeev-Kulish Hilbert space \cref{mQEDKFQ2} seriously as exact ``in'' and ``out'' states of the quantum field, so that one gets a genuine $S$-matrix relating them. The angular singularities represent genuine singularities in the physical properties of these states. Thus, although a Faddeev-Kulish Hilbert space construction can be carried out in massless QED in close parallel with massive QED, all of the states in the resulting Hilbert space in massless QED are singular and are not of physical relevance.

\subsection{Source-free Yang-Mills fields}
\label{subsec:YM}

In this subsection, we consider the scattering of a source-free Yang-Mills field. As we shall see, this provides a simple model that has features similar to both massless QED as well as the gravitational case to be considered in \cref{sec:Grav}.  

A Yang-Mills gauge field $A_{\mu}^{j}$ is a one-form field valued in a Lie-algebra $\mf{g}$ of a compact, semi-simple group $G$. 
The Lagrangian for this theory is 
\be 
\mc L = -\frac{1}{4} F^{\mu \nu,i}F_{\mu \nu, i}
\ee 
where the Yang-Mills field strength tensor is defined by
\be
F_{\mu \nu }^{i} \defn \partial_{\mu}A^{i}_{\nu}-\partial_{\nu}A^{i}_{\mu}+c^{i}{}_{jk}A_{\mu}^{j}A_{\nu}^{k}
\label{FYM}
\ee
where $c^{i}{}_{jk}$ is the structure tensor of the Lie algebra $\mf{g}$ and Lie algebra indices are raised and lowered with the (positive-definite) Cartan-Killing metric
\be
k_{ij}\defn -c^{l}{}_{ik}c^{k}{}_{jl} \, .
\ee
This theory is invariant under the action of the Yang-Mills gauge transformations 
\be 
A_{\mu}^{i}\mapsto A_{\mu}^{i}+\partial_{\mu}\lambda^{i}+c^{i}{}_{jk}A_{\mu}^{j}\lambda^{k} .
\ee 

We assume that in the asymptotic past and future, the nonlinear interactions of the Yang-Mills field with itself become negligible, and the Yang-Mills field behaves as a free field.\footnote{Of course, this property does not hold for the Yang-Mills fields occurring in nature on account of their interactions with other fields, which do not become negligible in the asymptotic past and future.} In that case, the Yang-Mills field behaves asymptotically at $\scri^-$ like a collection of decoupled electromagnetic fields. The points of the incoming classical phase space are again given by the specification of the pullback of $A_{\mu}^{i}$ to $\scri^{-}$. We again choose a gauge where $n^\mu A_{\mu}^{i}\vert_{\scri^{-}}=0$ and denote the pullback of $A_{\mu}^{i}$ to $\scri^{-}$ in our chosen gauge as $A^{i}_{A}$. 

The local field observables on phase space are again the smeared electric fields
\be 
E(s)=\int_{\scri^{-}}d^{3}x~E_{A}^{i}(x)s_{i}^{A}(x)
\ee 
where $s_{i}^{A}$ is a Lie-algebra valued test vector field on $\scri^{-}$ and 
\be 
E_{A}^{i}=-\pounds_{n}A^{i}_{A}=-\partial_{v}A_{A}^{i}
\ee 
is the pullback of $F^{i}_{\mu \nu}n^{\nu}$ to $\scri^{-}$. Note that
$E(s)$ generates the infinitesimal affine transformation $A^{i}_{A}\to A_{A}^{i}-2\pi \epsilon s_{A}^{i}$ on phase space.

In exact parallel to the electromagnetic case, the algebra $\ms{A}^{\text{YM}}_{\inn}$ is defined to be the free algebra generated by the smeared field $\op{E}(s)$ satisfying \ref{A31}--\ref{A33} in \cref{subsec:MaxMKGclass} where the symplectic form of the Yang-Mills field on $\scri^{-}$ is
\be 
\Omega_{\scri}^{\YM}(A_{1},A_{2})=-\frac{1}{4\pi}\int_{\scri^{-}}dv\dS~ \lb[E_{1}^{A,i} A_{2A,i}-E_{2}^{A,i} A_{1A,i}\rb] .
\ee 
The corresponding Hadamard regularity condition on the asymptotic states of the Yang-Mills field is that the $2$-point function has the form 
\be 
\label{eq:YM2pt}
\s(\op{E}_{A}^{i}(x_{1})\op{E}_{A}^{j}(x_{2}))=-\frac{k^{ij}q_{AB}\delta_{\bb{S}^{2}}(x^{A}_{1},x^{A}_{2})}{(v_{1}-v_{2}-i0^{+})^{2}}+S_{AB}^{ij}(x_{1},x_{2})
\ee 
where $S_{AB}^{ij}$ is a (state-dependent) smooth bi-tensor on $\scri^{-}$ that is symmetric under the simultaneous interchange of $x_{1},x_{2}$ and the indices $A,B$ and $i,j$. Additionally, the connected $n$-point functions for $n\neq 2$ are smooth. We also impose the same decay requirements of $S_{AB}^{ij}$ and all connected $n$-point functions for $n\neq 2$ as in the electromagnetic case in \cref{subsec:MaxMKGclass}. The $2$-point function of the vacuum state $\s_{0}$ takes the form of \cref{eq:YM2pt} with $S_{AB}^{ij}=0$. 

Apart from the extra Lie algebra index, there is no difference between Yang-Mills theory and electromagnetism in the above construction of the algebra of asymptotic local field observables and the regularity conditions on states. However, a significant difference with electromagnetism arises when we consider the extension of the algebra to include large gauge charges. In the Yang-Mills case, the infinitesimal action of a large gauge transformation is given by
\be 
A^{i}_{A}\mapsto A^{i}_{A}+\epsilon \lb[\ms{D}_{A}\lambda^{i}+c^{i}{}_{jk}A^{j}_{A}\lambda^{k}\rb]\, , \quad \quad E^{i}_{A}\mapsto E^{i}_{A}+\epsilon c^{i}{}_{jk}E^{j}_{A}\lambda^{k}. 
\ee 
In particular, the electric field $E^{i}_{A}$ is no longer gauge invariant. The ``charge'' that generates this infinitesimal gauge transformation is
\be 
\mc{Q}^{\text{YM}}_{i^{0}}(\lambda) \defn -\frac{1}{4\pi}\int_{\scri^{-}}d^{3}x~\big(2c^{i}{}_{jk}\lambda_{i}A^{A,j}E_{A}^{k}+\lambda^{i}\ms{D}^{A}E_{A,i}\big)
\ee 
where the subscript ``$i^0$'' again has been inserted to indicate that --- assuming that no additional fields with Yang-Mills charge are present --- $\mc{Q}^{\text{YM}}_{i^{0}}(\lambda)$ can be obtained by taking limits of surface integrals of the Yang-Mills electric field as one approaches $i^0$ along $\scri^-$, as can be shown by arguments similar to the massive and massless QED cases.

It is useful to separate the contributions to $\mc{Q}^{\text{YM}}_{i^{0}}(\lambda)$ into their linear and nonlinear parts. The linear part is the {\em memory} of the Yang-Mills field associated with large gauge transformation $\lambda$:
\be 
\label{eq:memYM}
\Delta^{\! \text{YM}}(\lambda) \defn -\int_{\scri^{-}}dv\dS~E^{i}_{A}(v,x^{B})\ms{D}^{A}\lambda_{i}(x^{B}).
\ee
Although the memory is no longer the generator of large gauge transformations, it is still an observable on the asymptotic phase space, since $\tfrac{1}{4\pi}\Delta^{\! \YM}(\lambda)$ generates the affine transformation 
\be 
A^{i}_{A}\mapsto A^{i}_{A}+\epsilon \ms{D}_{A}\lambda^{i} \, , \quad \quad E^i_{A}\mapsto E^i_{A}. 
\ee 
The nonlinear part of $\mc{Q}^{\text{YM}}_{i^{0}}(\lambda)$ is the {\em Yang-Mills charge-current flux} observable
\be 
\mc{J}^{\text{YM}}(\lambda)\defn\frac{1}{2\pi}\int_{\scri^{-}} dv \dS~ c^{i}{}_{jk}\lambda_{i}A^{A,j}E^{k}_{A} \, .
\ee 
By definition, we have
\be 
\label{eq:i0chargeYM}
\mc{Q}^{\text{YM}}_{i^{0}}(\lambda) = \mc{J}^{\text{YM}}(\lambda)+\frac{1}{4\pi}\Delta^{\! \text{YM}}(\lambda).
\ee 
This is closely analogous to \cref{qi0massless} except that now, the ``null memory'' $\mc{J}^{\YM}(\lambda)$ arises from the Yang-Mills field itself, not some additional massless charged field.

The Poisson brackets of $\mc{Q}^{\text{YM}}_{i^{0}}(\lambda)$ and $\Delta^{\! \text{YM}}(\lambda)$ with themselves and with the local fields $E(s)$ can be computed using \cref{pbdef} from the above phase space transformations that they generate. We obtain
\begin{subequations}\begin{align}
\label{qecom}
\{\mc{Q}^{\text{YM}}_{i^{0}}(\lambda),E(s)\}=E([\lambda,s]) &\eqsp \{ \Delta^{\! \text{YM}} (\lambda), E(s) \} = 0 \\
\{ \Delta^{\! \text{YM}} (\lambda_1), \Delta^{\! \text{YM}} (\lambda_2) \} = 0 &\eqsp \{ \mc{Q}^{\text{YM}}_{i^{0}}(\lambda_1),\Delta^{\! \text{YM}}(\lambda_{2})\}=\Delta^{\! \text{YM}}([\lambda_{1},\lambda_{2}]) \\
\{\mc{Q}^{\text{YM}}_{i^{0}}(\lambda_{1}),\mc{Q}^{\text{YM}}_{i^{0}}(\lambda_{2})\}&=\mc{Q}^{\text{YM}}_{i^{0}}([\lambda_{1},\lambda_{2}])
\end{align}\end{subequations}
where the bracket denotes the Lie bracket $[X,Y]^{i}=c^{i}{}_{jk}X^{j}Y^{k}$ between any two elements $X,Y$ of the Lie-algebra $\mf{g}$. 
Finally, the action of the infinitesimal Poincaré tranformations $F_{P}$ with $E(s),\Delta^{\!\text{YM}}(\lambda),\mc{Q}_{i^{0}}(\lambda)$ and themselves are again given by \crefrange{eq:PBFE}{lortrans}.

In exact parallel with \cref{subsec:chmem,subsec:MaxMlessKGext}, the algebra $\Alg_{\inn}^{\text{YM}}$ can be extended to $\Algexqp^{\text{YM}}$ by including $\cop_{i^{0}}^{\text{YM}}(\lambda)$, $\op{\Delta}^{\!\text{YM}}(\lambda)$ and the Poincar\'e generators $\op{F}_{\! P}$ in the algebra, with commutation relations corresponding to the above Poisson bracket relations. In addition, as before, we impose the further condition on states 
\be 
\omega(\op{\Delta}^{\!\text{YM}}(\lambda))=-\int_{\scri^{-}}dv\dS~\s(\op{E}_{A}^{i}(v,x^{B}))\ms{D}^{A}\lambda_{i}(x^{B}) \, ,
\label{mem1pt}
\ee 
which ensures that the expectation value of the memory observable corresponds to \cref{eq:memYM}.

The Fock representations $\Fock_{\Delta}^{\YM}$ of  $\Alg_{\inn}^{\YM}$ can be constructed in direct analogy to electromagnetic case. The GNS construction based upon the vacuum state $\omega_0$ again yields the standard Fock space $\Fock^{\YM}_{0}$, for which every state is an eigenstate of $\op{\Delta}^{\!\text{YM}}(\lambda)$ with vanishing eigenvalue. Representations of nonvanishing memory can be constructed in the same manner as discussed in \cref{subsubsec:MaxMKGFockrepscriminus}. The representation of $\Alg_{\inn}^{\YM}$ on zero-memory Fock space $\Fock^{\YM}_{0}$ can be extended to a representation of $\Algexqp^{\text{YM}}$. However, this is not possible for the representations of nonvanishing memory on account of the nontrivial commutation relations of $\op{\Delta}^{\!\YM}(\lambda)$ with both $\cop_{i^{0}}^{\YM}(\lambda)$ and with Lorentz generators. 

As in electromagnetism, the zero-memory Fock space $\Fock^{\YM}_{0}$ can be represented as a direct integral of ``improper'' plane wave states. As before, an arbitrary state $\ket{\Psi}\in \Fock_{0}^{\YM}$ can be expressed as 
\be 
\ket{\Psi}=\sum_{n}\int_{(C^{+})^{n}} d^{3}p_{1}\dots d^{3}p_{n}~\psi_{(n)i_{1},\dots,i_{n}}^{A_{1}\dots A_{n}}(p_{1},\dots,p_{n})\varepsilon_{A_{1}}^{i_{1}}\dots \varepsilon_{A_{n}}^{i_{n}} \ket{p_{1},\dots,p_{n}}
\ee 
where $\psi_{(n)}$ is a complex $L^{2}$ tensor-field and $\varepsilon^{i}_{A}$ denotes Lie-algebra valued ``polarization vectors''  which satisfy
\be 
k_{ij}\varepsilon^{i}_{A}\varepsilon^{j}_{B}= \frac{1}{2}q_{AB} \eqsp q^{AB}\varepsilon^{i}_{A}\varepsilon^{j}_{B}=\frac{1}{n} k^{ij}
\ee 
where $n$ is the dimension of the group $G$. The corresponding Fock space decomposition corresponds to the spectral decomposition the ``null memory operator'' $\op{\mc{J}}^{\text{YM}}(\lambda)$. Formally we have, 
\be\label{eq:JYMmom}
    & \op{\mc{J}}^{\text{YM}}(\lambda)\varepsilon_{A_{1}}^{i_{1}}\dots \varepsilon_{A_{n}}^{i_{n}} \ket{p_{1},\dots, p_{n}} = \\[1.2ex] 
    &\qquad\qquad \big(c^{i_{1}}{}_{jk}\lambda^{j}(x^{A}_{p_{1}})\varepsilon^{k}_{A_{1}}\dots \varepsilon_{A_{n}}^{i_{n}}+\dots + c^{i_{n}}{}_{jk}\lambda^{j}(x^{A}_{p_{n}}) \varepsilon_{A_{1}}^{i_{1}}\dots \varepsilon^{k}_{A_{n}}\ket{p_{1},\dots ,p_{n}}
\ee
In the ``unsmeared'' form, the formal action of $\op{\mc{J}}^{\text{YM}}(x^{A})$ on plane wave states is given by 
\be\label{eq:Jmom}
    &\op{\mc{J}}_{j}^{\text{YM}}(x^{A})\varepsilon_{A_{1}}^{i_{1}}\dots \varepsilon_{A_{n}}^{i_{n}} \ket{p_{1},\dots, p_{n}}= \\[1.2ex]
    &\qquad\qquad \big(\delta_{\bb{S}^{2}}(x^{A},x^{A}_{p_{1}})c^{i_{1}}{}_{jk}\varepsilon_{A_{1}}^{k}\dots \varepsilon_{A_{n}}^{i_{n}}+\dots+\delta_{\bb{S}^{2}}(x^{A},x^{A}_{p_{n}})c^{i_{n}}{}_{jk} \varepsilon_{A_{1}}^{i_1}\dots \varepsilon_{A_{n}}^{k}\big)\ket{p_{1},\dots, p_{n}}.
\ee

We turn now to the issue of whether analogs of the Faddeev-Kulish ``in'' and ``out'' Hilbert spaces can be constructed.\footnote{Dressed states in non-abelian gauge theories have been previously considered in, e.g. \cite{Anupam:2019oyi,Catani:1984dp,Greco:1978te}.} As the in the case of Maxwell fields, for solutions to the Yang-Mills equations which are suitably regular at spatial infinity, the charges obtained from the limit along past and future null infinity to spatial infinity match by  
\be
\mc{Q}^{\text{YM},\inn}_{i^{0}}(\lambda)=\mc{Q}^{\text{YM},\out}_{i^{0}}(\lambda \circ \Upsilon) 
\label{chconsYM}
\ee
where $\Upsilon$ is the antipodal map on $\bb{S}^{2}$. Therefore, we are again led to seek ``in'' and ``out'' Hilbert spaces composed of eigenstates of the charge observable
\be 
\label{eq:opi0charge}
\cop_{i^{0}}^{\text{YM}}(\lambda)=\currop^{\text{YM}}(\lambda)+\frac{1}{4\pi}\op{\Delta}^{\!\text{YM}}(\lambda). 
\ee 
It should be noted first, that, in contrast to the abelian case, the charge operator now satisfies the nontrivial commutation relation $[\cop_{i^{0}}^{\text{YM}}(\lambda_{1}),\cop_{i^{0}}^{\text{YM}}(\lambda_{2})]=\cop_{i^{0}}^{\text{YM}}([\lambda_{1},\lambda_{2}])$. For a semisimple Lie group as considered here, it follows that there cannot exist any eigenstate of $\cop_{i^{0}}^{\text{YM}}(\lambda)$ for all $\lambda^i$ unless the eigenvalues vanish for all $\lambda^{i}$. (In massive and massless QED, eigenstates of $\cop^{\YM}_{i^{0}}(\lambda)$ of nonzero eigenvalue exist, although we restricted to vanishing eigenvalue in order to have an infinitesimal action of the Lorentz group.) In the case of massive and massless QED in \cref{subsec:MaxZMKGKFrep,subsec:MaxMKGKFreps}, the analog of $\currop^{\text{YM}}(\lambda)$ was played by the charge or current flux of an additional scalar field. In those cases, we can choose an (improper) eigenstate of this scalar field observable and then ``dress'' it with electromagnetic field states belonging to a corresponding memory representation. However, in the present case, $\currop^{\text{YM}}(\lambda)$ and $\op{\Delta}^{\!\text{YM}}(\lambda)$ arise from the same Yang-Mills field, so we cannot independently choose the eigenstate of $\currop^{\text{YM}}(\lambda)$ and the memory representation to which it belongs. 

To gain insight into the nature of the difficulty caused by the fact that $\currop^{\text{YM}}(\lambda)$ and $\op{\Delta}^{\!\text{YM}}(\lambda)$ arise from the same field, let us attempt to construct eigenstates of $\cop^{\YM}_{i^{0}}(\lambda)$ with vanishing eigenvalue by following a similar procedure to that used in massless QED. As in massless QED, we can start with an improper plane wave momentum eigenstate $\varepsilon_{A_{1}}^{i_{1}}\dots \varepsilon_{A_{n}}^{i_{n}} \ket{p_{1}\dots p_{n}}$ in the zero memory incoming Fock space $\Fock_{0}^{\YM}$. This state is an eigenstate of $\op{\mc{J}}^{\text{YM}}(\lambda)$ with eigenvalue given by \cref{eq:Jmom}. We now wish to ``dress'' this state with ``soft YM particles'' belonging to the memory representation with $-\Delta^{\!\text{YM}} (\lambda)/4\pi$ equal to this eigenvalue, so as to produce an eigenstate of $\cop^{\YM}_{i^{0}}(\lambda)$ with vanishing eigenvalue. As in massless QED, on account of the $\delta$-functions on $\bb{S}^{2}$ appearing in \cref{eq:Jmom}, the required memory will be singular, and the dressed states will have infinite expected total energy flux. Thus, as in massless QED, the states constructed in this manner will be unphysical. However, a further major difficulty occurs in the Yang-Mills case because the ``dressing'' now also contributes to the Yang-Mills charge-current flux. Since the dressing is singular the Yang-Mills charge-current flux of the ``dressing'' is infinite and so the resulting ``dressed state'' cannot be defined. Furthermore, even if the dressing could be defined, the resulting state would no longer be an eigenstate of $\op{\mc{J}}^{\text{YM}}(\lambda)$ and, hence, is not an eigenstate of $\cop^{\YM}_{i^{0}}(\lambda)$. Thus, the states produced by the Faddeev-Kulish ``dressing'' procedure are not only unphysical but they do even yield the desired eigenstate property that motivated their construction. 

Thus, in order to implement the strategy for constructing ``in'' and ``out'' Hilbert spaces based upon the conservation law \cref{chconsYM}, we must seek eigenstates of $\cop_{i^{0}}^{\rm YM}(\lambda)$ of vanishing eigenvalue by some procedure other than ``dressing.'' To see the nature of the restrictions on states imposed by the eigenstate condition, we note that, by definition, for any eigenstate $\s$ of $\cop_{i^{0}}^{\rm YM}(\lambda)$ with vanishing eigenvalue, we have 
\be 
\label{eq:QYME1}
\s(\cop_{i^{0}}^{\text{YM}}(\lambda)\op{E}(s))= \s(\op{E}(s)\cop_{i^{0}}^{\text{YM}}(\lambda)) = 0 .
\ee 
However, the commutation relation \cref{qecom} then implies
\be 
\label{eq:1ptYME}
c^{i}{}_{jk}\s(\op{E}_{A}^{j}(x))\lambda^{k}=0 \text{ for all $\lambda^{k}(x^{A})$}.
\ee 
which, for a semi-simple Lie algebra, implies, in turn, that
\be 
\s(\op{E}_{A}^{j}(x))=0 \, .
\ee 
Thus, the  $1$-point function of any eigenstate must vanish. Note that it then follows from \cref{mem1pt} that --- in contrast with massless QED --- the expected memory must vanish. It also then follows that the expected Yang-Mills charge-current flux must vanish. By similar arguments, the $2$-point must satisfy
\be 
\label{eq:casimir}
c^{i_1}{}_{jk}\s(\op{E}_{A_{1}}^{k}(x_{1}) \op{E}_{A_{2}}^{i_{2}}(x_{2})) + c^{i_{2}}{}_{jk}\s(\op{E}^{i_{1}}_{A_{1}}(x_{1})\op{E}^{k}_{A_{2}}(x_{2}))=0.
\ee 
This implies that $\s(\op{E}_{A_{1}}^{i_1}(x_{1}) \op{E}_{A_{2}}^{i_{2}}(x_{2}))$ must be proportional to $k^{i_1 i_2}$. This condition is satisfied by choices of $S_{AB}^{ij}(x_{1},x_{2})$ in \cref{eq:YM2pt} of the form $S_{AB}^{ij}(x_{1},x_{2}) = S'_{AB} (x_{1},x_{2}) k^{ij}$. Nevertheless, this is an extremely restrictive condition on the $2$-point function. More generally, the $n$-point correlation functions of $\s$ must be proportional to Casimirs\footnote{For a semi-simple Lie algebra $\mf{g}$, the number of independent Casimirs is finite and is equal to the rank of $\mf{g}$. For $\mf g = \mf{su}(n)$, the number of independent Casimirs is $n-1$.} of the Lie algebra $\mf{g}$. Thus, although there exist nontrivial algebraic eigenstates of $\cop_{i^{0}}^{\rm YM}(\lambda)$, it is clear that there are insufficiently many states to obtain a Hilbert large enough to carry representatives of all ``hard'' scattering processes. 

In summary, in Yang-Mills theory, the Faddeev-Kulish ``dressing'' procedure fails to produce eigenstates of $\cop_{i^{0}}^{\rm YM}(\lambda)$. Although eigenstates of $\cop_{i^{0}}^{\rm YM}(\lambda)$ do exist, there are insufficiently many of them for scattering theory. Thus, the attempt to construct ``in'' and ``out'' Hilbert spaces composed of eigenstates of charges fails. In the next section, we will see that in the gravitational case, this failure is even more dramatic, since there are no eigenstates of the large gauge (i.e., supertranslation) charges at all except for the vacuum state.

\section{Vacuum general relativity}
\label{sec:Grav}

In this section we turn our attention to the asymptotic quantum theory of full nonlinear general relativity at null infinity. There are many nontrivial, unresolved issues concerning the formulation of quantum gravity in the bulk. However, as has been emphasized by Ashtekar \cite{PhysRevLett.46.573,asymp-quant}, in asymptotically flat spacetimes the asymptotic phase space of general relativity at null infinity is an affine manifold similar to that of electromagnetism. Consequently, one can quantize the asymptotic degrees of freedom in exact parallel with the electromagnetic case.  The asymptotic symmetries of general relativity are the BMS transformations, which enlarge the Poincar\'e group by the inclusion of supertranslations. The supertranslations play a role in the asymptotic quantization of general relativity that is closely analogous to the role played by large gauge transformations in electromagnetism.

We present the asymptotic quantization algebra of local field observables for general relativity in \cref{sec:GR-quant}. The extension of this algebra to include the charges that generate BMS transformations is given in \cref{sec:GR-quant2}. We then show in \cref{subsec:GravKFreps} that an analog of Faddeev-Kulish ``in'' and ``out'' Hilbert spaces does not exist in quantum gravity.

\subsection{Asymptotic quantization of general relativity}
\label{sec:GR-quant}

As discussed in \cite{AS-symp}, the points in the asymptotic phase space of general relativity at past null infinity can be specified by an equivalence class of derivative operators intrinsic to \(\scri^-\). For our purposes, it is convenient to instead adopt the following equivalent formulation. Choose a Bondi advanced time coordinate $v$ and consider the foliation of \(\scri^-\) by the cross-sections with \(v = \text{constant}\). This foliation determines a unique null vector \(l^\mu\) at \(\scri^-\) which is normal to the cross-sections and at \(\scri^-\) satisfies
\be
    l^\mu n_\mu = -1 \eqsp l^\mu l_\mu = 0.
\ee
Then, the points of the asymptotic phase space are specified by the \emph{shear} of \(l^\mu\) which is defined by\footnote{Alternatively, one can define a symmetric tracefree tensor, \(C_{AB}\), as the angular components of the physical metric at order \(1/r\) in Bondi coordinates in the bulk spacetime. This is related to the shear that we have defined by \(C_{AB} = -2 \sigma_{AB}\); see \cite{GPS}.}
\be\label{eq:shear-defn}
    \sigma_{\mu \nu} = ({q_\mu}^\alpha {q_\nu}^\beta - \tfrac{1}{2} q_{\mu \nu}q^{\alpha \beta}) \nabla_\alpha l_\beta
\ee
where $q_{\mu \nu}$ is the metric on the cross-sections. Since $\sigma_{\mu \nu}$ is orthogonal to $l^\mu$ and $n^\mu$, we can write it as \(\sigma_{AB}\). In general relativity, \(\sigma_{AB}\) is the analogue of the vector potential \(A_A\) in the electromagnetic case. The analogue of the electric field \(E_A\) is given by the \emph{News tensor} \(N_{AB}\) defined as
\be\label{eq:News-defn}
    N_{AB} \defn 2 \Lie_n \sigma_{AB} = 2 \partial_v \sigma_{AB}.
\ee
The asymptotic symplectic form is then given by
\be
    \Omega_{\scri}^\GR(\sigma_1,\sigma_2) = \frac{1}{16\pi} \int_{\scri^{-}} d^{3}x~ \lb[ N_1^{AB} \sigma_{2AB} - N_2^{AB} \sigma_{1AB} \rb].
\ee

The local field observables are the smeared News
\be
    N(s) \defn \int_{\scri^{-}} d^{3}x~ N_{AB}(x) s^{AB}(x) \,.
\ee
where $s^{AB}$ is a real test tensor field.
The smeared News generates the affine transformation \(\sigma_{AB} \mapsto \sigma_{AB} + \epsilon 8 \pi s_{AB}\) on phase space. The Poisson brackets of the smeared News are computed to be
\be
    \pb{ N(s_1)}{N(s_2)} = - 64 \pi^2 \Omega_{\scri}^\GR(s_1,s_2) 1= 8 \pi \int_{\scri^{-}} d^{3}x~\lb[ s_{1AB} \partial_v s_2^{AB} - s_{2AB} \partial_v s_1^{AB}  \rb].
\ee

In exact parallel with electromagnetism, the asymptotic quantization algebra of local field observables, $\Alg_{\inn}^{\GR}$, is defined to be the unital $\ast$-algebra generated by the elements $\op{N}(s)$, $\op{N}(s)^\ast$ and $\op{1}$, factored by the following relations: 

\begin{enumerate}[label=(C.{\Roman*})]
\item $\op{N}(c_{1}s_{1}+c_{2}s_{2})=c_{1}\op{N}(s_{1})+c_{2}\op{N}(s_{2})$ for any $s_{1}^{AB},s_{2}^{AB}$ and any $c_{1},c_{2}\in \bb{R}$ \label{A51}
\item $\op{N}(s)^{\ast}=\op{N}(s)$ for all $s^{AB}$  \label{A52}
\item $[\op{N}(s_{1}),\op{N}(s_{2})]=-64\pi^{2}i\Omega_{\scri}^{\GR}(s_{1},s_{2}) \1$ \label{A53}
\end{enumerate}

The Hadamard regularity condition on asymptotic states \(\s\) on the News algebra $\Nalg_{\inn}$ analogous to \cref{sympsmear} is that the $2$-point function has the form 
\be \label{hadformNews}
   \s(\op{N}_{AB}(x_{1})\op{N}_{CD}(x_{2}))= - 8 \frac{\lb(q_{A(C} q_{D)B} - \half q_{AB} q_{CD}\rb) \delta_{\mathbb{S}^{2}}(x^{A}_{1},x^{A}_{2})}{(v_{1}-v_{2}-i0^{+})^{2}} + S_{ABCD}(x_{1},x_{2})
\ee
where $S_{ABCD}$ is a (state-dependent) bi-tensor on $\scri^{-}$ that is symmetric in $A,B$ and in $C,D$, satisfies \(q^{AB}S_{ABCD} = q^{CD} S_{ABCD} = 0\) and is symmetric under the simulataneous interchange of $x_{1}$ with $x_{2}$ and the pair of indices $A,B$ with the pair $C,D$. As before, we also require that $S_{ABCD}$ and the connected $n$-point functions for $n\neq 2$ of a Hadamard state on $\scri^{-}$ are smooth and decay as $O((\sum_i v_i^2)^{-1/2 - \epsilon})$ for some $\epsilon > 0$. 

\subsection{Extension of the asymptotic quantization algebra to include BMS charges}
\label{sec:GR-quant2}

The gauge symmetries of general relativity are the diffeomorphisms on spacetime. However, the transformations induced by diffeomorphisms that preserve the asymptotic structure of spacetime but do not vanish at null infinity are not degeneracies of the symplectic form and must be treated as symmetries. The group of such diffeomorphisms is known as the Bondi-Metzner-Sachs (BMS) group. For a given choice of Bondi advanced time coordinate $v$ on $\scri^-$, the vector field $\xi^\mu$ that generates an arbitrary infinitesimal BMS transformation takes the form
\be\label{eq:bms-vector}
    \xi^\mu = (f + \tfrac{1}{2}v\ms D_A X^A) n^\mu + X^\mu \, .
\ee
Here \(f(x^A)\) is an arbitrary smooth function on \(\bb S^2\) with conformal weight $+1$, and \(X^\mu\) is a vector field tangent to the cross-sections of \(\scri\) that is a conformal Killing vector field on the \(2\)-sphere. By a slight abuse of the notational conventions\footnote{By the conventions of \cref{subsec:notation}, $X^A$ would denote an equivalence class of vector fields $X^\mu$ modulo multiples of $n^\mu$. Here, since we have made a choice of Bondi advanced time coordinate $v$, we use $X^A$ to denote the particular representative that is tangent to the cross-sections of constant $v$.} stated at the end of \cref{subsec:notation}, we will denote this vector field as $X^A$. The transformations with $X^A = 0$ are referred to as {\em supertranslations} and the supertranslations with $f$ given by a linear combination of $\ell = 0,1$ spherical harmonics are the ordinary {\em translations}. The transformations generated by $X^A$ are Lorentz transformations. However, it should be noted that the decomposition of $\xi^\mu$ into a supertranslation and a Lorentz transformation depends on the choice of Bondi advanced time coordinate $v$, i.e., if $\xi^\mu$ is a ``pure Lorentz transformation'' ($f=0$) for one choice of $v$, it would correspond to a Lorentz transformation (with the same $X^A$) plus a supertranslation for other choices of $v$.

The action of an infinitesimal BMS transformation of the form \cref{eq:bms-vector} on phase space is given by the following affine transformation
\be
\label{eq:BMStrans}
    \sigma_{AB} &\mapsto \sigma_{AB} + \epsilon \lb[\tfrac{1}{2} (f + \tfrac{1}{2}v \ms D_C X^C) N_{AB} + (\ms D_A \ms D_B - \tfrac{1}{2} q_{AB} \ms D^2 ) f + \Lie_X \sigma_{AB} - \tfrac{1}{2} (\ms D_C X^C) \sigma_{AB} \rb]\\
    N_{AB} &\mapsto N_{AB} + \epsilon \lb[(f + \tfrac{1}{2}v \ms D_C X^C)\partial_v N_{AB} + \Lie_X N_{AB} \rb].
\ee
Note that $N_{AB}$ is {\em not} invariant under BMS symmetries. The charge observable that generates this infinitesimal BMS transformation is given by (see \cite{GPS})
\be\label{eq:bms-flux}
    {\mc Q}^{\GR}_{i^0} {(f,X)} = \frac{1}{16\pi} \int_{\scri^{-}} dv\dS~ N^{AB} &\bigg[ \tfrac{1}{2} (f + \tfrac{1}{2}v \ms D_C X^C) N_{AB} + \ms D_A \ms D_B f \\
    & \qquad + \Lie_X \sigma_{AB} - \tfrac{1}{2} (\ms D_C X^C) \sigma_{AB} \bigg].
\ee
As we shall now explain, we have inserted the subscript ``$i^0$'' on ${\mc Q}^{\rm GR}_{i^0} {(f,X)}$ for reasons analogous our use of this notation in massive and massless QED. As shown in \cite{WZ,GPS}, the right-hand-side of \cref{eq:bms-flux} can be written as\footnote{Note that \cref{qi0gr} has a relative overall sign compared to \cref{eq:QGRu1u2} in \cref{sec:intro} due to the fact that we are now working at $\scri^-$ rather than $\scri^+$.}
\be
  &\frac{1}{16\pi} \int_{\scri^{-}} dv\dS~ N^{AB} \lb[ \tfrac{1}{2} (f + \tfrac{1}{2}v \ms D_C X^C) N_{AB} + \ms D_A \ms D_B f + \Lie_X \sigma_{AB} - \tfrac{1}{2} (\ms D_C X^C) \sigma_{AB} \rb] \\
  =~ & \lim_{v \to \infty} \mc Q^{\GR}_v(f,X) - \lim_{v \to - \infty} \mc Q^{\GR}_{v}(f,X) 
  \label{qi0gr}
\ee
with
\be
    \mc Q^{\GR}_v(f,X) &= \frac{1}{8\pi}\int_{S(v)} \dS~ \bigg[ \tfrac{1}{2} (f+\tfrac{1}{2}v \ms D_A X^A) \sigma^{AB} N_{AB} + X^A \sigma_{AB} \ms D_C \sigma^{BC} - \tfrac{1}{4} \sigma_{AB} \sigma^{AB} \ms D_C X^C \\
    &\qquad\qquad\quad \quad   + \Omega^{-1} C_{\mu\nu\lambda\rho} \xi^\mu l^\nu n^\lambda l^\rho \bigg]
\ee
where the integral on the right side is taken over the cross-section, $S(v)\cong \bb{S}^{2}$, of advanced time $v$ on $\scri^-$, and \(C_{\mu\nu\lambda\rho}\) is the Weyl tensor of the conformally-completed spacetime. If massive fields (or black/white holes) are present, they would, in general, contribute to $\lim_{v \to - \infty} \mc Q^{\GR}_{v}(f,X)$ in a manner similar to massive QED. We will assume, for simplicity, that this is not the case and thus that\footnote{We emphasize that the conclusions of \cref{subsec:GravKFreps} and, in particular, \cref{thm:eigensupcharge} do not depend upon this assumption.} $\lim_{v \to - \infty} \mc Q^{\GR}_{v}(f,X) = 0$. In that case, \cref{qi0gr} shows that ${\mc Q}^{\GR}_{i^0} {(f,X)}$ can be obtained as a limit of a surface integral of local quantities as one approaches spatial infinity, $i^0$, along $\scri^-$. Thus, the subscript ``$i^0$'' is appropriate in \cref{eq:bms-flux}.

As in the Yang-Mills case, it is useful to separate the contributions to ${\mc Q}_{i^0}^{\rm GR} {(f,X)}$ into their linear and nonlinear parts. The linear term arises only for supertranslations and defines the gravitational memory observable
\be 
\label{eq:memGR}
\Delta^{\! \GR}(f) \defn \frac{1}{2}\int_{\scri^{-}}dv\dS~N_{AB}(v,x^{C}) \ms{D}^{A}\ms{D}^{B}f(x^{C})
\ee 
which, we note, vanishes if $f$ is a linear combination of $\ell=0,1$ spherical harmonics. On the asymptotic phase space, $\tfrac{1}{8\pi}\Delta^{\! \GR}(f)$ generates the affine transformation 
\be
\sigma_{AB} &\mapsto \sigma_{AB} + \epsilon  (\ms D_A \ms D_B - \half q_{AB} \ms D^2) f \\
    N_{AB} &\mapsto  N_{AB} .
\ee
For $X^A = 0$, the supertranslation charge ${\mc Q}^{\rm GR}_{i^0} {(f)}$, which generates supertranslations by \cref{eq:BMStrans} with \(X^A = 0\), is given by 
\be\label{eq:QFmem}
 {\mc Q}^{\rm GR}_{i^0} {(f)} = {\mathcal J}^\GR(f) + \frac{1}{8\pi} {\Delta}^{\! \GR}(f)
\ee
where
\be
{\mathcal J}^\GR(f) \defn \frac{1}{32\pi} \int_{\scri^{-}} dv\dS~ f N^{AB} N_{AB} 
\label{nullmem}
\ee
is called the {\em null memory}. If massless fields with stress energy $T_{\mu \nu }$ are present they will, in general contribute to the null memory by the simple substitution $N^{AB}N_{AB} \to N^{AB}N_{AB} + 32 \pi \Omega^{-2}T_{\mu \nu}n^{\mu}n^{\nu}$ in \cref{nullmem}. However, for simplicity, we shall consider only case of vacuum gravitational fields in this section. 

The Poisson bracket of ${\mc Q}^{\rm GR}_{i^0} {(f,X)}$ with the local News observable is given by
\be
    \pb{{\mc Q}^{\rm GR}_{i^0} {(f,X)}}{N(s)} = N(s')
    \label{newsch}
\ee
where \(s'_{AB} = (f + \tfrac{1}{2}v \ms D_C X^C) \partial_v s_{AB} + \Lie_X s_{AB} - \tfrac{1}{2} (\ms D_C X^C) s_{AB} \). We also have
\begin{subequations}\begin{align}
    \pb{{\mc Q}^{\GR}_{i^0} {(f_1)}}{{\mc Q}^{\GR}_{i^0} {(f_2)}} = 0 \eqsp& \pb{{\mc Q}^{\GR}_{i^0} {(X_1)}}{{\mc Q}^{\GR}_{i^0} {(X_2)}} = {\mc Q}^{\rm GR}_{i^0} {([X_1,X_2])}, \\
    \pb{{\mc Q}^{\rm GR}_{i^0} {(X)}}{{\mc Q}^{\rm GR}_{i^0} {(f)}} &=  {\mc Q}^{\GR}_{i^0} (\Lie_X f - \half (\ms D_A X^A)  f) 
\end{align}\end{subequations}
where \([X_1,X_2]\) is the Lie bracket of $X^A_1$ and $X^A_2$. The memory observable has vanishing Poisson brackets with the News and with itself
\be
\pb{\Delta^{\! \GR}(f)}{N(s)} = 0, \quad \quad \pb{\Delta^{\! \GR}(f_1)}{\Delta^{\! \GR}(f_2)} = 0.
\ee
Finally, we have
\be
  \pb{{\mc Q}^{\rm GR}_{i^0} {(f_1)}}{\Delta^{\! \GR}(f_2)} = 0, \quad \quad   \pb{{\mc Q}^{\rm GR}_{i^0} {(X)}}{\Delta^{\! \GR}(f)} = \Delta^{\! \GR}(\Lie_X f - \half (\ms D_A X^A) f).
\ee
Thus, the memory observable is supertranslation-invariant but not Lorentz-invariant.

In exact parallel with massive and massless QED and Yang-Mills theory, we now can extend the algebra,  $\Alg_{\inn}^{\GR}$, of asymptotic local field observables to an algebra $\Algex^{\GR}$ by including $\cop_{i^{0}}^{\GR}(f,X)$ and $\op{\Delta}^{\! \GR}(f)$ in the algebra, with commutation relations corresponding to the above Poisson bracket relations. (The Poincar\'e generators are, of course, already included in the BMS charges $\cop_{i^{0}}^{\GR}(f,X)$, so there is no need for a further extension of the algebra.) In parallel with \cref{mem1ptfun,mem1pt} we impose on states the condition
\be
    \s(\op\Delta^{\! \GR}(f)) = \frac{1}{2} \int_{\scri^{-}} dv \dS~ \s(\op N_{AB}(v,x^{A}))\ms D^A \ms D^B f.
\ee

The Fock representations of  $\Alg_{\inn}^{\GR}$ can be constructed in direct analogy to electromagnetic case. The GNS construction based upon the vacuum state $\omega_0$ again yields the standard Fock space, ${\mathscr F}^{\GR}_0$, for which every state is an eigenstate of $\op{\Delta}^{\! \GR}(f)$ with vanishing eigenvalue. Again, representations of nonvanishing memory, ${\mathscr F}^{\GR}_\Delta$, can be constructed in the same manner as discussed in \cref{subsubsec:MaxMKGFockrepscriminus}. The representation of $\Alg_{\inn}^{\GR}$ on the zero-memory Fock space can be extended to a representation of $\Algex^{\GR}$. The representations of $\Alg_{\inn}^{\GR}$ on the Fock spaces of nonzero memory can be extended to include representatives of the generators of supertranslations, $\cop_{i^{0}}^{\GR}(f)$. However, the representations of nonzero memory cannot be extended to include representatives of the generators of infinitesimal Lorentz transformations, $\cop_{i^{0}}^{\GR}(X)$, on account of the nontrivial commutation relation of $\op{\Delta}^{\! \GR}(f)$ with $\cop_{i^{0}}^{\GR}(X)$. In particular, angular momentum is not well-defined on the Fock spaces of nonzero memory. 

\subsection{Faddeev-Kulish representations do not exist in quantum gravity}
\label{subsec:GravKFreps}

For solutions of the vacuum Einstein equation that satisfy the Ashtekar-Hansen \cite{AH} asymptotic flatness conditions together with an additional null regularity condition at spatial infinity, it was shown in \cite{KP-GR-match,PS-GR-match} that the charges \(\mc Q^{\GR}_{i^0}(f,X)\) obtained from the limit along past null infinity are matched antipodally to the similarly defined charges obtained from the limit along future null infinity.\footnote{For linearized gravity around a Minkowski background, the matching of the supertranslation charges is also shown in \cite{Tro,Mohamed_2021} (see also \cite{Henneaux:2018hdj}).} In particular the supertranslation charges satisfy
\be\label{gravcons}
    \mc Q_{i^0}^{\GR,\inn}(f) = \mc Q_{i^0}^{\GR,\out} (f \circ \Upsilon)
\ee
where, as before, \(\Upsilon\) is the antipodal map on \(\bb S^2\). This is an exact analog of \cref{eq:chargecons} in electromagnetism. As previously explained in \cref{subsec:MaxMKGKFreps}, this conservation law provides a potential means of constructing ``in'' and ``out'' Hilbert spaces satisfying the desired properties \ref{enum:S1}--\ref{enum:S5} given in \cref{sec:intro}. Namely, if we can construct an ``in'' Hilbert space composed entirely of eigenstates of the supertranslation charges, it will evolve to an ``out'' Hilbert space composed of eigenvectors of corresponding eigenvalue. In order to have a continuous action of the Lorentz group, we must choose the eigenvalues of all of the charges to vanish. If such ``in'' and ``out'' Hilbert spaces of vanishing charges are separable and contain sufficiently many states to account for all ``hard'' scattering processes, then properties \ref{enum:S1}--\ref{enum:S5} should hold.

In massive QED, this strategy was successfully implemented by the Faddeev-Kulish construction described in \cref{subsec:MaxMKGKFreps}. In this construction, one ``dresses'' each momentum eigenstate of the incoming massive charged particles with an electromagnetic state belonging to the memory representation whose memory cancels the large gauge charges of the incoming charged particle state, so as to produce an eigenstate of vanishing eigenvalue of all of the total large gauge charges. As shown in \cref{subsec:MaxZMKGKFrep}, in massless QED, the same ``dressing'' construction can be given. However, the required memory in this case is singular --- so singular that the ``soft photon dressing'' has infinite energy flux. As shown in \cref{subsec:YM}, the situation is worse in Yang-Mills theory. Not only is a singular ``dressing'' required, but the ``dressing'' does not provide eigenstates of charge. Although eigenstates of charge can be constructed by other means, there are insufficiently many of them for scattering theory. We turn now to the analysis of the situation in quantum gravity.

First, as already mentioned above, in order to have a continuous action of the Lorentz group on the Hilbert space, it is necessary to restrict to eigenstates of the charges of vanishing eigenvalue. In massive QED, this required the vanishing of all large gauge charges, including the total ordinary electric charge. The vanishing of large gauge charges for $\ell \geq 1$ is achieved by the ``dressing,'' but the vanishing of the total ordinary electric charge is an unwanted restriction on the scattering states. Nevertheless, arguably, this is not a genuine restriction since one could always put additional particles ``behind the moon'' to make the total charge vanish. However, in the gravitational case, the corresponding requirement for a continuous action of the Lorentz group is for all of the supertranslation charges ${\mc Q}^{\rm GR}_{i^0} {(f)}$ to vanish, include the charges associated with ordinary translations. In other words, the states must be eigenstates of $4$-momentum of eigenvalue zero. But the vacuum state is the only such state; one cannot cancel the $4$-momentum of a state of interest by putting additional particles ``behind the moon.'' Thus, the Faddeev-Kulish construction fails for this elementary reason at this initial stage.

Nevertheless, one could give up on having a continuous action of the Lorentz group and seek eigenstates of ${\mc Q}^{\rm GR}_{i^0} {(f)}$ of nonzero eigenvalue. It is instructive to see what happens if one attempts to construct such eigenstates by a ``dressing'' procedure. 

First, consider {\em linearized gravity} with an additional massless quantum field source, where the null memory is due to the massless source rather than to gravitational radiation. As previously noted below \cref{nullmem}, the massless field will contribute a null memory of the form
\be
{\mathcal J}^{\GR, {\rm source}}(f) = \int_{\scri^{-}} dv\dS~ f (\Omega^{-2}T_{\mu \nu}n^{\mu}n^{\nu}) \, .
\ee
In a manner similar to massless QED and Yang-Mills theory, the (improper) plane wave states of the massless source field are formal eigenstates of null memory. In order to produce an eigenstate of $\cop^{\GR}_{i^{0}}(f)$ with eigenvalue $\mc{Q}^{\GR}_{i^{0}}(f)$ for all $f$ in linearized gravity, we must ``dress'' a source particle momentum eigenstate $\ket{p}$ by choosing a memory representation of the gravitational field such that 
\be 
\label{eq:memGR2}
\ms{D}^{A}\ms{D}^{B}\Delta^{\! \GR}_{AB}(x^{A};p)=8\pi \omega \delta_{\bb{S}^{2}}(x^{A},x^{A}_{p})-8\pi \mc{Q}^{\GR}_{i^{0}}(x^{A})
\ee 
where \(\omega\) is the frequency associated with the null momentum \(p=(\omega,x^{A}_{p})\). This equation differs from the corresponding equation (\ref{lapalpha2}) in massless QED in that now there are two angular derivatives of memory rather than one. This difference results in milder angular singularities in the solution, namely, $|\Delta^{\! \GR}_{AB}| \sim |\log(|x^{A}-x^{A}_{p}|)|$ rather than $|\Delta_A| \sim 1/|x^{A}-x^{A}_{p}|$ as in massless QED. In other words, the collinear divergences in quantum gravity are less severe than in massless QED and Yang-Mills theory. Although the required memories are still singular, they are square integrable, and the corresponding ``dressed states'' --- which were previously constructed in \cite{Akhoury_2011} --- do not have an infinite energy flux. In this respect, the situation in linearized gravity is {\em better} than in massless QED and Yang-Mills theory, although since we cannot construct eigenstates of vanishing charges, the action of the Lorentz group is undefined and therefore the angular momentum is undefined for all such ``dressed'' states in linearized gravity.

However, in {\em nonlinear gravity}, as in Yang-Mills theory, the ``dressing'' will now contribute to the null memory, so the resulting dressed state is no longer an eigenstate of \(\currop^\GR(f)\) and hence is not an eigenstate\footnote{In contrast to the Yang-Mills case, the dressing contribution to the null memory is finite and so the ``dressed state'' can have a well-defined expected charge. Nevertheless, it cannot be an eigenstate of $\cop^{\GR}_{i^{0}}(f)$.} of $\cop^{\GR}_{i^{0}}(f)$. Thus, the ``dressing'' construction does not yield the desired eigenstate property that motivated the procedure. In order to implement the strategy for constructing ``in'' and ``out'' Hilbert spaces based upon the conservation law \cref{gravcons}, we must seek eigenstates of the supertranslation charges $\cop^{\GR}_{i^{0}}(f)$ by some other means. In Yang-Mills theory, we were able to find some eigenstates of large gauge charges, but insufficiently many to do scattering theory. However, one of the key results of this paper is that in gravity, there are no nontrivial eigenstates at all.\footnote{Note that there is no state in any non-zero memory representation that has vanishing null memory \cref{nullmem}; zero is merely the lower bound of the continuous spectrum of the null memory operator, as emphasized by Ashtekar \cite{asymp-quant}. Consequently, in contrast to claims in \cite{Laddha:2020kvp}, memory vacua are not eigenstates of the charges \(\cop^{\GR}_{i^0}(f)\) at spatial infinity.} This is shown by the following theorem:
\begin{thm}
\label{thm:eigensupcharge}
Let $f$ be any smooth function on \(\bb S^2\) whose support is all of \(\bb S^2\), i.e., $f$ does not vanish identically on any open subset of \(\bb S^2\). Suppose that the state $\omega$ is Hadamard, satisfies our decay conditions, and is an eigenstate of the supertranslation charge $\cop^{\GR}_{i^{0}}(f)$. Then $\omega = \s_{0}$, where $\s_0$ is the BMS-invariant vacuum state. 
\begin{proof}
Since $\omega$ is an eigenstate of $\cop^{\GR}_{i^{0}}(f)$, we have
\be 
\label{1pteigengrav1}
\s (\cop^{\GR}_{i^{0}}(f)\op{N}(s)) = \s (\op{N}(s)\cop^{\GR}_{i^{0}}(f)) = \kappa~ \s (\op{N}(s)). 
\ee 
where $\kappa$ denotes the eigenvalue, which is real since since $\cop^{\GR}_{i^{0}}(f)$ is self-adjoint. Thus, we have
\be
\label{1pteigengrav2}
0 = \s ([\cop^{\GR}_{i^{0}}(f),\op{N}(s)]) = i\s (\op{N}(f\partial_{v}s))
\ee
where the commutation relation corresponding to \cref{newsch} was used. Since this holds for all $s^{AB}$, we have for all $x = (v,x^A)$
\be
f(x^A) \frac{\partial}{\partial v} \left[ \omega (\op{N}_{AB} (x)) \right] = 0
\ee
Since $f$ does not vanish on open sets, it follows that \(\s (\op{N}_{AB}(x))\) is constant in \(v\). The decay conditions then imply that the $1$-point function $\s (\op{N}_{AB}(x))$ vanishes identically. 

By similar arguments starting with
\be
\s (\cop^{\GR}_{i^{0}}(f)\op{N}(s_{1})\dots \op{N}(s_{n})) = \s (\op{N}(s_{1})\dots \op{N}(s_{n})\cop^{\GR}_{i^{0}}(f))= \kappa~\s (\op{N}(s_{1})\dots \op{N}(s_{n}))
\ee
we find that the $n$-point functions satisfy
\be
    \sum_{i=1}^{n} \partial_{v_i} \s (\op{N}_{A_{1}B_{1}}(x_{1})\dots \op{N}_{A_{n}B_{n}}(x_{n})) = 0.
\label{nptcond}
\ee
It then follows that $S_{ABCD}$ in \cref{hadformNews} and the truncated \(n\)-point functions also satisfy this equation. But $S_{ABCD}$ and the truncated \(n\)-point functions are required to decay as $O((\sum_i v_{i}^2)^{-\half-\epsilon})$. It follows that $S_{ABCD}$ and all truncated $n$-point functions of $\omega$ vanish, i.e., $\omega = \omega_0$. 
\end{proof}
\end{thm}

We emphasize that the implications of \cref{thm:eigensupcharge} are quite strong in that constructions based upon the use of \cref{gravcons} require eigenstates of $\cop^{\GR}_{i^{0}}(f)$ for all $f$. Note that \cref{nptcond} is in close parallel to \cref{eq:casimir} in the Yang-Mills case. However, nontrivial solutions to \cref{eq:casimir} do exist, whereas the vacuum state is the only state that satisfies \cref{nptcond}. Thus, in the gravitational case, the attempt to construct ``in'' and ``out'' Hilbert spaces by using charge eigenstates fails in a much more catastrophic manner.

\section{Non-Faddeev-Kulish representations}
\label{sec:NKFreps}

As we have just seen, the Faddeev-Kulish ``dressing'' procedure cannot be used to construct a Hilbert space of ``in'' and ``out'' states in quantum gravity, nor is there any other procedure that can produce eigenstates of the supertranslation charges $\mc{Q}^{\GR}_{i^{0}}$. Nevertheless, as described at the end of \cref{sec:GR-quant2}, there is an ample supply of ``in'' and ``out'' states given by the memory Fock spaces $\Fock^{\rm GR}_\Delta$, with $\Delta^{\!\GR}_{AB}$ an arbitrary smooth, symmetric tracefree tensor on $\bb S^2$. Is there is some other way of assembling these states into Hilbert spaces in such a way that the desired conditions \ref{enum:S1}--\ref{enum:S5} given in \cref{sec:intro} can be satisfied? In this section, we explore this possibility. 

An obvious candidate for the ``in'' Hilbert space would be the direct sum over all of the ``in'' memory Fock spaces $\Fock^{\rm GR, in}_\Delta$, i.e., 
\be
\Fock_{\rm DS}^{\inn} = \bigoplus_\Delta \Fock^{\rm GR, in}_\Delta
\ee
where $\Delta^{\! \GR}_{AB}$ ranges over all (say, smooth) symmetric tracefree tensors on ${\bb S}^2$, where the ``${\rm DS}$'' subscript stands for ``direct sum.'' $\Fock_{\rm DS}^{\out}$ would then be defined similarly. Clearly, $\Fock_{\rm DS}^{\inn}$ and $\Fock_{\rm DS}^{\out}$ would then allow all possible memories. However, this choice has many serious deficiencies.\footnote{The first two of these deficiencies are analogous to what would occur if one attempted to take the Hilbert space of one-dimensional Schr\"odinger quantum mechanics to be $\oplus_{x \in {\bb R}} H_x$ where $H_x$ is a one dimensional Hilbert space representing an eigenstate of the position operator with eigenvalue $x$. This Hilbert space is nonseparable and does not admit a strongly continuous action of translations --- so the momentum operator cannot be defined.} First, since there are uncountably many choices of $\Delta^{\! \GR}_{AB}$, this Hilbert space is clearly nonseparable. Second, although the BMS group acts naturally on $\Fock_{\rm DS}^{\inn}$, Lorentz transformations act nontrivially on memory and a ``small'' Lorentz transformation will map a vector in the sector $\Fock^{\rm GR}_\Delta$ into an entirely different sector $\Fock^{\rm GR}_{\Delta'}$. Since all states in different memory sectors are orthogonal to each other, Lorentz transformations do not act in a strongly continuous manner on $\Fock_{\rm DS}^{\inn}$. Thus, infinitesimal generators of Lorentz transformations --- in particular, angular momentum --- cannot be defined. However, by far the most serious deficiency is that it is clear that states in $\Fock_{\rm DS}^{\inn}$ will not evolve to states in the similarly defined ``out'' Hilbert space $\Fock_{\rm DS}^{\out}$, so condition \ref{enum:S2} of the \cref{sec:intro} will not be satisfied. To see this, we note that ---  since the norm of the direct sum is the sum over the norms in each $\Fock_{\Delta}^{\GR,\inn}$ and any uncountable sum of strictly positive numbers is infinite --- for any vector in $\Fock_{\rm DS}^{\inn}$ the probability of having a given value of memory can be nonvanishing for only a countable number of memories. Thus, the possible memories of any state in $\Fock_{\rm DS}^{\inn}$ are discrete. However, it seems clear that most states in $\Fock_{\rm DS}^{\inn}$ will evolve to ``out'' states where the memory is continuously distributed. Such states cannot lie in $\Fock_{\rm DS}^{\out}$.

A more promising candidate would be to take a direct integral of the Fock spaces $\Fock^{\rm GR, in}_\Delta$ with respect to a measure that is continuously distributed in $\Delta^{\!\GR}_{AB}$. To do so, we first need to make a precise choice of the space, $\ms{M}$, of memories. Then we need to specify a $\sigma$-algebra of measurable subsets of $\ms{M}$. Then, we need to define a measure on $\ms{M}$, i.e., a map, $\mu$, from measurable subsets to nonnegative real numbers such that $\mu(\emptyset)=0$ and $\mu$ is ``countably additive'', that is, for any countable collection of disjoint measurable sets $\{\ms{O}_{i}\}$ we have that $\mu (\union_{i}\ms{O}_{i})=\sum_{i}\mu(\ms{O}_{i})$. Given such a measure, $\mu$, we can construct a direct integral Hilbert space $\Fock_{\rm DI}^{\inn}(\mu)$ from the memory Fock spaces $\Fock^{\rm GR, in}_\Delta$ as follows: A vector $\ket{\Psi} \in \Fock_{\rm DI}^{\inn}(\mu)$ consists of the specification of a measurable family of vectors $\ket{\psi(\Delta)} \in \Fock^{\rm GR, in}_\Delta$ for all $\Delta^{\! \GR}_{AB}$, where $\ket{\Psi}$  and $\ket{\Psi'}$ are considered equivalent if $\ket{\psi(\Delta)}$ and $\ket{\psi'(\Delta)}$ differ only on a set of measure zero. The states in $\Fock_{\rm DI}^{\inn}(\mu)$ are required to have finite norm 
\be 
\label{eq:dirintnorm}
\norm{\Psi}^2 =  \int_{\ms{M}}d\mu~ \norm{\psi(\Delta)}^2 < \infty 
\ee 
and the inner product of two states is then defined by
\be
\braket{\Psi_1|\Psi_2} =  \int_{\ms{M}}d\mu~ \braket{\psi_1 (\Delta) | \psi_2 (\Delta)}.
\ee

The direct sum Hilbert space $\Fock_{\rm DS}^{\inn}$ is a special case of the direct integral Hilbert space wherein the $\sigma$-algebra is taken to be all subsets of $\ms{M}$ and $\mu$ is taken to be the ``discrete measure'' that assigns unit measure to any subset consisting of a single point. As already stated above, this yields a Hilbert space that is nonseparable and has other unacceptable properties. However, choices of ``continuous measures'' can yield a separable Hilbert space and have the possibility of satisfying the other properties desired for scattering theory. If $\ms{M}$ were a finite dimensional vector space, there is an essentially unique notion of Lebesgue measure and this would provide a natural choice of measure. However, $\ms{M}$ is infinite dimensional, so there is no notion of Lebesgue measure (see sec.~A.4 of \cite{GJ}). Since there is a direct correspondence between memories and supertranslations \cite{Hollands_2016,Satishchandran_2019} and the supertranslations comprise a group, one might try to use
an ``invariant Haar measure'' on memories. However, the supertranslation group is not locally compact and therefore the Haar measure does not exist. Nevertheless, there are well-defined notions of Gaussian measures.\footnote{In path integral formulations of Euclidean QFT of some field \(\phi\), it is common to write the measure in the path integral as \(D\phi~ e^{-S_0(\phi)}\) where \(S_0(\phi)\) is the Euclidean action of a free field and \(D\phi\) is a ``Lebesgue measure'' on the space of fields. However, \(D\phi\) does not really exist and it is the full quantity \(D\phi~ e^{-S_0(\phi)}\) which is a genuine Gaussian measure; the covariance of this measure is the Euclidean Green's function determined by the action \(S_0\). This is also true in path integral formulations of quantum mechanics where the measure over the space of paths is the (Gaussian) Wiener measure \cite{GJ}. \label{fn:path-integral}} We can obtain a natural class of Gaussian measures in the following manner (see, e.g., \cite{Bogachev,Kuo,Gelfand4} for further details).

We start with the topological vector space of smooth functions \(f\) with conformal weight \(1\) on ${\bb S}^2$ with the nuclear topology. The trace-free part of $\ms{D}_{A}\ms{D}_{B}f$ for such $f$ provides a space of test tensor fields for memory (see \cref{eq:memGR}). We take $\ms{M}$ to be the topological dual space. We choose the $\sigma$-algebra of subsets to be generated by the ``cylindrical sets'' (see \cite{Bogachev,Gelfand4}). By the Bochner-Minlos theorem (see sec.~A.6 of \cite{GJ}), a Gaussian measure on $\ms{M}$ (centered at zero) is then determined by specifying any positive, symmetric, bilinear map $K(f_1,f_2)$ (the ``covariance matrix'') on the space of test functions. A necessary condition for two Gaussian measures with covariance $K$ and $K^{\prime}$, respectively, to be equivalent (i.e., such that they agree on which subsets of $\ms{M}$ have measure zero) is that they define equivalent norms on the space of test functions. In other words, $K$ and $K'$ are equivalent if there is a positive constant $c$ such that 
\be 
\label{eq:equivnorm}
c^{-1}K(f,f)\leq K^{\prime}(f,f)\leq c K(f,f)
\ee 
for all test functions $f$. Thus, there exists a very large class of inequivalent Gaussian measures that can be constructed by this procedure. 

Thus, the key issue for the construction of a Gaussian measure on the space, $\ms{M}$, of memories --- and, thereby, direct integral Hilbert spaces of ``in'' and ``out'' states --- is the choice of covariance matrix $K$. A key criterion is that $K$ be Lorentz invariant, since the resulting Gaussian measure $\mu$ will then be Lorentz invariant, and the Lorentz group will then act naturally on $\Fock_{\rm DI}^{\inn}(\mu)$. However, there is a unique choice of Lorentz invariant covariance matrix $K$ (see Ch.~III.4 of \cite{Gelfand5}). This covariance matrix is simply the $L^2$ inner product on the space of smooth test tensors $\ms{D}_{A}\ms{D}_{B}f$ on ${\bb S}^2$
\be
K(f_{1},f_{2}) = \int_{\bb{S}^{2}\times \bb{S}^{2}}d\Omega_{1}d\Omega_{2} ~K_{ABCD}(x_{1}^{A},x_{2}^{B})\ms{D}^{A}\ms{D}^{B}f_1(x^{A}_{1})\ms{D}^{C}\ms{D}^{D}f_2(x_{2}^{B}) \,
\ee 
with integral kernel 
\begin{equation}\label{eq:inv-K}
    K_{ABCD}(x^{A}_{1},x_{2}^{B})=\delta_{\bb{S}^{2}}(x_{1}^{A},x_{2}^{B})\lb(q_{A(C}q_{D)B}-\tfrac{1}{2}q_{AB}q_{CD}\rb).
\end{equation}
The direct integral Hilbert space $\Fock_{\rm DI}^{\inn}(\mu)$ obtained from the Gaussian measure determined by $K$ is a separable Hilbert space.

However, there is a very serious problem with attempting to use this Hilbert space for scattering theory. For any Gaussian measure constructed in the manner described above using a covariance matrix $K$ there is a subset of $\ms{M}$, determined by $K$, known as the ``Cameron-Martin space'' on which $\mu$ has zero measure (see theorem 2.4.7 in \cite{Bogachev}). For the case of a Gaussian measure with covariance given by \cref{eq:inv-K}, the  Cameron-Martin space is the space of square-integrable memories.  This means that ``almost all'' of the memories, $\Delta^{\GR}_{AB}$, that contribute to $\Fock_{\rm DI}^{\inn}(\mu)$ fail to be square-integrable.\footnote{This is analogous to the statement that in the case of Wiener measure, the differentiable paths are of measure zero, while in the Euclidean path integral the field configurations with finite action are of measure zero.} However, as argued in \cref{subsec:MaxZMKGKFrep}, states with a non-square-integrable memory that satisfy our fall-off conditions cannot be Hadamard and have divergent expected total energy flux. Thus, all of the states in $\Fock_{\rm DI}^{\inn}(\mu)$ are unphysical.

Thus, the direct integral Hilbert space obtained from the Gaussian measure constructed from the Lorentz invariant covariance matrix \cref{eq:inv-K} does not yield an acceptable candidate for ``in'' and ``out'' Hilbert spaces. One could try instead using a covariance matrix $K'$ corresponding, e.g., to a Sobolev norm, so that the states in $\Fock_{\rm DI}^{\inn}(\mu')$ would have physically acceptable memories.\footnote{See \cite{Herdegen_1997} for a construction of a direct integral Hilbert space in the electromagnetic case with respect to Gaussian measures defined on square integrable memories. Such representations have a well-defined action of translations, but the action of Lorentz is not well-defined.} However, one would then have to give up on having a natural Lorentz group action. More significantly, there would be no reason to expect that states in $\Fock_{\rm DI}^{\inn}(\mu')$ would evolve to states in the similarly constructed $\Fock_{\rm DI}^{\rm out}(\mu')$. Of course, we also could make different choices of the precise specification of $\ms{M}$, different choices of the $\sigma$-algebra, and one could also try to use non-Gaussian measures. We certainly have not proven that no such choice could work. But we see no reason to believe that there is any such choice that would work to construct ``in'' and ``out'' Hilbert spaces with the desired properties for scattering theory.

\section{Algebraic scattering theory}
\label{sec:AlgScatt}

As we have seen in \cref{subsec:MaxMKGKFreps}, the Faddeev-Kulish construction in massive QED gives a basically satisfactory way of defining ``in'' and ``out'' Hilbert spaces in such a way that a genuine $S$-matrix should exist. However, as we found in \cref{subsec:MaxZMKGKFrep}, the analogous construction in massless QED does not work, as the required ``soft photon dressing'' gives all states an infinite expected total energy flux due to collinear divergences. As discussed in \cref{subsec:YM}, these problems persist in Yang-Mills theory, but an additional serious difficulty arises in that case due to the fact that the ``soft dressing'' itself will carry a large gauge charge-current flux, which will spoil the property that the ``dressing'' is designed to achieve. As we found in \cref{sec:Grav}, in the gravitational case the problems caused by collinear divergences are not as severe, but the problem arising from the fact that any soft graviton dressing will contribute to supertranslation fluxes is much more severe, and we proved in \cref{subsec:GravKFreps} that no analog of the Faddeev-Kulish ``in'' and ``out'' Hilbert spaces can exist in quantum gravity. Finally, we explored alternatives to the Faddeev-Kulish construction in \cref{sec:NKFreps} and found that several natural attempts do not work. It is our strong belief that in the gravitational case, no definition of ``in'' and ``out'' Hilbert spaces will satisfy conditions \ref{enum:S1}--\ref{enum:S5} of \cref{sec:intro}.

It should be emphasized that there is no difficulty in the construction of ``in'' and ``out'' states. As we found, we can construct Fock space representations of $\Alg_{\inn}^{\GR}$ for the ``in'' and ``out'' states with arbitrary choices of memory $\Delta^{\! \GR}_{AB}$. As we have noted in \cref{subsec:GravKFreps}, the representations with non-vanishing memory cannot be extended as representations of the full algebra $\Algex^{\GR}$. However, one can obtain ``in'' states by starting with any choice of smooth memory $\Delta^{\!\GR}(f)$ and considering its ``Lorentz orbit'' i.e., the space of all memories obtained by acting by Lorentz transformations on  chosen memory $\Delta^{\! \GR}(f)$. This ``orbit space'' is \emph{finite} dimensional and can be equipped with a Lorentz invariant measure (see, e.g. a similar analysis for supertranslation charges by McCarthy \cite{McCarthy1}). A direct integral of the memory Fock spaces over this orbit space yields a representation of $\Algex^{\GR}$. Applying this procedure to all smooth memories yields an enormous supply of physically acceptable states. This supply of states is certainly ample enough to encompass all of the ``in'' states that one might wish to consider, and we expect that it also would be ample enough to encompass all of the ``out'' states that arise from the dynamical evolution of these ``in'' states. Thus, the difficulties that we have elucidated in this paper do not arise from any problems with constructing ``in'' and ``out'' states nor do they arise from any problems with dynamical evolution through the bulk. They arise solely from the attempt to assemble all of the ``in'' and ``out'' states of interest into a single (separable) Hilbert space.

However, there is no reason to try to force the ``in'' and ``out'' states to live in a single Hilbert space. The algebra of asymptotic observables is entirely well-defined. As reviewed in \cref{sec:revhad}, in the algebraic viewpoint, a state is simply a positive, linear map on the algebra of observables. There is no need to specify a Hilbert space in order to define a state. The regularity conditions that we have imposed upon asymptotic states --- namely the Hadamard condition and decay conditions --- also do not require the specification of a Hilbert space. However, given a state, the GNS construction allows us to represent that state as a vector in a Hilbert space representation of the algebra. Thus, Hilbert spaces of asymptotic states may be viewed as somewhat analogous to coordinate patches on a manifold.\footnote{However, it should be kept in mind that this analogy is not perfect in that Hilbert space representations are much more rigid than coordinate patches. In particular, it is important that coordinate patches have nontrivial overlap regions, whereas irreducible Hilbert space representations will not overlap unless they coincide.} Given any point of a manifold, one can choose a coordinate patch in which it lies, and it is often very convenient to do so. Similarly, given any state on an algebra, one can choose a Hilbert space in which it lies, and it is often very convenient to do so. However, in the case of a manifold of nontrivial topology, it would not be reasonable to demand that a single coordinate patch represent all points of interest in the manifold. Similarly, in the case of scattering theory, it does not appear reasonable to demand that a single Hilbert space represent all scattering states of interest.

What would scattering theory look like in a framework where no ``in'' and ``out'' Hilbert spaces are specified at the outset? In the algebraic viewpoint, one would specify an ``in'' state $\omega_{\inn}$ as a positive linear map on the ``in'' algebra of asymptotic observables. This would consist of specifying the correlation functions of all of these observables. Of course, there is nothing stopping one from considering an ``in'' state that corresponds to a vector in the standard zero memory Fock representation $\ms{F}^{\inn}_0$ --- but one would not be forced to do so in this framework. Similarly, in massive QED, one would be allowed to ``dress'' the incoming charged particles with incoming electromagnetic states in the corresponding memory representation as in the Faddeev-Kulish construction, but it also would be allowed to consider ``bare'' incoming charged particles. Given $\omega_{\inn}$, one then computes the corresponding outgoing state $\omega_{\rm out}$ by obtaining all of its correlation functions of the ``out'' observables. Of course, this is much easier said than done, since one would not have the simplicity of the LSZ reduction, which relies, in particular, on the ability to express any ``in'' or ``out'' state in terms of local field operators acting on the Poincar\'e invariant vacuum state --- which would not be the case for states of nonzero memory. Nevertheless, if one wishes to know any particular ``out'' correlation function, it seems clear that to any finite order in perturbation theory, it must be possible to evolve this correlation function backwards into the past and express it in terms of ``in'' correlation functions, all of which would have been given in the specification of $\omega_{\inn}$. In this manner, we should, in principle, be able to determine a convex-linear\footnote{Given two algebraic states $\s$ and $\s^{\prime}$ any convex linear combination $\lambda \s + (1-\lambda)\s^{\prime}$ where $0\leq \lambda \leq 1$, gives a new state.} superscattering matrix $\$$ such that
\be
\omega_{\out} = \$ \omega_{\inn}.
\ee
Here, we have adopted the terminology ``superscattering matrix'' and the notation $\$$ from Hawking \cite{Hawking_1976} even though there are substantial differences in our motivation and framework from his. Hawking was concerned with generalizing the usual framework of scattering theory to allow pure states to evolve to mixed states (``information loss''), but he was not concerned with infrared issues and he assumed that all states lie in the folium of a single Hilbert space representation containing the Poincar\'e invariant vacuum. We are not concerned here with information loss but are similarly generalizing the framework so as to allow $\$$ to map between all regular algebraic states, not necessarily belonging to the folium of a single Hilbert space representation. In this framework, conservation of probability would be expressed by the requirement that if $\omega_{\inn}$ is any normalized ``in'' state (i.e., $\omega_{\inn} (\op{1}) = 1$), then $\omega_{\out} = \$ \omega_{\inn}$ also is normalized. If there is no information loss, then $\$$ would   take any pure algebraic ``in'' state to a pure algebraic ``out'' state. 

We note that the notion of an algebraic state is, in principal, sufficient to answer all physical questions regarding the field observables. In particular, the specification of a state $\s$ yields the expected value of all powers of $\op{N}(s)$. Since the conditions of the Hamburger moment problem (see e.g. \cite{Reed_1975,Akhizer_1965}) hold for a free field,\footnote{However, nonlinear observables such as the stress tensor in the bulk do not satisfy the required conditions on moments. Determining the probability distribution for general, nonlinear observables remains an open problem.} these moments determine the probability distribution for observing the values of this field observable. Therefore, despite the absence of a pre-chosen Hilbert space, one can determine the probability distribution of field observables.

Of course, if one is interested only in calculating the types of quantities that might be measured in collider experiments, there is no need to develop a new framework for scattering theory that properly treats infrared effects, since quantities like inclusive cross-sections surely can be calculated much more efficiently by present means than in a framework in which one takes proper account of the far infrared degrees of freedom. Nevertheless, we believe it would be of interest to further develop the ``algebraic scattering'' framework that we have sketched above.

\section*{Acknowledgements}
We would like to thank Pawe\l{} Duch for very helpful discussions on the issues and difficulties that arise when using Gaussian measures to define a direct integral Hilbert space of memory representations. This research was supported, in part by NSF grant PHY-2105878 to the University of Chicago and NSF grant PHY-2107939 to the University of California, Santa Barbara.\\



\bibliographystyle{JHEP}
\bibliography{asymp-quantization}

\end{document}